\documentclass[manuscript,screen,nonacm]{acmart}
\acmJournal{TOMM} 
\usepackage{caption}
\usepackage{subcaption}
\usepackage{bm}
\usepackage{upgreek}
\usepackage{kotex}
\usepackage{hyperref}
\usepackage{amsmath,amsfonts}
\usepackage{algorithmic}
\usepackage{algorithm}
\usepackage{array}
\usepackage{scalerel}
\usepackage{tikz}
\usepackage[dvipsnames]{xcolor}

\newcommand{\etal}{\textit{et al}.}

\newcommand{\minus}{\scalebox{0.75}[1.0]{$-$}}

\makeatletter
\DeclareRobustCommand{\cev}[1]{%
  \mathpalette\do@cev{#1}%
}
\newcommand{\do@cev}[2]{%
  \fix@cev{#1}{+}%
  \reflectbox{$\m@th#1\vec{\reflectbox{$\fix@cev{#1}{-}\m@th#1#2\fix@cev{#1}{+}$}}$}%
  \fix@cev{#1}{-}%
}
\newcommand{\fix@cev}[2]{%
  \ifx#1\displaystyle
    \mkern#23mu
  \else
    \ifx#1\textstyle
      \mkern#23mu
    \else
      \ifx#1\scriptstyle
        \mkern#22mu
      \else
        \mkern#22mu
      \fi
    \fi
  \fi
}

\makeatother

\newcommand{\dotieconcat}[2]{% auxiliary macro, don't use it directly
  \text{\raisebox{.8ex}{$\frown$}}%
}

\newcommand{\floor}[1]{\left\lfloor #1 \right\rfloor}
\newcommand{\ceil}[1]{\left\lceil #1 \right\rceil}

\AtBeginDocument{%
  }

\setcopyright{acmlicensed}
\copyrightyear{2025}
\acmYear{2025}
\acmDOI{XXXXXXX.XXXXXXX}

\begin{document}

\title{DeepHQ: Learned Hierarchical Quantizer for Progressive Deep Image Coding}

\author{Jooyoung Lee}
\email{leejy1003@etri.re.kr}
\orcid{0000-0003-0753-0699}
\affiliation{
  \institution{Electronics and Telecommunications Research Institute}
  \city{Daejeon}
  \country{Republic of Korea}
}

\author{Se Yoon Jeong}
\email{jsy@etri.re.kr}
\orcid{0000-0002-1675-4814}
\affiliation{
  \institution{Electronics and Telecommunications Research Institute}
  \city{Daejeon}
  \country{Republic of Korea}
}

\author{Munchurl Kim*}
\authornote{Corresponding author}
\email{mkimee@kaist.ac.kr}
\orcid{0000-0003-0146-5419}
\affiliation{%
  \institution{Korea Advanced Institute of Science and Technology}
  \city{Daejeon}
  \country{Republic of Korea}
}

\renewcommand{\shortauthors}{...}

\begingroup
\renewcommand\thefootnote{}\footnotetext{
© ACM 2025. This is the author’s version of the work. 
It is posted here for your personal use. Not for redistribution.  
The definitive Version of Record was published in \textit{ACM Transactions on Multimedia Computing, Communications, and Applications (TOMM)},\url{https://doi.org/10.1145/3773994}
}
\addtocounter{footnote}{-1}
\endgroup

\renewcommand{\shortauthors}{Lee et al.}

\begin{abstract}
  Research on entropy model-based learned image compression (LIC) has been actively progressing, leading to rapid advancements in coding efficiency. Beyond improvements in coding efficiency, LIC methods have also been explored for practical codec development. Despite these advancements, research on learned progressive image coding (PIC) remains in its early stages. PIC aims to encode multiple quality levels into a single bitstream, improving bitstream versatility and achieving higher compression efficiency than simulcast compression.
  Existing learned PIC methods hierarchically quantize transformed latent representations with varying quantization step sizes. More specifically, these approaches progressively compress the additional information needed for quality improvement, considering that a wider quantization interval for lower-quality compression includes multiple narrower subintervals for higher-quality compression. However, they rely on handcrafted quantization hierarchies, leading to suboptimal compression efficiency. In this paper, we propose a learned PIC method that first exploits learned quantization step sizes for each quantization layer. We also incorporate selective compression, ensuring that only essential representation components are retained in each quantization layer. Our experimental results demonstrate that the proposed method significantly enhances coding efficiency compared to the existing approaches while also reducing decoding time and model size. 
  % The source code is provided in the supplementary material.
  The source code is publicly available at \url{https://github.com/JooyoungLeeETRI/DeepHQ}
\end{abstract}





\maketitle

\section{Introduction}
\begin{figure}[!t]
    \captionsetup{belowskip=0pt}
    \captionsetup[subfigure]{font=scriptsize,labelfont=scriptsize}
    \centering
    \subfloat[]{\includegraphics[width=0.27\linewidth]{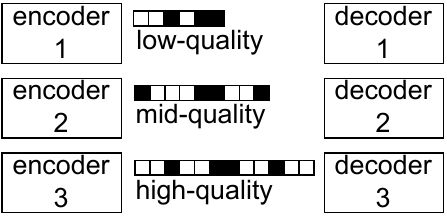}}
    % \vspace{0.2cm}
    \hspace{0.05\linewidth}
    \subfloat[]{\includegraphics[width=0.29\linewidth]{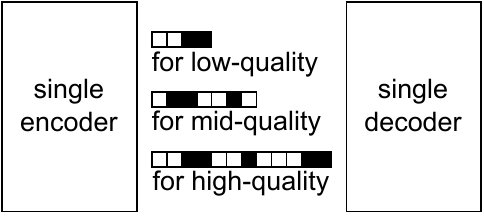}}
    % \vspace{0.2cm}
    \hspace{0.05\linewidth}
    \subfloat[]{\includegraphics[width=0.33\linewidth]{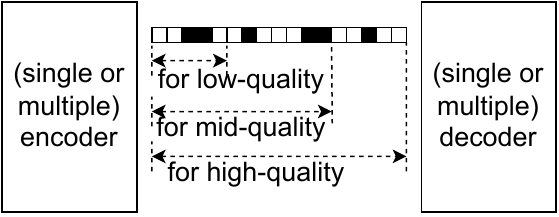}}
    \vspace{-0.2cm}\caption{Illustrations of (a) fixed-rate image coding models, (b) a variable-rate image coding model, and (c) a progressive image coding (PIC) model.}
    \label{fig:concept}
    \vspace{-0.3cm}
\end{figure}

Recently, learned image compression (LIC) methods~\cite{Toderici16,Toderici17, Johnston2018,Balle17,Theis17,Rippel2017,Balle18,Minnen2018,Lee2019,Zhou19,Chen2021,cheng2020image,Lee2020,Minnen2020,he2021checkerboard,zhu2021transformer,qian2022entroformer,he2022elic,liu2023tcm} has rapidly advanced and demonstrated performance surpassing traditional codecs, such as BPG~\cite{BPG} and JPEG2000~\cite{Taubman2001}. Currently, various research efforts are underway not only to improve performance but also to enhance usability (or functionality) from a practical perspective. One notable research area is learned progressive image coding (PIC), which aims to enable the versatile utilization of a single bitstream to accommodate various transmission and consumption environments. A progressive compression model compresses an input image into various qualities in the form of a single bitstream, as depicted in Fig.~\ref{fig:concept}-(c). Therefore, the progressive compression model offers high compression efficiency in an overall sense compared to the simulcast compression case of fixed-rate image compression models (Fig.~\ref{fig:concept}-(a)) or variable-rate image compression models (Fig.~\ref{fig:concept}-(b)) where an image is encoded into multiple separate bitstreams, each corresponding to a single quality level.
%It should be noted that the variable-rate image compression model in Fig.~\ref{fig:concept}-(b) refers to a \textit{single model} that supports compression at various compression qualities, whereas a progressive compression model (or models) in Fig.~\ref{fig:concept}-(c) encodes an input image into a \textit{single bitstream} that can be reconstructed back to images of various qualities.

\begin{figure}[!t]
    \captionsetup{belowskip=2pt}
    \captionsetup[subfigure]{font=scriptsize, labelfont=scriptsize,aboveskip=2pt}
    \centering
    \begin{subfigure}{0.51\linewidth}
        \centering
        \includegraphics[width=\linewidth]{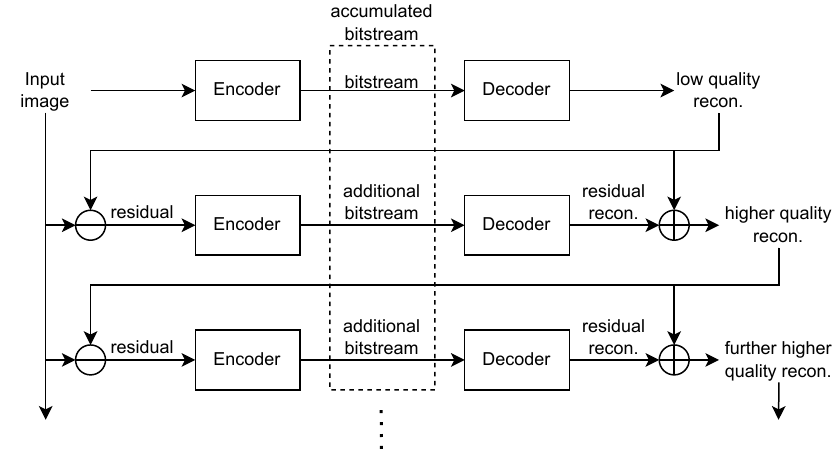}
        \caption{}
    \end{subfigure}
    \hfill
    \begin{subfigure}{0.45\linewidth}
        \centering
        \raisebox{3mm}{\includegraphics[width=\linewidth]{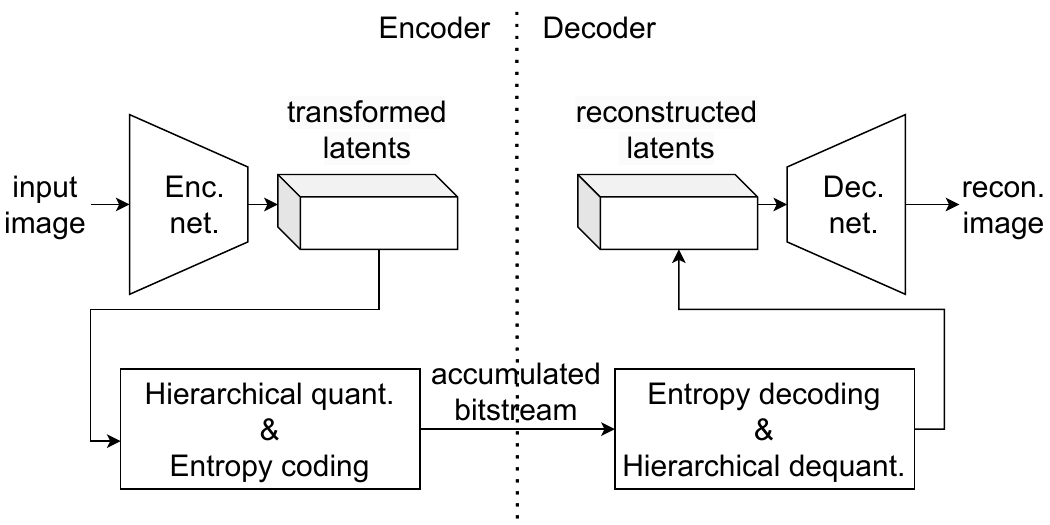}}
        \caption{}
    \end{subfigure}
    
    \vspace{-0.3cm}
    \caption{Illustrations of (a) a recurrent residual PIC scheme in the pixel domain and (b) a hierarchical quantization-based PIC scheme in latent space (feature domain).}
    \label{fig:residual_and_hierarchical_coding}
    \vspace{-0.2cm}
\end{figure}

In the early stage of learned PIC, as shown in Fig.~\ref{fig:residual_and_hierarchical_coding}-(a), some methods~\cite{Toderici16,Toderici17,Johnston2018} stacked multiple en/decoding stages, each compressing the residual signal between the input and the reconstruction of the previous compression stage. The compressed bitstream is progressively accumulated, and its corresponding reconstruction quality gets enhanced as the number of stacked en/decoding stages increases. However, this iterative residual coding increases complexity due to its repeated recurrent processes.
On the other hand, a few recent approaches~\cite{Lu_progressive_2021,Lee_2022_CVPR,Li2023deadzone,2023_CVPR_jeon} adopt hierarchical quantization for a single transformed latent representation (a feature map) with progressively decreasing quantization step sizes, as shown in Figs.~\ref{fig:residual_and_hierarchical_coding}-(b) and \ref{fig:concept_hierarchical_quantization}.
% , rather than using the recurrent residual coding. 
Specifically, for each component value in the transformed latent representation,
the hierarchical quantization allows a wider quantization interval for lower-quality compression and its nested narrower (finer) quantization subintervals for higher-quality compression, as shown in Fig.~\ref{fig:concept_hierarchical_quantization}. By doing so, only the information required for finer quantization is progressively added to the bitstream as the compression quality gets enhanced.

Although the hierarchical quantization-based PIC scheme in latent space (Fig.~\ref{fig:residual_and_hierarchical_coding}-(b)) improves coding efficiency and reduces overall complexity compared with the previous recurrent residual PIC schemes~\cite{Lu_progressive_2021,Lee_2022_CVPR,Li2023deadzone,2023_CVPR_jeon} in pixel domain (Fig.~\ref{fig:residual_and_hierarchical_coding}-(a)), the existing PIC schemes still have the following two drawbacks:
i) They use the handcrafted quantization hierarchies for all representation components, where a quantization interval is divided into three subintervals for finer quantization in its next quantization layer. Such a fixed quantization structure disregarding the characteristics of individual representation components may lead to suboptimal rate-distortion (R-D) performance (See Sec.~\ref{sec:background} for further details);
ii) They encode all representation components at all quantization layers into the bitstream, which implies that, as reported in the field of variable-rate image coding~\cite{lee2022selective}, encoding all representation components regardless of the target compression quality can lead to suboptimal performance in both compression efficiency and complexity.

On this basis, we introduce a novel learned hierarchical quantizer, called DeepHQ, which exploits learned quantization step sizes for each quantization layer in PIC. To this end, (i) we determine the quantization step sizes by applying distinct learnable step sizes to each quantization layer for a single latent representation transformed by the encoder of a single existing compression network. After computing the rate and distortion for each layer, training is performed through joint optimization over all layers (Sec.\ref{sec:training}). (ii) For PIC, we introduce layer-wise quantization and dequantization processes based on the learned quantization step sizes, including a nested interval determination process and a probability estimation scheme for entropy coding of quantized representations (Sec.\ref{sec:hierarchical_quantization}). iii)
In addition, our DeepHQ further improves the compression efficiency and reduces the model complexity by compressing only the essential representation elements for each quantization layer (Sec.~\ref{sec:selective}). Building upon the concept of learning quantization step sizes inspired by variable-rate image compression methods~\cite{Cui2021,lee2022selective}, the key contribution of this paper is their hierarchical exploitation for PIC, realized through (ii) a progressive nested quantization process and (iii) selective coding.

Through extensive experiments on various datasets, our DeepHQ achieves 11.97\% average bit savings over the current state-of-the-art (SOTA) model~\cite{2023_CVPR_jeon}, while requiring only 14.19\% of its model size and 8.72\% of its decoding time.
It should also be noted that our DeepHQ model utilizes only a single trained model for progressive coding across all bit-rate ranges, making it both a progressive coding model and a variable-rate model. In contrast, the competing SOTA method~\cite{2023_CVPR_jeon} employs individually trained multiple refinement sub-networks, each dedicated to one of the predefined bit rate ranges, leading to a much larger whole model size. Our contributions are summarized as follows:
\begin{itemize}
\vspace{1mm}
\item We \textit{firstly} propose a learned hierarchical quantizer with the learned quantization step sizes via learning, called DeepHQ, for learned PIC, resultantly offering superior compression efficiency.
\vspace{1mm}
\item We further improve the progressive coding efficiency and significantly reduce the model complexity by incorporating the learned hierarchical quantization and the selective coding of latent representations into progressive neural image coding. 
\item Our DeepHQ (w/ a single base compression model) achieves 11.97\% higher coding efficiency than the best state-of-the-art progressive coding method (w/ multiple subnetworks with different target bitrate ranges), only with 14.19\% of its model size and 8.72\% of the decoding time, on average.
\end{itemize}

\section{Related work}
\label{sec:relatedwork}
\noindent \textbf{Learned image compression (non-progressive)}.
The LIC field is broadly divided into i) dimension reduction-based approaches~\cite{Toderici16,Toderici17,Johnston2018} aiming to pack as much information as possible into a small representation, and ii) entropy minimization approaches~\cite{Balle17,Theis17,Balle18,Minnen2018,Lee2019,Zhou19,Chen2021,cheng2020image,Lee2020,Minnen2020,he2021checkerboard,zhu2021transformer,qian2022entroformer,he2022elic,liu2023tcm} aiming to minimize (cross) entropy of latent representations while also minimizing distortions of reconstructions.
In the early stage, research was actively conducted in both areas, but currently, most approaches are based on entropy minimization due to its performance advantages.

Balle~\etal~\cite{Balle17} and Theis~\etal~\cite{Theis17} proposed the first entropy minimization-based image compression methods. They utilized entropy models (distribution approximation models) for latent representations to calculate the rate term (cross-entropy of latent representations), and they performed joint optimization in an end-to-end manner to simultaneously minimize rate and distortion. For the entropy models, Balle~\etal~\cite{Balle17} adopted linear spline models while Theis~\etal~\cite{Theis17} used Gaussian scale mixture models. In contrast to the first two models~\cite{Balle17, Theis17} that directly optimize the distribution parameters, Balle~\etal~\cite{Balle18} proposed the Hyperprior model in which model parameters are not directly learned but are instead adaptively compressed through the hyper-en/decoder networks. 

Subsequently, Minnen~\etal~\cite{Minnen2018} and Lee~\etal~\cite{Lee2019} regarded the spatial correlation existing within the latent representation as redundancy from a compression perspective and proposed autoregressive models to mitigate it. Specifically, they utilized the previously reconstructed neighboring latent representation components in a raster scanning order environment to predict the distribution parameters of the current element, ensuring a higher distribution approximation accuracy. 
To further improve coding efficiency, Cheng~\etal~\cite{cheng2020image} and Lee~\etal~\cite{Lee2020} utilized Gaussian mixture models instead of single Gaussian models and deeper en/decoder networks. 
In these autoregressive schemes, Chen~\etal\cite{Chen2021} first exploited the non-local attention blocks for en/decoder networks and hyper en/decoder networks, and Li~\etal\cite{Li2023nonlocal} introduced a special non-local operation for context modeling by employing the global similarity.
To mitigate the high complexity of decoding in the early-stage autoregressive models~\cite{Minnen2018,Lee2019,cheng2020image,Lee2020,Chen2021} while keeping the advantages of autoregression as much as possible, some approaches~\cite{li2020,Minnen2020,he2021checkerboard} used unique forms of autoregression methods. Li~\etal~\cite{li2020} introduced a 3-D zigzag scanning order and a 3-D code-dividing technique that enables better parallel entropy decoding. Minnen~\etal~\cite{Minnen2020} divided the latent representation into a few slices along the channel direction and performed autoregression between these slices. He~\etal~\cite{he2021checkerboard} divided the latent representation into two subsets in a spatial checkerboard pattern and predicted the model parameters of one subset based on the other subset. Subsequently, Xu~\etal~\cite{Xu2025universal} proposed a method to optimize latent representation at the encoding stage in a learned compression network, Jin~\etal~\cite{Jin2025regional} introduced a correspondence structure between each part of a bitstream and the spatial domain of its reconstructed image, enabling regional decoding, and Jiang~\etal\cite{jiang2023mlic} proposed Multi-Reference Entropy Model (MEM) which
captures local spatial, global spatial, and channel contexts simultaneously.
More recently, some studies~\cite{zhu2021transformer,qian2022entroformer,Kim_2022_CVPR,liu2023tcm,Koyuncu2024,Jiang2025mlic} have proposed replacing the traditionally dominant CNN-based architectures with Transformer~\cite{vaswani2017transformeer}. Zhu~\etal~\cite{zhu2021transformer} replaced all convolutions in en/decoder networks with Swin Transformer~\cite{liu2021swin} blocks and Qian~\etal~\cite{qian2022entroformer} utilizes a self-attention stack to replace the hyper en/decoder networks. Kim~\etal~\cite{Kim_2022_CVPR} proposed an entropy model called Information Transformer that exploits both global and local dependencies to replace the hyper-encoder and -decoder. Liu~\etal~\cite{liu2023tcm} utilized a parallel Transformer-CNN Mixture (TCM) block to incorporate the advantages of CNN and transformers. Koyuncu~\etal~\cite{Koyuncu2024} introduced a computationally efficient transformer-based autoregressive context model called eContextformer. Jiang~\etal~\cite{Jiang2025mlic} proposed an advanced multi-reference entropy model (MEM) that incorporates linear-complexity global spatial context modeling, enabling efficient compression of high-resolution images without sacrificing rate-distortion performance.

\noindent \textbf{Learned progressive image coding}.
Although various research efforts are ongoing to improve the practicality of LIC, learned PIC is still relatively under-explored. Initially, a few methods~\cite{Toderici16,Toderici17,Johnston2018} repeatedly compress and reconstruct the residual between the lower-quality reconstruction and the original input, thus progressively enhancing the compression quality as the number of iterations increases. 
Park~\etal~\cite{Park2023color} introduced a scalable color quantization method, but they mainly focussed on the scalability of color bit-depth rather than the quality-scalability.
Recently, 
Lu~\etal~\cite{Lu_progressive_2021} and Lee~\etal~\cite{Lee_2022_CVPR} adopted hierarchical quantization, in which they perform an encoding transformation only once and apply progressively decreasing quantization step sizes to the transformed latent representations as the compression quality get improved. Both approaches utilized a handcrafted quantization hierarchy with fixed reduction ratios of quantization step sizes between the quantization layers. In addition, both methods adopt fined-grained component-wise progressive coding where representation elements are sequentially compressed. Similarly, Li~\etal~\cite{Li2023deadzone} introduced a learned progressive coding model based on a handcrafted quantization hierarchy using dead-zone quantizers.
More recently, Jeon~\etal~\cite{2023_CVPR_jeon} proposed an extended method called context-based trit-plane coding (CTC)~\cite{2023_CVPR_jeon} that improves the coding efficiency of DPICT~\cite{Lee_2022_CVPR} by adding two types of separate network modules, the context-based rate reduction (CRR) and context-based distortion reduction (CDR), that refine the estimated distribution parameters and reconstructed latent representations, respectively. However, the architectures of the CRR and CDR modules are highly complex. Furthermore, these two modules utilize a total of six models, each of which is dedicated to one of the three predefined bit-rate ranges, thus causing an extremely high number ($\sim$400 million) of model parameters.

\noindent \textbf{Partial compression of latent representations}.
Meanwhile, some LIC approaches~\cite{li2018learning,mentzer2018conditional1,lee2022selective} adopted partial coding of latent representations to improve coding efficiency and to reduce computational complexity at the same time. Li~\etal~\cite{li2018learning} and Mentzer~\etal~\cite{mentzer2018conditional1} adopted 2-D importance maps to represent the spatial importance of representations, which allows for spatially different bit allocations in different regions. According to the 2-D importance map, they determine the number of channels to be involved in the compressed bit-streams. Whereas, Lee~\etal~\cite{lee2022selective} introduced a more generalized 3-D importance map that represents component-wise inherent importance of representations for variable-rate image coding. According to the target quality, the 3-D importance map is adjusted with a learned and dedicated adjustment vector to determine essential representation elements.

\section{Background and motivations for learned quantization}
\label{sec:background}
\begin{figure}[!t]
\captionsetup{belowskip=5pt}
\begin{center}
\includegraphics[width=0.7\linewidth]{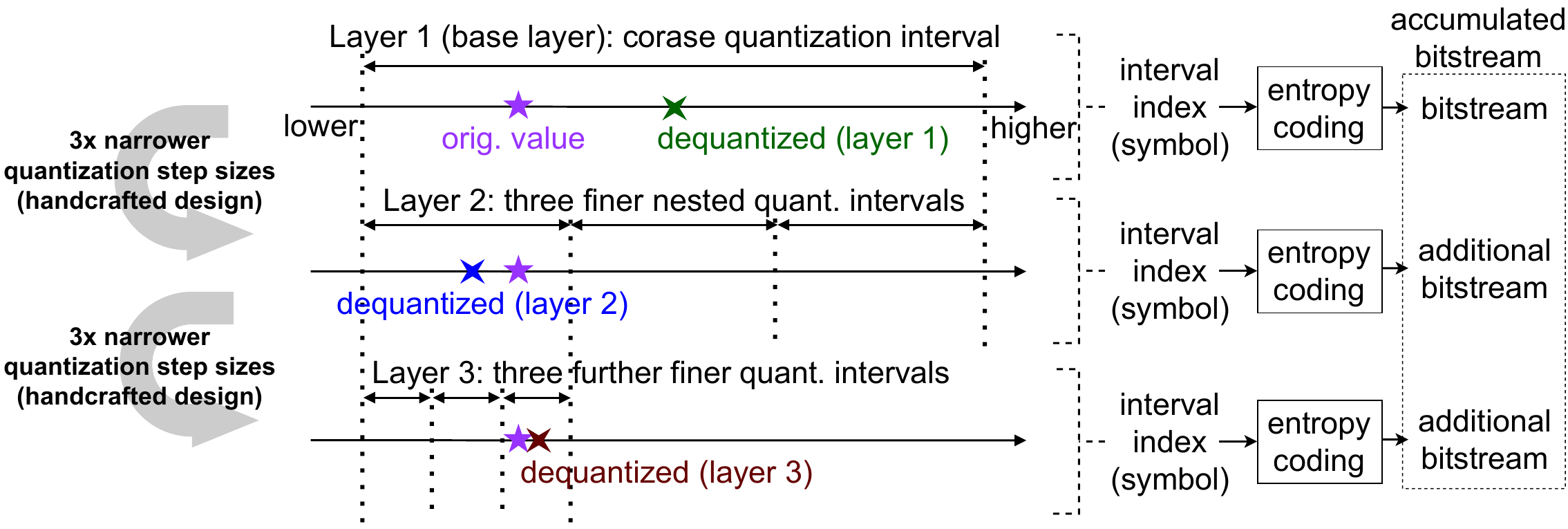}
\end{center}
\vspace{-0.3cm}\caption{Illustration of the existing handcrafted hierarchical quantization process.
}
\label{fig:concept_hierarchical_quantization}
\vspace{-0.5cm}
\end{figure}

In the recent learned PIC approaches~\cite{Lu_progressive_2021,Lee_2022_CVPR,Li2023deadzone,2023_CVPR_jeon}, an encoder network transforms input image $\bm x$ into latent representation $\bm y$, and $\bm y$ is hierarchically quantized with different quantization step sizes in a coarse-to-fine manner as quantization layers recursively get deeper (higher) from the first layer (Layer 1 in Fig.~\ref{fig:concept_hierarchical_quantization}.) with the coarsest quantization step sizes. 
Specifically, the quantization interval of the lower quantization layer containing the original value is divided into three parts for its upper quantization layer, as shown in Fig.~\ref{fig:concept_hierarchical_quantization}. The quantized interval index is converted into a bitstream through entropy coding, and this process repeats from the base quantization layer to higher layers, generating additional bitstreams.
The decoding process also proceeds in the order of lower-to-higher quantization layers, but the tasks of entropy-decoding and dequantization in each quantization layer are performed in the reverse order of the encoding process.
For entropy coding, where the probability estimations for quantization intervals in each quantization layer are necessary, the entropy model $p(\bm y)$ is utilized. The entropy model $p(\bm y)$ is a learnable approximation model for the distribution of $\bm y$, where the distribution parameters can be estimated via a neural network~\cite{Balle18,Minnen2018,Lee2019,cheng2020image} or be directly learned~\cite{Balle17,Theis17} in the end-to-end neural image compression fields. As the entropy model $p(\bm y)$, the existing methods~\cite{Lu_progressive_2021,Lee_2022_CVPR,Li2023deadzone,2023_CVPR_jeon} adopt the hyper en/decoder model~\cite{Balle18} where $\bm y$ is transformed (and compressed) into side information $\bm z$ via the hyper-encoder network. From $\bm z$, the estimated Gaussian distribution parameters $\bm \mu$ and $\bm \sigma$ of $p(\bm y)$ are reconstructed via the hyper-decoder network.

As aforementioned, the existing hierarchical quantization-based PIC methods~\cite{Lu_progressive_2021,Lee_2022_CVPR,Li2023deadzone,2023_CVPR_jeon} adopt the approach to dividing each quantization interval into three subintervals for all components of the latent representation $\bm y$.
However, such a fixed number and structure of quantization subintervals, which are independent of the target reconstruction qualities and the signal characteristics of representations, may lead to suboptimal performance.
For example, finer quantization is preferable at lower layers for components containing global and structural information, whereas higher layers may benefit more from finer quantization for texture details. This motivated us to develop DeepHQ, which accommodates various learned quantization step sizes for different components across layers.

Note that all existing works~\cite{Lu_progressive_2021,Lee_2022_CVPR,Li2023deadzone,2023_CVPR_jeon} on PIC, including our DeepHQ, were developed targeting non-autoregressive models. This is because supporting autoregressive component-wise PIC functionality within each quantization layer would not only incur a substantial computational overhead—since the decoder would need to predict probability distributions every time a component to be transmitted is processed—but also require designing a completely new component-wise probability prediction model with a structure entirely different from the original autoregressive model of the base compression codec. Research on such autoregressive component-wise PIC techniques is beyond the scope of this work and is considered a subject for future study.

\begin{figure}[t]
    \captionsetup[subfigure]{font=scriptsize, labelfont=scriptsize,aboveskip=2pt}
    \setlength\columnsep{0pt}
    \captionsetup{belowskip=10pt}

    \centering
    \begin{subfigure}{0.75\textwidth}
        \centering
        \includegraphics[width=\linewidth, clip, trim={0cm 0cm 0cm 0.25cm}]{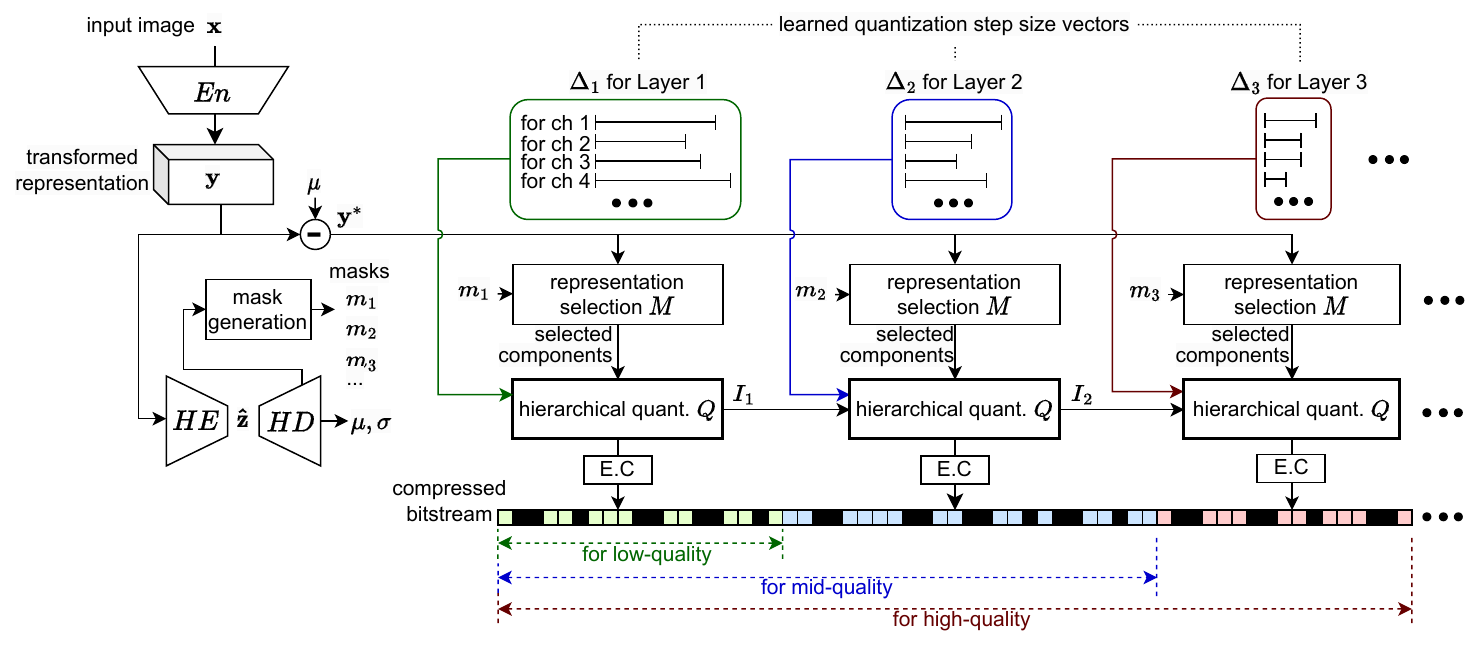}
        \vspace*{-0.3cm}
        \hspace*{30mm}
        \caption{}
    \end{subfigure}
    \vspace*{-0.3cm}

    \begin{subfigure}{0.7\textwidth}
        \centering
        \includegraphics[width=\linewidth, clip, trim={0cm 0cm 0cm 0.25cm}]{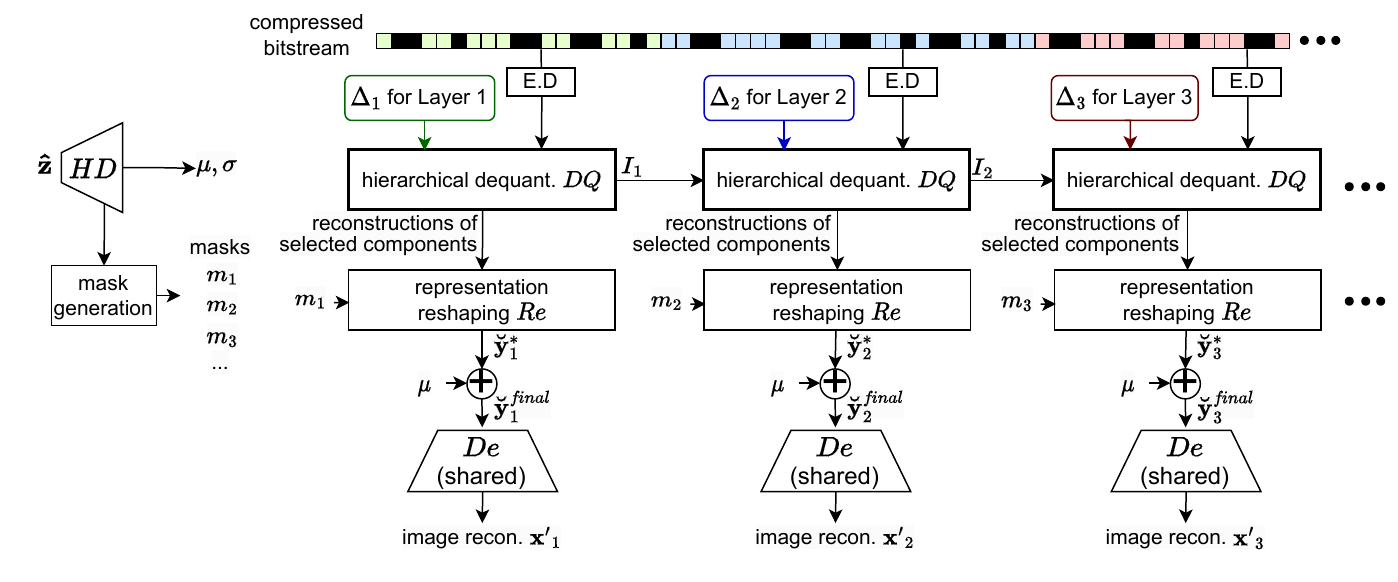}
        \vspace*{-0.5cm}
        \hspace*{30mm}
        \caption{}
    \end{subfigure}    
    \vspace{-0.5cm}
    \caption{(a) Overall encoding procedure of DeepHQ. Only essential representation components of a representation $\bm y^*$ are selected for each quantization layer, and then the selected representation components are hierarchically quantized and entropy-coded utilizing quantization step sizes learned for each quantization layer. (b) Overall decoding procedure of DeepHQ. The hierarchical dequantization process and the reshaping of restored representation components are conducted in response to the operation of the encoder. The detailed operation flowcharts of the two key elements, $Q$ and $DQ$, highlighted with bold boxes, are provided in Fig.~\ref{fig:q_and_dq}. Encoder, Decoder, hyper-encoder, and hyper-decoder networks are denoted as $En$, $De$, $HE$, and $HD$, respectively. Representation selection mask $m(\bm{\hat z}, l)$ in Eq.~\ref{eq:selection_process} is abbreviated as $m_l$. Note that the compression and decompression processes for hyperprior representation $\bm {\hat z}$ are omitted for briefness, for which we adopt the Hyperprior model~\cite{Balle18}.}
    \label{fig:overall_architecture}
    \vspace{-0.5cm}
\end{figure}

\begin{figure*}[t]
    \captionsetup{belowskip=5pt}
    \captionsetup[subfigure]{font=scriptsize, labelfont=scriptsize}
    \setlength\columnsep{0pt}
    \centering
    \begin{subfigure}{0.4\textwidth}
        \centering
        \includegraphics[width=\linewidth, clip, trim={0cm -0.2cm 0cm -0.3cm}]{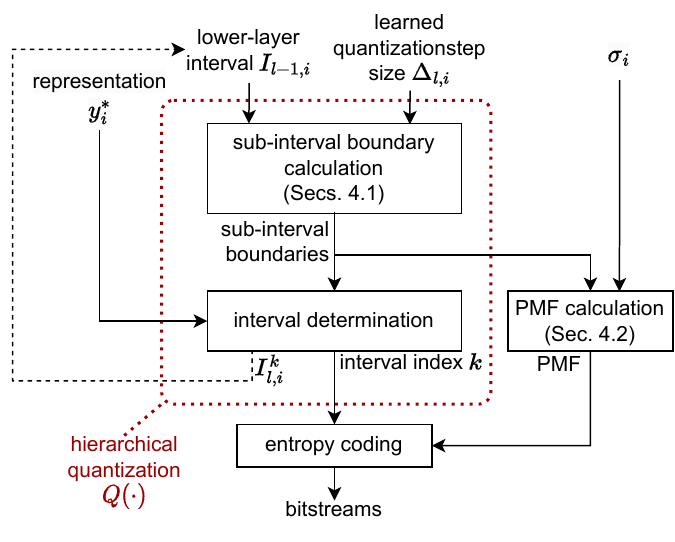}
        \vspace*{-0.7cm}
        \hspace*{0mm}
        \caption{}
    \end{subfigure}
    \hspace{0.1\textwidth}
    \begin{subfigure}{0.4\textwidth}
        \centering
        \includegraphics[width=\linewidth, clip, trim={0cm -0.6cm 0cm -0.5cm}]{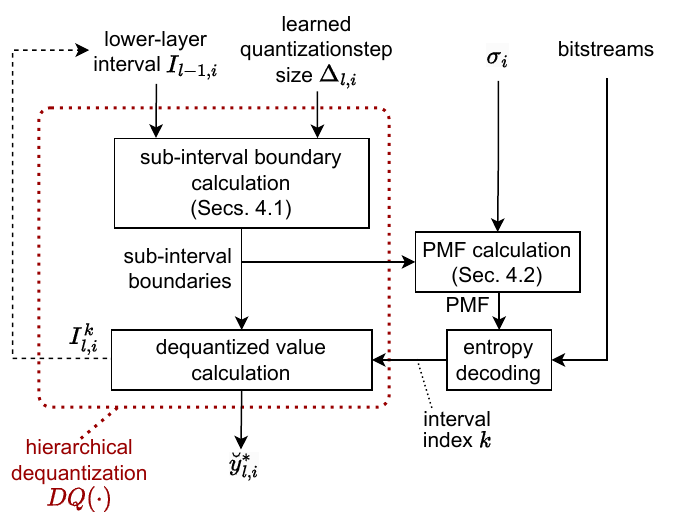}
        \vspace*{-0.7cm}
        \hspace*{0mm}
        \caption{}
    \end{subfigure}
    \vspace{-0.5cm}
    \caption{(a) Detailed flowchart of the hierarchical quantization process $Q$. The quantized interval information is recursively used in its upper layer to determine the finer quantization subintervals. (b) Detailed flowchart of the hierarchical dequantization process $DQ$.}
    \label{fig:q_and_dq}
    \vspace{-0.5cm}
\end{figure*}

\section{Hierarchical quantization with learned step sizes}
\label{sec:hierarchical_quantization}
The proposed DeepHQ has two distinctive features: i) our DeepHQ utilizes learned quantization step sizes for each quantization layer and representation component and ii) our DeepHQ encodes only essential representation components in each layer. In this section, we first introduce the i) hierarchical quantization, and then we extend our method with the ii) selective compression of representations in Sec.~\ref{sec:selective}.

Our DeepHQ model consists of $L{=}8$ quantization layers, each using a dedicated set of quantization step sizes, denoted as $\bm{\Updelta}_l=\{\Updelta_{l,c}\ |\ c=\{1, 2, ..., C_y\}\}$, where $C_y$ represents the total number of channels in $\bm y$. That is, each $\bm{\Updelta}_l$ contains the channel-wise quantization step sizes for the $\bm{y}=En(\bm x)$, where $En(\cdot)$ denotes the encoding transform function (via the encoder network) and $\bm x$ represents the input image. 
As shown in Fig.~\ref{fig:overall_architecture}, the proposed DeepHQ performs progressive encoding and decoding in the order of lower-to-higher quantization layers.
Generally, via the optimization described in Sec.~\ref{sec:training}, larger $\bm{\Updelta}_l$ values are learned for the lower quantization layer (lower-quality compression).
The hierarchical quantization process at the \textit{l}-th quantization layer on the encoder side is represented as follows:
\begin{align}
\label{eq:Q_function}
\bm{k} = Q(\bm y^*, \bm{\Updelta}_l, \bm{I}_{l-1}), \text{\ \ \ \ \ with } \bm y^* = \bm y - \bm \mu,
\end{align}
where $\bm{y}^*$ is an unbiased representation obtained by shifting the representation $\bm y$ by the estimated $\bm \mu$, $\bm{I}_{l\minus1}$ represents the lower-level ($l\minus1$) quantization intervals of the $\bm{y}^*$, $\bm{k}$ denotes the indexes of the current level ($l$) quantization intervals corresponding to the original values in $\bm{y}^*$, and $Q(\cdot)$ is the hierarchical quantization process that will be described below. Note that the resulting indexes in $\bm{k}$ are entropy-coded into the bitstream in a lossless manner.
The hierarchical dequantization process at the \textit{l}-th quantization layer on the decoder side is represented as follows:
\begin{align}
\label{eq:DQ_function}
\bm{\breve y}^*_l = DQ(\bm{k}, \bm{\Updelta}_l, \bm{I}_{l-1}),
\end{align}
where $\bm{\breve y}^{*}_l$ denotes the dequantized reconstruction for $\bm{y}^*$, and $DQ(\cdot)$ is the hierarchical dequantization process. Note that this process follows the entropy-decoding process to reconstruct $\bm{k}$. Subsequently, the decoder determines the final representation $\bm{\breve y}^\textit{final}_l$ that is fed into the decoder network as follows:
\begin{align}
\label{eq:y_breve_final}
\bm{x'}_l = De(\bm{\breve y}^\textit{final}_l), \text{\ \ \ \ \ with  } \bm{\breve y}^\textit{final}_l = \bm{\breve y}^*_l + \bm \mu,
\end{align}
where $\bm{x'}_l$ is a reconstruction image of the \textit{l}-th quantization layer, and $De(\cdot)$ is the decoding transform function (via the decoder network). $Q(\cdot)$ and $DQ(\cdot)$ are key elements of our work that progressively quantize and dequantize $\bm y^*$ more finely as the quantization layer increases, recursively utilizing its lower-layer intervals $\bm{I}_{l-1}$ of $\bm y^*$, along with the learned quantization step sizes in $\bm{\Updelta}_l$, as shown in Fig.~\ref{fig:q_and_dq}. It should be noted that the processes in Eqs.~\ref{eq:Q_function} and ~\ref{eq:DQ_function} omit the selective compression process for brevity, while the full process, including selective compression, is described in Eq.~\ref{eq:selection_process}. 
In addition, from now on, we represent the quantization processes component-wise for simplicity and better understanding, but in practice, we entirely use array operations.

As shown in Fig.~\ref{fig:q_and_dq}, the $Q(\cdot)$ in the encoder and the $DQ(\cdot)$ in the decoder share the same processes for determining the subinterval boundaries (Sec.~\ref{sec:boundary_calculation}) and computing the approximated probability mass functions (PMFs) (Sec.~\ref{sec:PMF_calculation}), both of which will be described in detail later in this section. With the determined subinterval boundaries, the $Q(\cdot)$ simply determines the subinterval index $k$ as follows:
\begin{align}
\label{eq:interval_index_calculation}
k \text{ such that } y^*_i \in I^k_{l,i}, &\text{\ \ \ \ \ with } I^k_{l,i} = [B^{(k)}_{l,i}, B^{(k+1)}_{l,i}),
\end{align}
where $y^*_i$ is the \textit{i}-th component of $\bm y^*$ and $k$ represents the index of the quantization subinterval $I^k_{l,i}$ for $y^*_i$ in the \textit{l}-th quantization layer and two boundaries of $I^k_{l,i}$ are denoted as $B^{(k)}_{l,i}$ and $B^{(k+1)}_{l,i}$, respectively. 
The subinterval index $k$ is entropy-coded into the bitstream with its approximate PMF.
On the decoder side, the bitstream is entropy-decoded with the same PMF to obtain the subinterval index $k$, and then $DQ(\cdot)$ dequantizes the reconstructed subinterval index $k$, as follows: 
\begin{align}
\label{eq:recon_interval}
\breve y^*_{l,i} {=} (B^{(k)}_{l,i} + B^{(k+1)}_{l,i}) / 2,
\end{align}
where the subinterval boundaries are determined in the same manner as in the encoder. We further describe how we determine the subinterval boundaries (Sec.~\ref{sec:boundary_calculation}) and the PMF for the quantization subintervals (Sec.~\ref{sec:PMF_calculation}) as follows:
\subsection{Hierarchical subinterval boundary calculation}
\label{sec:boundary_calculation}
An important issue when using the learned quantization step sizes for progressive coding is that, in most cases, the quantization step size of a lower layer is not an integer multiple of the quantization step size of an upper layer, which leads to boundary misalignment between adjacent quantization layers. To address this, we utilize the boundary clipping and boundary adjustment techniques as follows.

\begin{figure}[t!]
\begin{center}
\includegraphics[width=0.6\linewidth,clip, trim={0cm 0cm 2.5cm 0cm}]{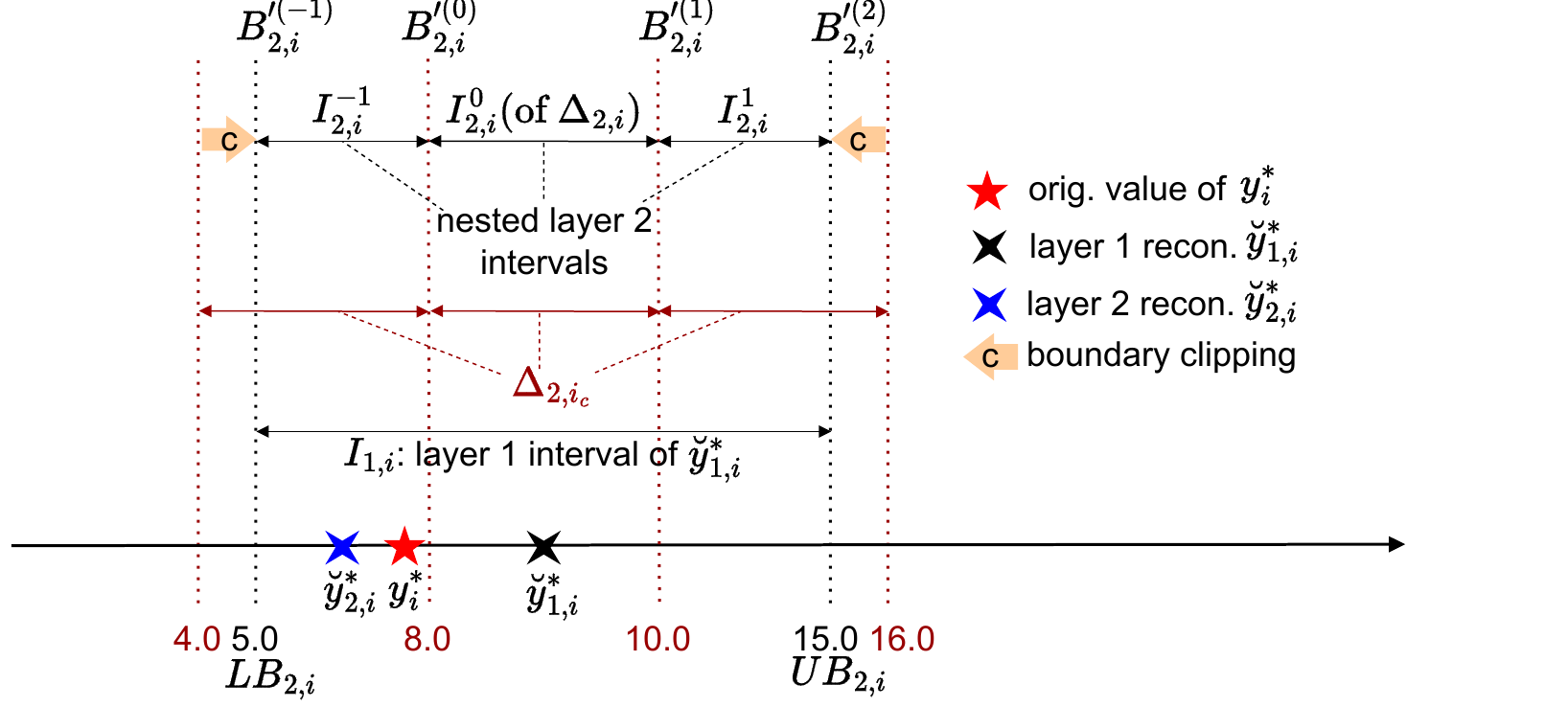}
\end{center}
\vspace{-0.2cm}\caption{Illustration of the interim boundary ($\bm {B'}_{l,i}$) determination.}
\label{fig:vanilla}
\vspace{-0.5cm}
\end{figure}

\noindent \textbf{Interim boundary calculation}. We first determine the interim subinterval boundary set $\bm {B'}_{l,i}$ for $y^*_i$ at the \textit{l}-th quantization layer using the leanred step sizes as follows:
\begin{align}
\label{eq:boundary_calculation}
  \bm {B'}_{l,i} = \{\text{clip}((j-0.5) &\times \Updelta_{l,i_c} \ + \ \breve y^*_{l-1,i},\ LB_{l,i},\ UB_{l,i})\ \ |\ \ j \in \mathbb{Z}, -J \le j \le J\}\\
  &\text{with  } LB_{l,i} = \text{min}(I_{l-1,i}), \ \ UB_{l,i} = \text{max}(I_{l-1,i}),\notag
\end{align}
where clip($\cdot$) denotes the clipping process and $i_c$ represents the channel index of the representation $y^*_i$. With Eq.~\ref{eq:boundary_calculation}, the distances between adjacent values in $\bm B'_{l,i}$ are basically equal to $\Updelta_{l,i_c}$.
However, in case a boundary falls outside of the lower layer interval $I_{l\minus1,i}$, the clipping process is performed to remove redundancy between quantization layers from the compression perspective, as shown in Fig.~\ref{fig:vanilla}. Note that some intervals among the \textit{nominal} $2J+1$ subintervals can have zero width due to the clipping operation in Eq.~\ref{eq:boundary_calculation}. Considering subintervals with widths greater than zero to be \textit{valid} subintervals, the number of the valid subintervals, denoted as $N^{\bm {B'}_{l,i}}$, is less than or equal to $2J+1$.
 
With Eq.~\ref{eq:boundary_calculation}, the interim subinterval boundaries of the upper layer are symmetrically arranged around the lower layer reconstruction $\breve y^*_{l-1,i}$. We empirically found that this boundary arrangement shows much better coding efficiency than the mid-tread (zero-centered) arrangement (See Fig.~\ref{fig:ablation} for the results).
For the first ($l{=}1$) quantization layer, we set $LB_{1,i}$, $UB_{1,i}$, and its virtual lower-layer reconstruction $\breve y^*_{0,i}$ values to $\minus \Updelta_{1,i_c} \times J$, $\Updelta_{1,i_c} \times J$, and 0, respectively. We empirically set $J$ to $\text{absmax}(\bm{y^*} / \bm{\Updelta}_1)$ so that it can cover all valid subintervals. 

\begin{figure}[!t]
    \captionsetup[subfigure]{font=scriptsize,labelfont=scriptsize}
    \centering
    \subfloat[]{\includegraphics[width=0.52\linewidth]{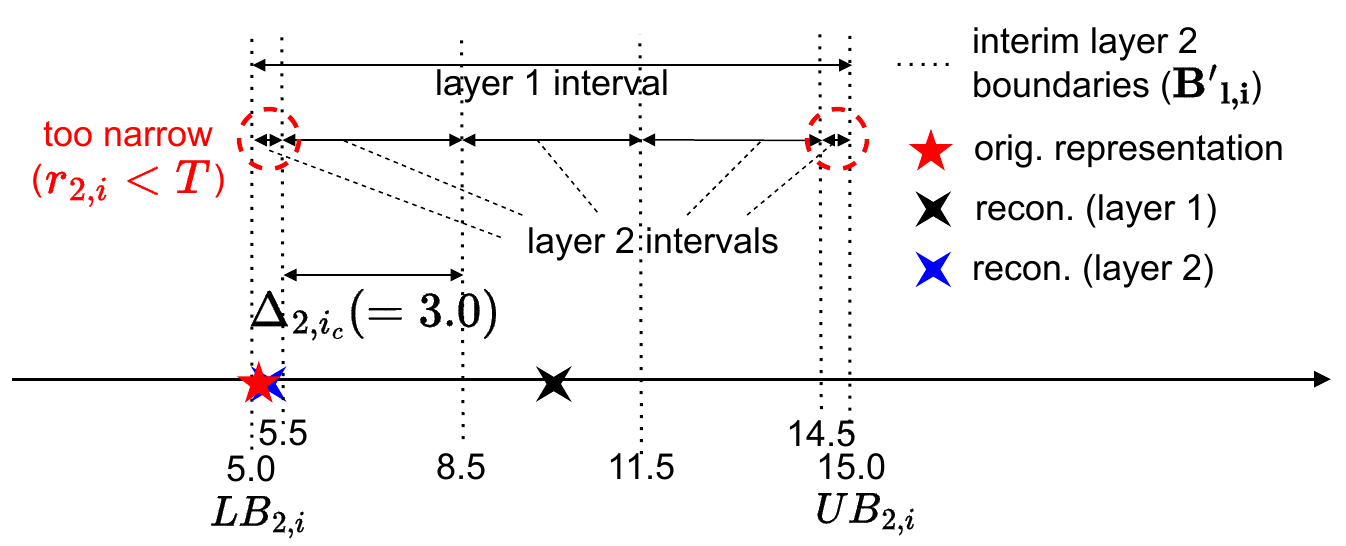}\vspace{-0.2cm}}
    \hfil
    \subfloat[]{\includegraphics[width=0.52\linewidth, trim={0cm 0cm 0cm -0.4cm}]{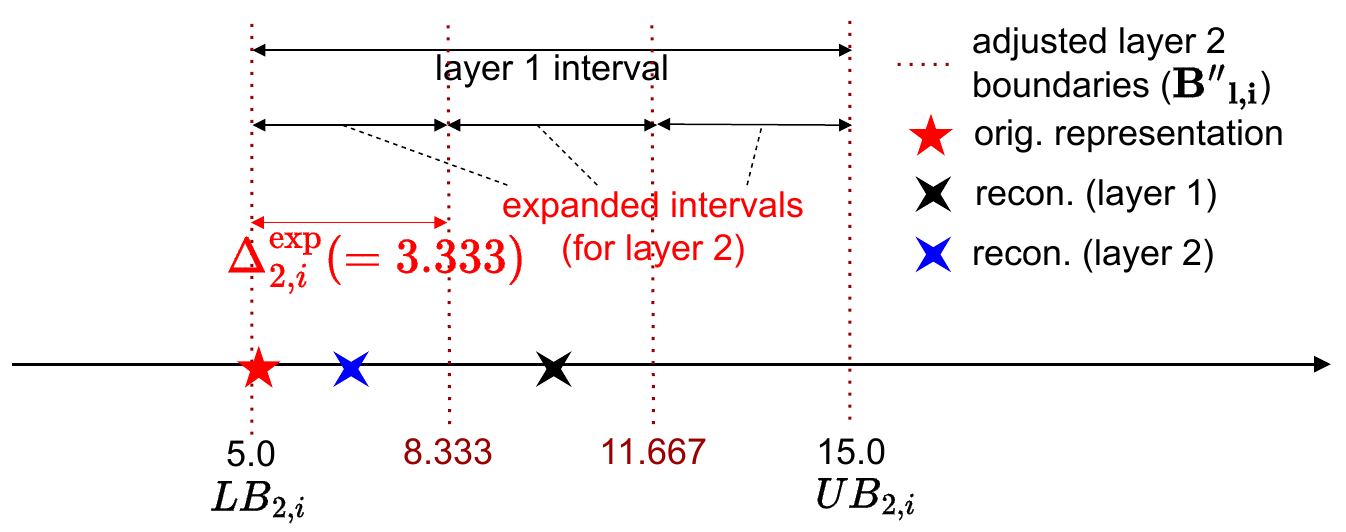}\vspace{-0.2cm}}
    \vspace{-0.3cm}
    \caption{Examples of the quantization boundaries (a) without and (b) with the boundary adjustment.}
    \label{fig:boundary_adjustment}
    \vspace{-0.2cm}
\end{figure}

\noindent \textbf{Boundary adjustment}. In the interim subinterval boundary set $\bm {B'}_{l,i}$, some subinterval fragments, the first and last valid subintervals with clipped step sizes, often tend to be significantly narrower than the learned normal quantization step sizes $\Updelta_{l,i_c}$.
Although these narrow subintervals, as shown in Fig.~\ref{fig:boundary_adjustment}-(a), can reduce quantization error to some extent, they come at the cost of a higher bit rate, significantly degrading overall compression efficiency in an R-D sense. To mitigate it, as shown in Fig.~\ref{fig:boundary_adjustment}-(b), our DeppHQ adaptively performs the boundary adjustment that can avoid severely narrowed subintervals. Specifically, when the ratio between the step size of the first (last) subinterval and the normal step size $\Updelta_{l,i_c}$ is below a certain threshold $T$, we adaptively exploit the expanded boundaries $\bm {B''}_{l,i}$ as follows:
\begin{align}
\label{eq:inteval_expansion}
  &\bm{B}_{l,i} = \begin{cases}
    \bm {B''}_{l,i}, & \text{if $r_{l,i} < T$}.\\%[1.5ex]
    \bm {B'}_{l,i}, & \text{otherwise.}\\%[1.5ex]
    \end{cases},
\end{align}
where the final boundary set $\bm{B}_{l,i}$ is selectively determined depending on $r_{l,i}$, the ratio of the first (last) valid subinterval width in $\bm{B'}_{l,i}$ compared to $\Updelta_{l,i_c}$. If $r_{l,i}$ is smaller than the threshold $T$, the expanded boundary set $\bm {B''}_{l,i}$ is used instead of $\bm{B'}_{l,i}$. The expanded boundary set $\bm {B''}_{l,i}$ is determined in the same manner as in Eq.~\ref{eq:boundary_calculation}. However, it is computed using the expanded step size $\Updelta^\text{exp}_{l,i}$ instead of $\Updelta_{l,i_c}$. The value of $\Updelta^\text{exp}_{l,i}$ is determined as follows:
\begin{align}
\label{eq:expanded_boundaries}
\Updelta^\text{exp}_{l,i} = \frac{UB_{l,i} - LB_{l,i}}{N^{\bm{B'}_{l,i}}-2},
\end{align}
where $N^{\bm{B'}_{l,i}}$ denotes the number of valid subintervals in the first interim boundary set $\bm{B'}_{l,i}$ (e.g. $N^{\bm{B}_{l,i}}{=}5$ in Fig.~\ref{fig:boundary_adjustment}-(a)).
Note that $\Updelta_{l,i_c}$ is shared across all components within the same channel of $\bm y^*$, whereas $\Updelta^\text{exp}_{l,i}$ is adaptively determined for each $y^*_i$.
In Eq.~\ref{eq:expanded_boundaries}, the denominator indicates the adjusted number of quantization subintervals inside the range from $LB_{l,i}$ to $UB_{l,i}$ (e.g. $N^{\bm{B'}_{l,i}}\minus2{=}3$ in Fig.~\ref{fig:boundary_adjustment}-(b)).
Accordingly, the range between $LB_{l,i}$ and $UB_{l,i}$ has an integer number of $\Updelta^\text{exp}_{l,i}$-sized subintervals. This adaptive boundary adjustment significantly improves coding efficiency, as seen in Sec.~\ref{sec:quantitative_results}  (See Fig.~\ref{fig:ablation}). We set the threshold $T$ to $0.3$ from the experiments with various $T$ values (See Appendix~\ref{sec:threshold}).

\subsection{PMF calculation for entropy coding and decoding}
\label{sec:PMF_calculation}
For the entropy coding and decoding of the subinterval index $k$, the PMF for $y^*_i$ in the quantization layer $l$ is determined as follows:
\begin{align}
\label{eq:pmf_deephq}
P(y^*_i \in I^k_{l,i}\;|&\;y^*_i \in I_{l\minus1,i}) = \frac{\Phi(B^{(k+1)}_{l,i}) - \Phi(B^{(k)}_{l,i})}{\Phi(UB_{l,i}) - \Phi(LB_{l,i})},
\end{align}
where $P(\cdot)$ represents the conditional probability that $y^*_i$ falls into the subinterval $I^k_{l,i}$ at the $l$-\textit{th} quantization layer when $y^*_i$ is in the interval $I_{l\minus1,i}$ at the lower quantization layer, and $\Phi(\cdot)$ represents the cumulative distribution function (CDF) determined based on the distribution parameters estimated by the hyper-decoder network. In this work, zero-mean Gaussian based on $\sigma_i$ is used because $y^*_i{=}y_i\minus\mu_i$ is an unbiased representation.
In Eq.~\ref{eq:pmf_deephq}, the denominator denotes the probability over all possible ranges where $y^*_i$ can be located in the current quantization layer, and the numerator represents the probability for each subinterval.

\begin{figure*}[!t]
    \captionsetup[subfigure]{labelformat=empty}
    \captionsetup[subfigure]{font=scriptsize, labelfont=scriptsize}
    \setlength\columnsep{0pt}
    \captionsetup{belowskip=5pt}
    \centering
    
    % first row
    \begin{subfigure}{0.120\textwidth}
        \centering
        \includegraphics[scale=0.23, width=\linewidth]{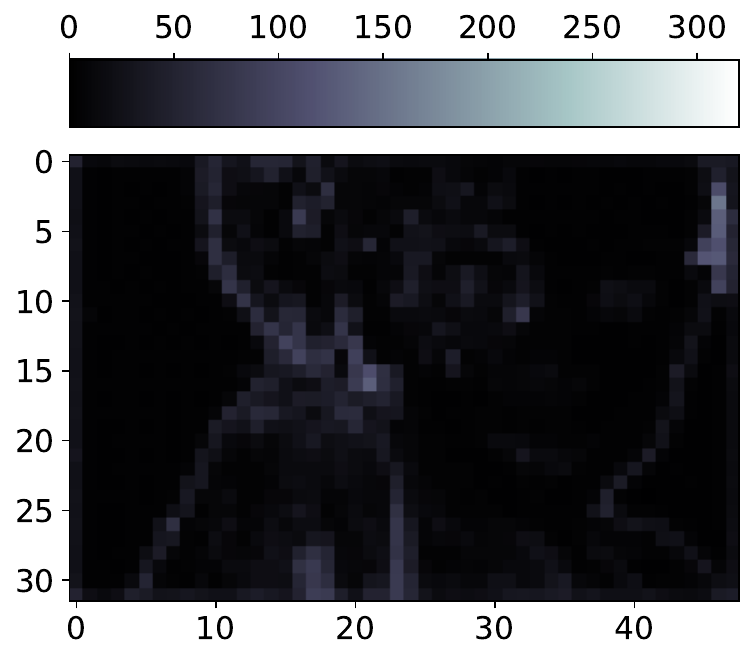}
        \vspace*{-5mm}
        \caption{}
    \end{subfigure}
    \begin{subfigure}{0.120\textwidth}
        \centering
        \includegraphics[scale=0.23, width=\linewidth]{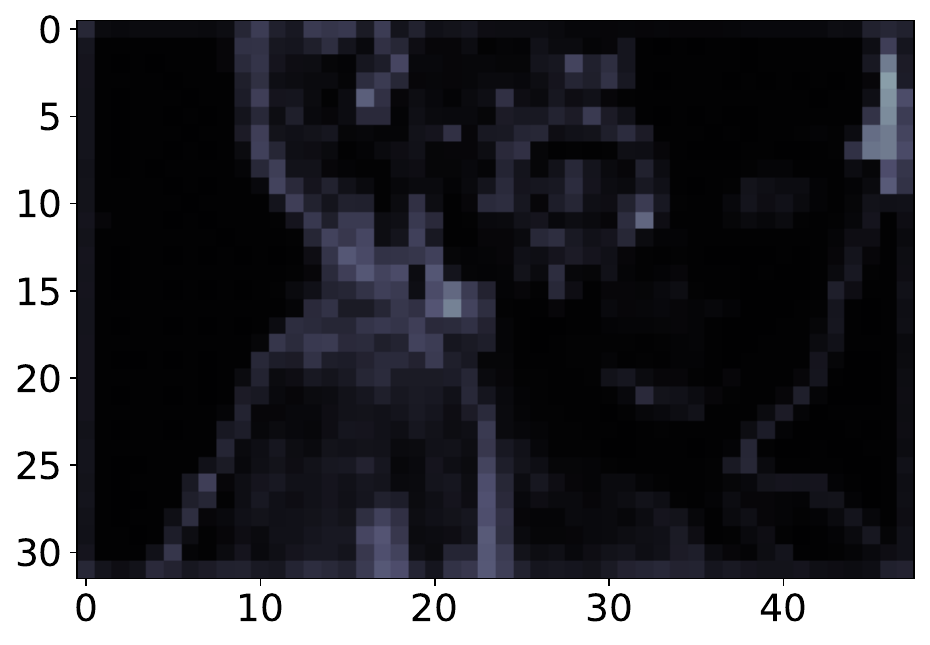}
        \vspace*{-5mm}
        \caption{}
    \end{subfigure}
    \begin{subfigure}{0.120\textwidth}
        \centering
        \includegraphics[scale=0.23, width=\linewidth]{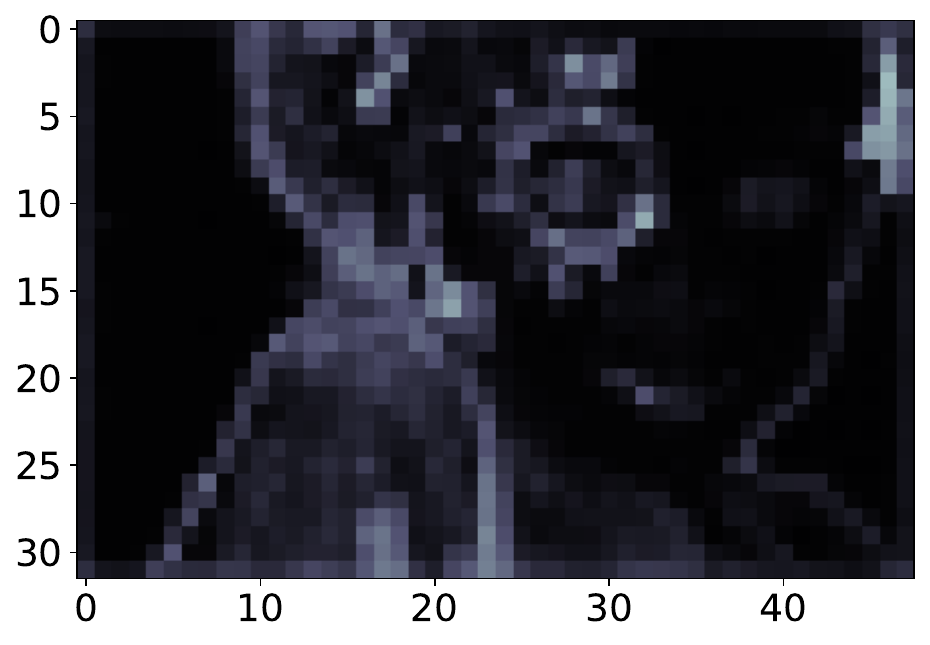}
        \vspace*{-5mm}
        \caption{}
    \end{subfigure}
    \begin{subfigure}{0.120\textwidth}
        \centering
        \includegraphics[scale=0.23, width=\linewidth]{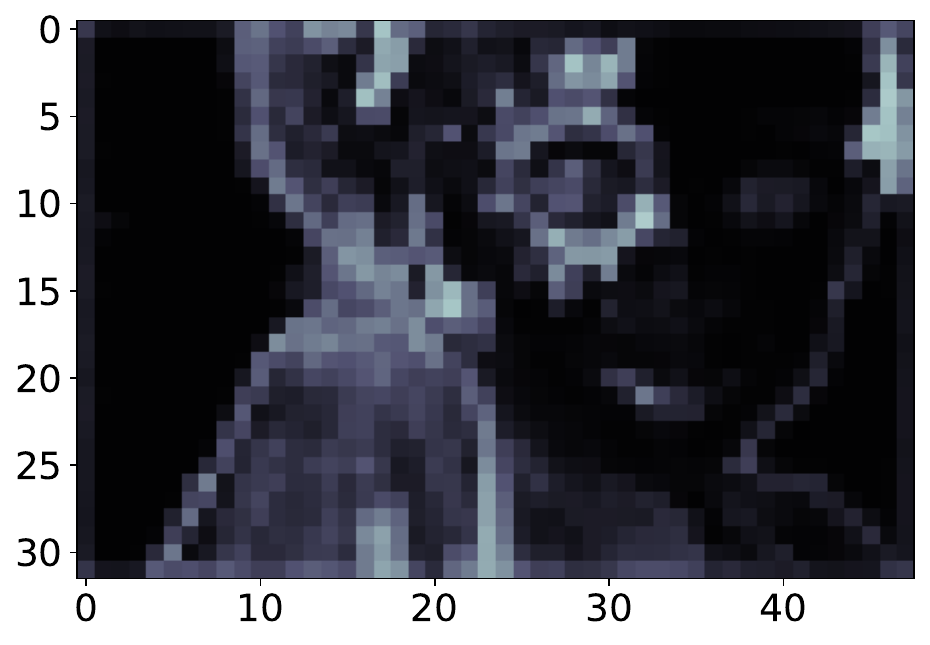}
        \vspace*{-5mm}
        \caption{}
    \end{subfigure}
    \begin{subfigure}{0.120\textwidth}
        \centering
        \includegraphics[scale=0.23, width=\linewidth]{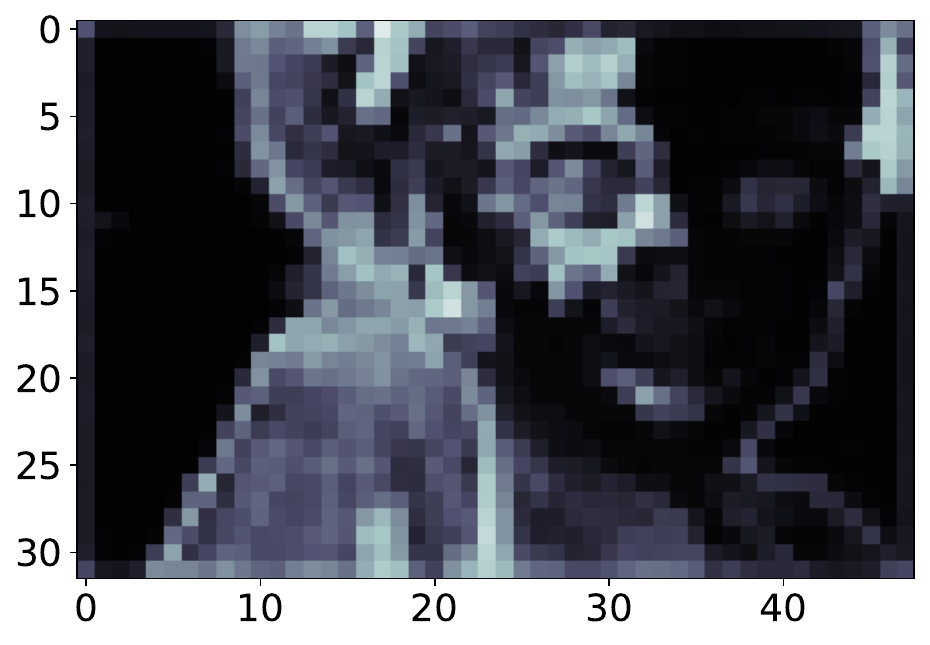}
        \vspace*{-5mm}
        \caption{}
    \end{subfigure}
    \begin{subfigure}{0.120\textwidth}
        \centering
        \includegraphics[scale=0.23, width=\linewidth]{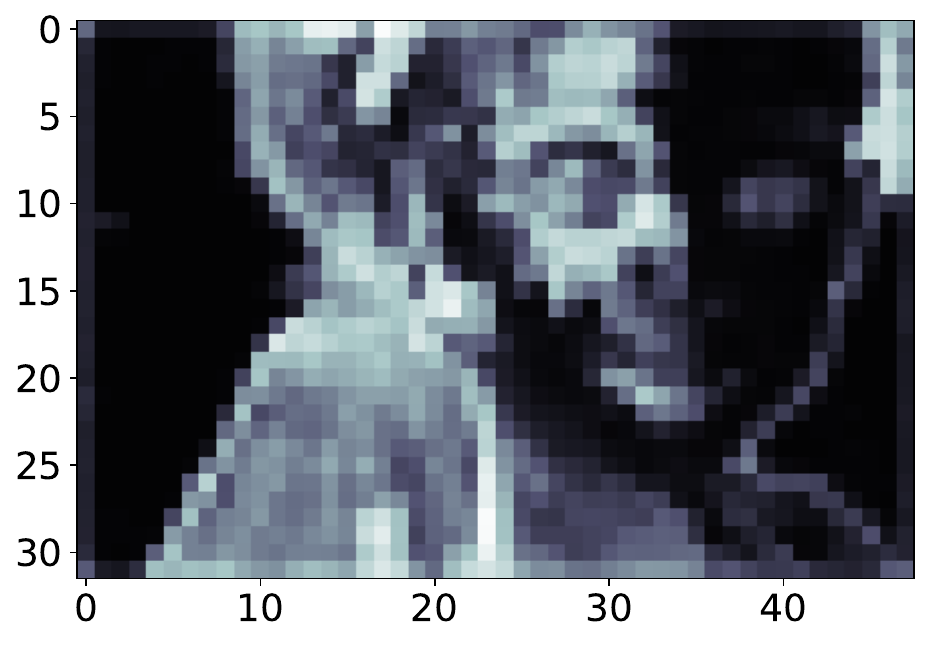}
        \vspace*{-5mm}
        \caption{}
    \end{subfigure}
    \begin{subfigure}{0.120\textwidth}
        \centering
        \includegraphics[scale=0.23, width=\linewidth]{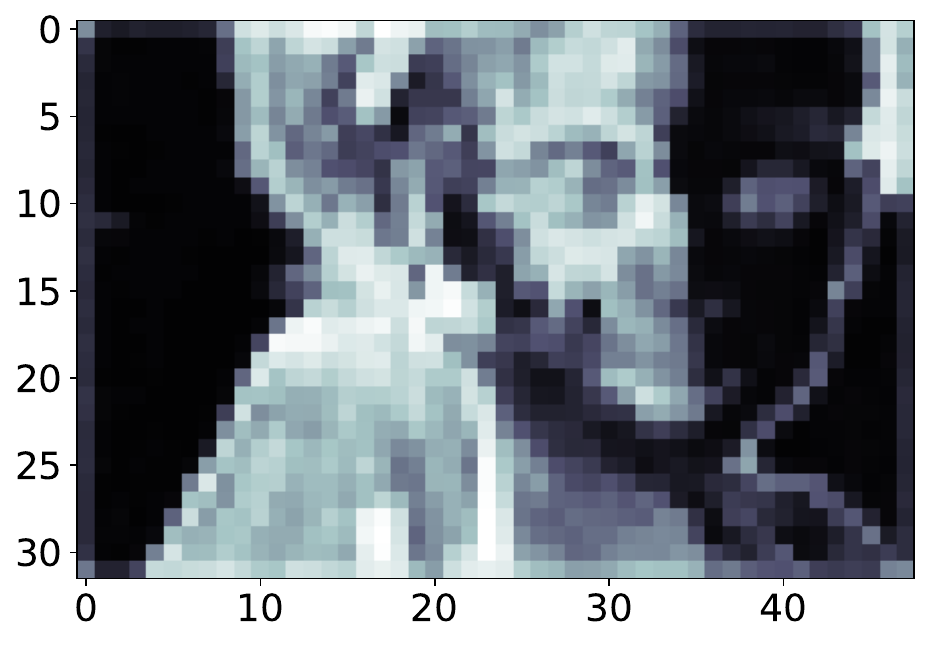}
        \vspace*{-5mm}
        \caption{}
    \end{subfigure}
    \begin{subfigure}{0.120\textwidth}
        \centering
        \includegraphics[scale=0.23, width=\linewidth]{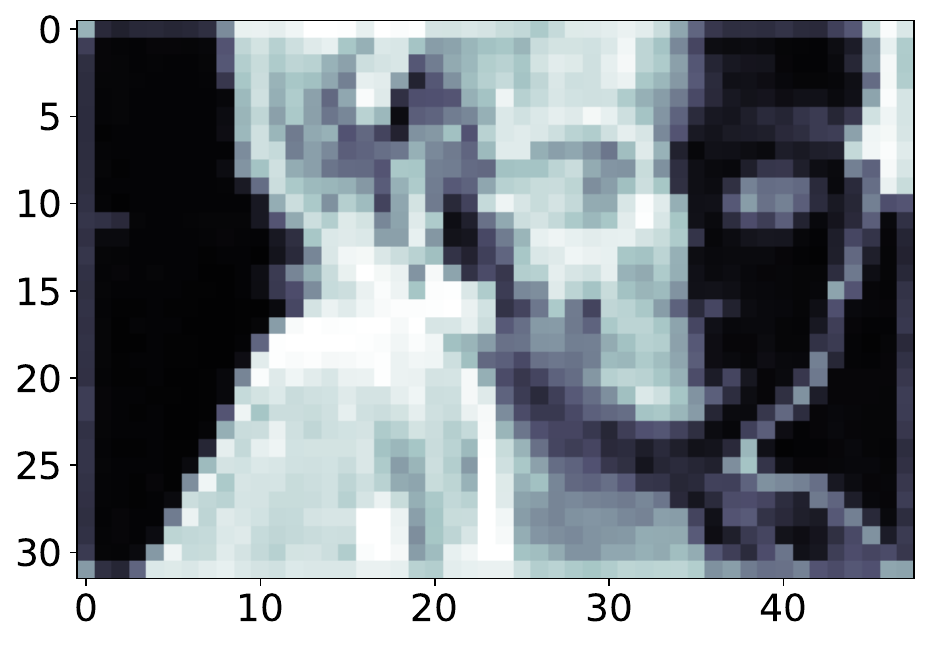}
        \vspace*{-5mm}
        \caption{}
    \end{subfigure}
    \vspace{-1.0cm}

    % second row
    \begin{subfigure}{0.120\textwidth}
        \centering
        \includegraphics[width=\linewidth, clip, trim={23cm 0cm 23cm 0cm}]{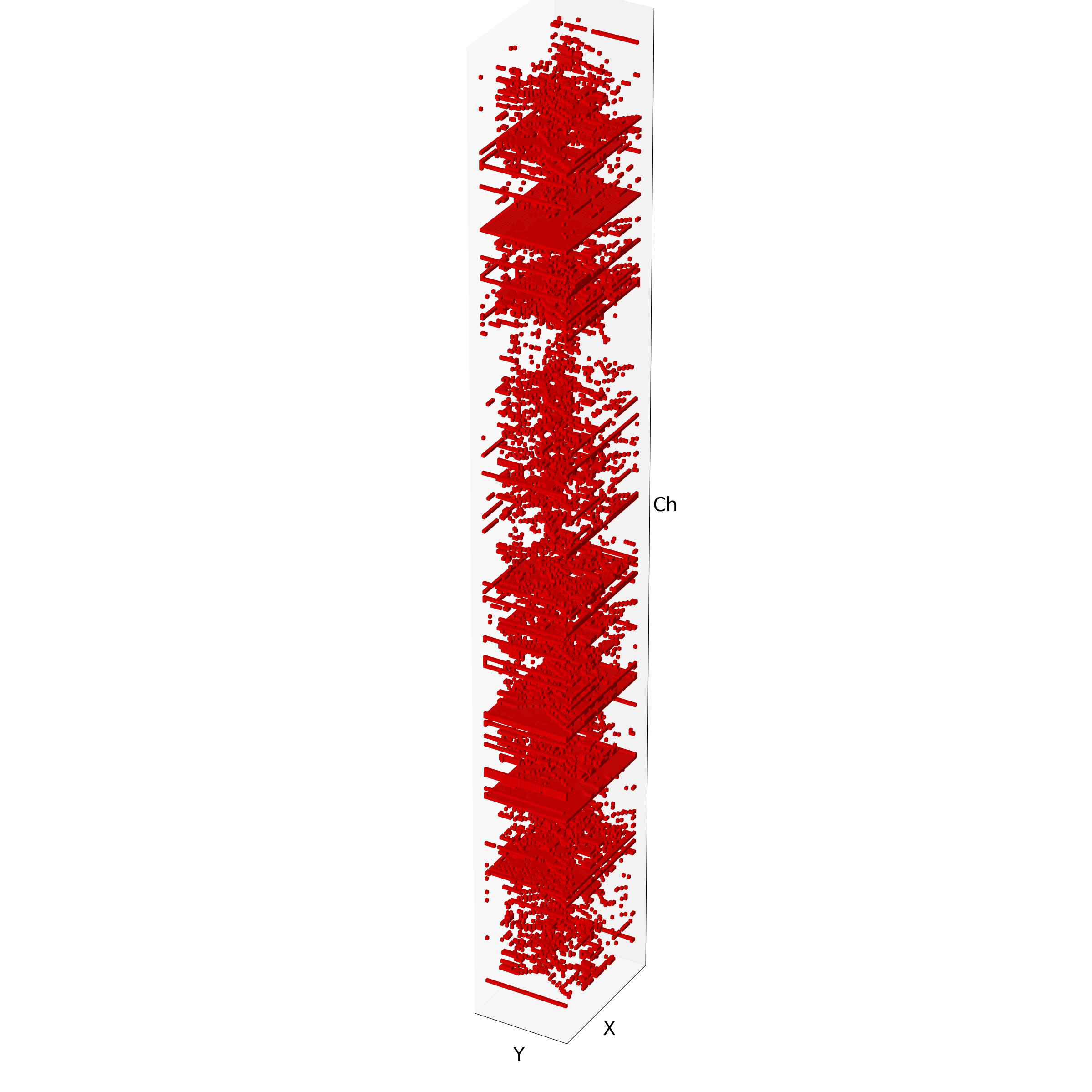}
        \vspace*{-0.7cm}
        \caption{\textit{layer 1\\4.97\%}}
    \end{subfigure}
    \begin{subfigure}{0.120\textwidth}
        \centering
        \includegraphics[width=\linewidth, clip, trim={23cm 0cm 23cm 0cm}]{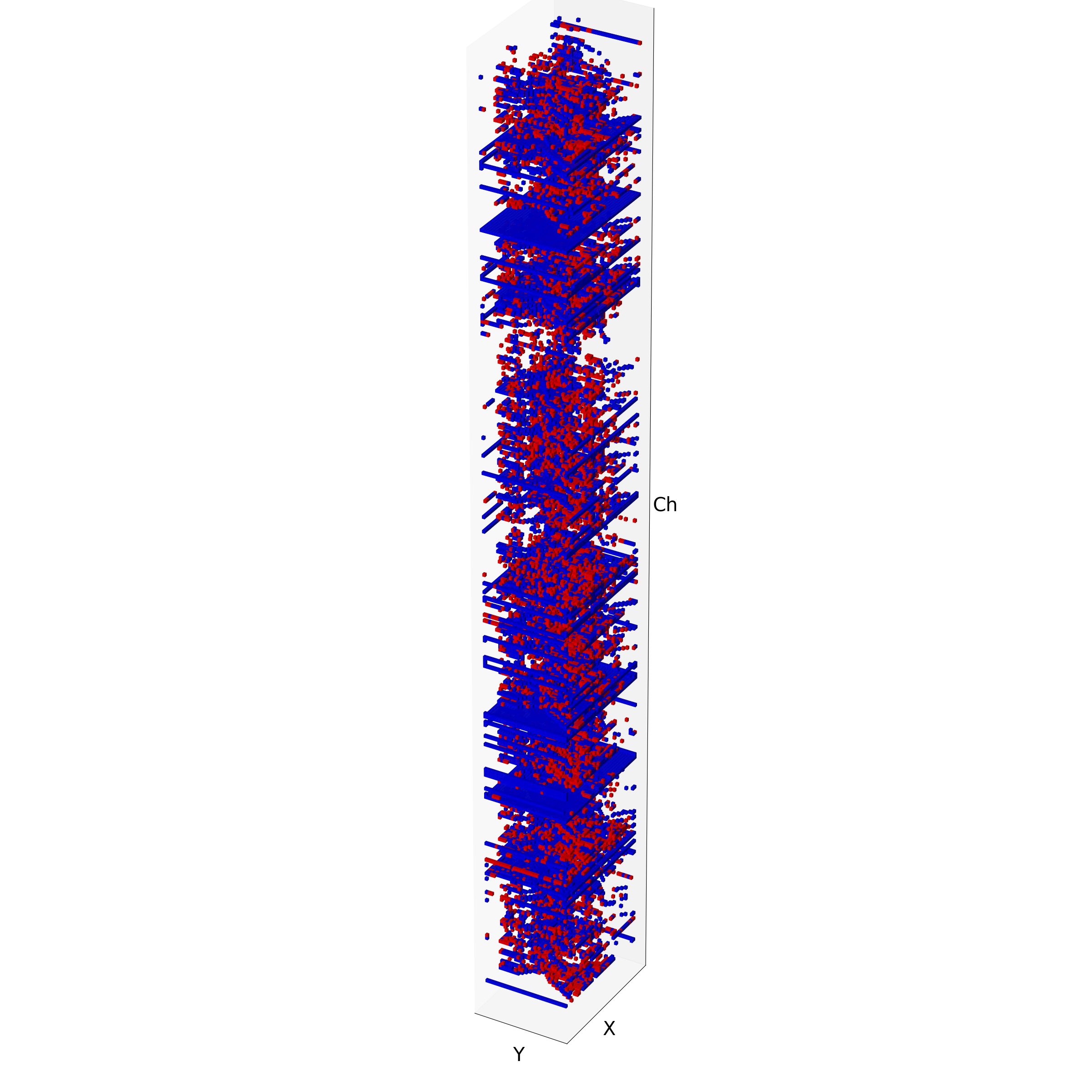}
        \vspace*{-0.7cm}
        \caption{\textit{layer 2\\7.23\%}}
    \end{subfigure}
    \begin{subfigure}{0.120\textwidth}
        \centering
        \includegraphics[width=\linewidth, clip, trim={23cm 0cm 23cm 0cm}]{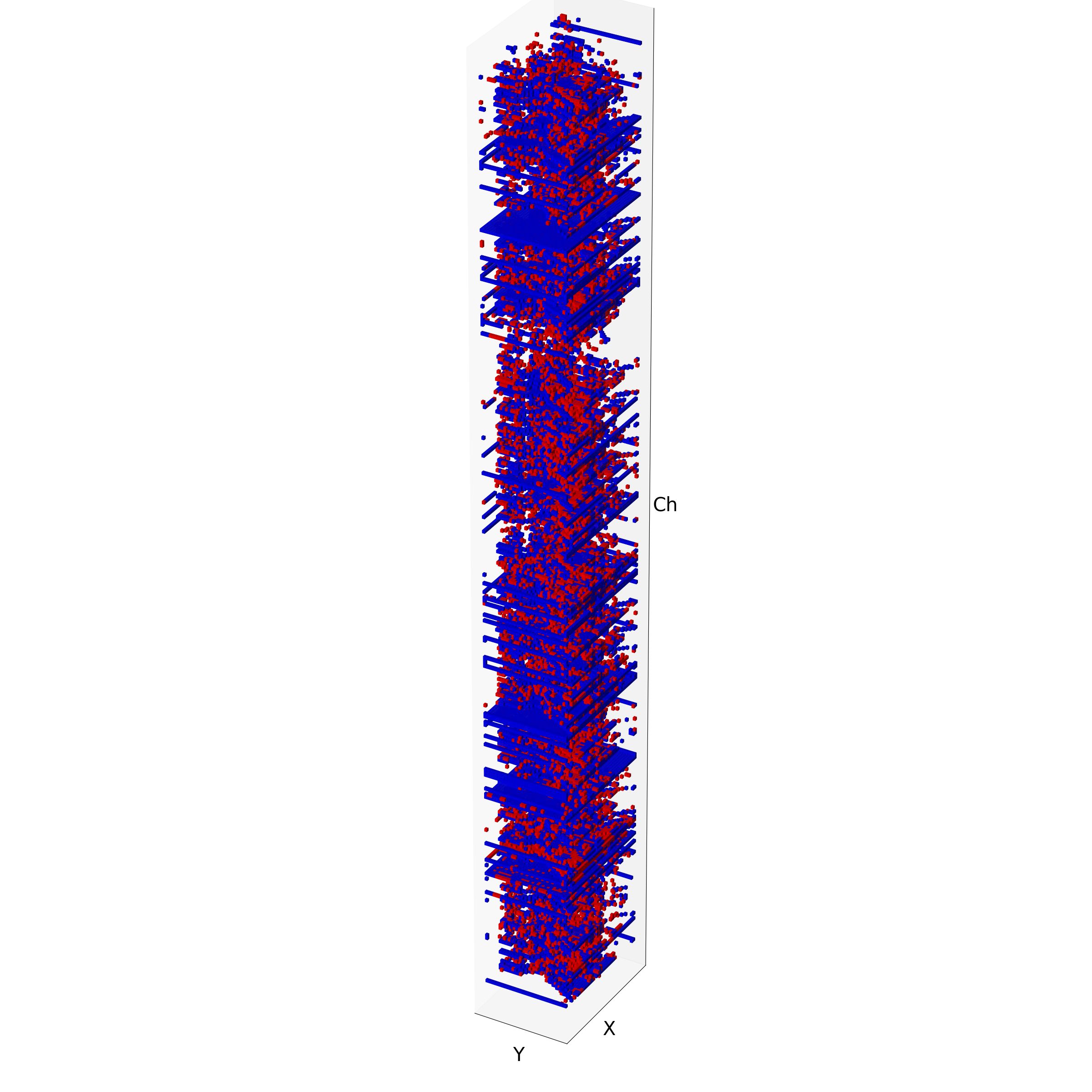}
        \vspace*{-0.7cm}
        \caption{\textit{layer 3\\10.65\%}}
    \end{subfigure}
    \begin{subfigure}{0.120\textwidth}
        \centering
        \includegraphics[width=\linewidth, clip, trim={23cm 0cm 23cm 0cm}]{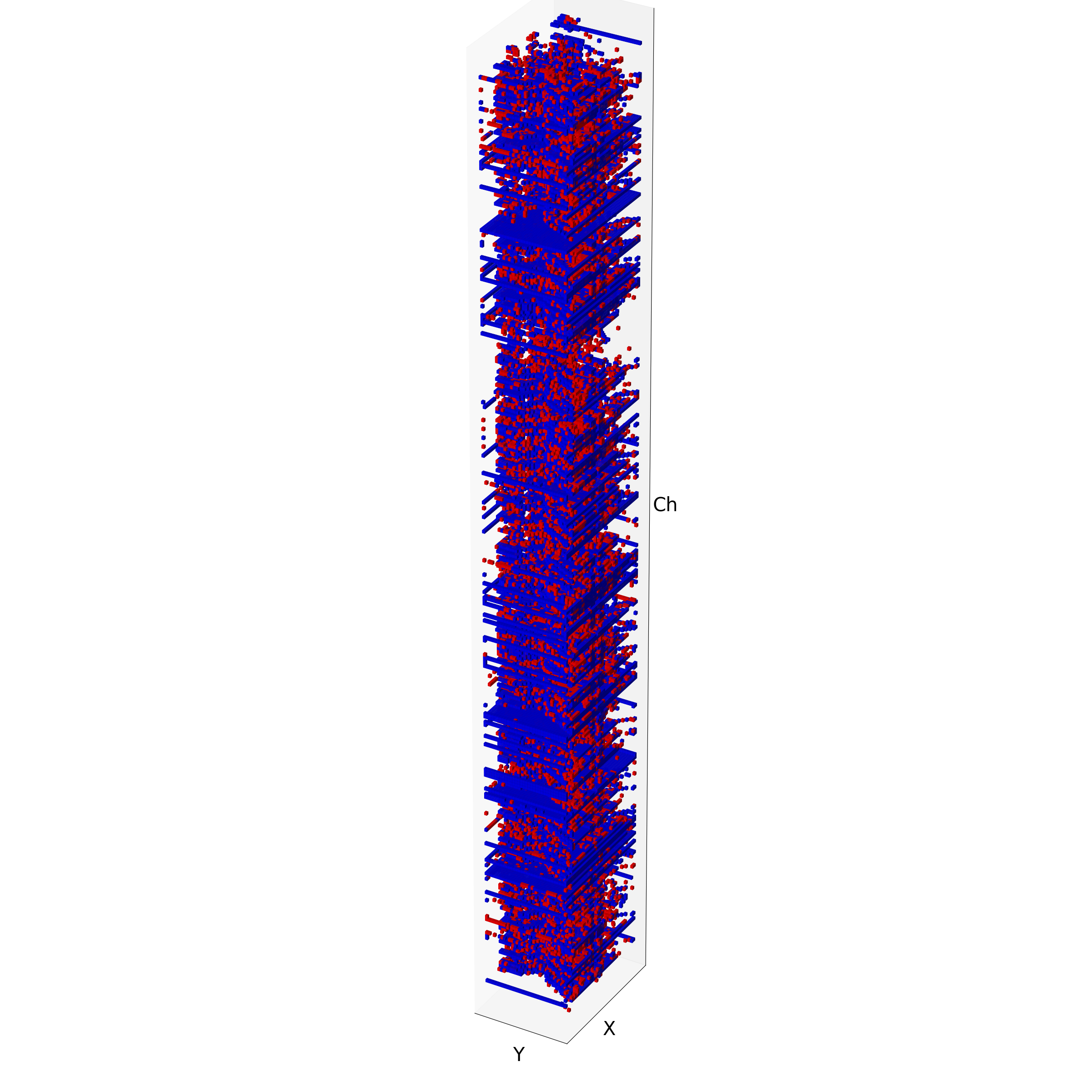}
        \vspace*{-0.7cm}
        \caption{\textit{layer 4\\15.24\%}}
    \end{subfigure}
    \begin{subfigure}{0.120\textwidth}
        \centering
        \includegraphics[width=\linewidth, clip, trim={23cm 0cm 23cm 0cm}]{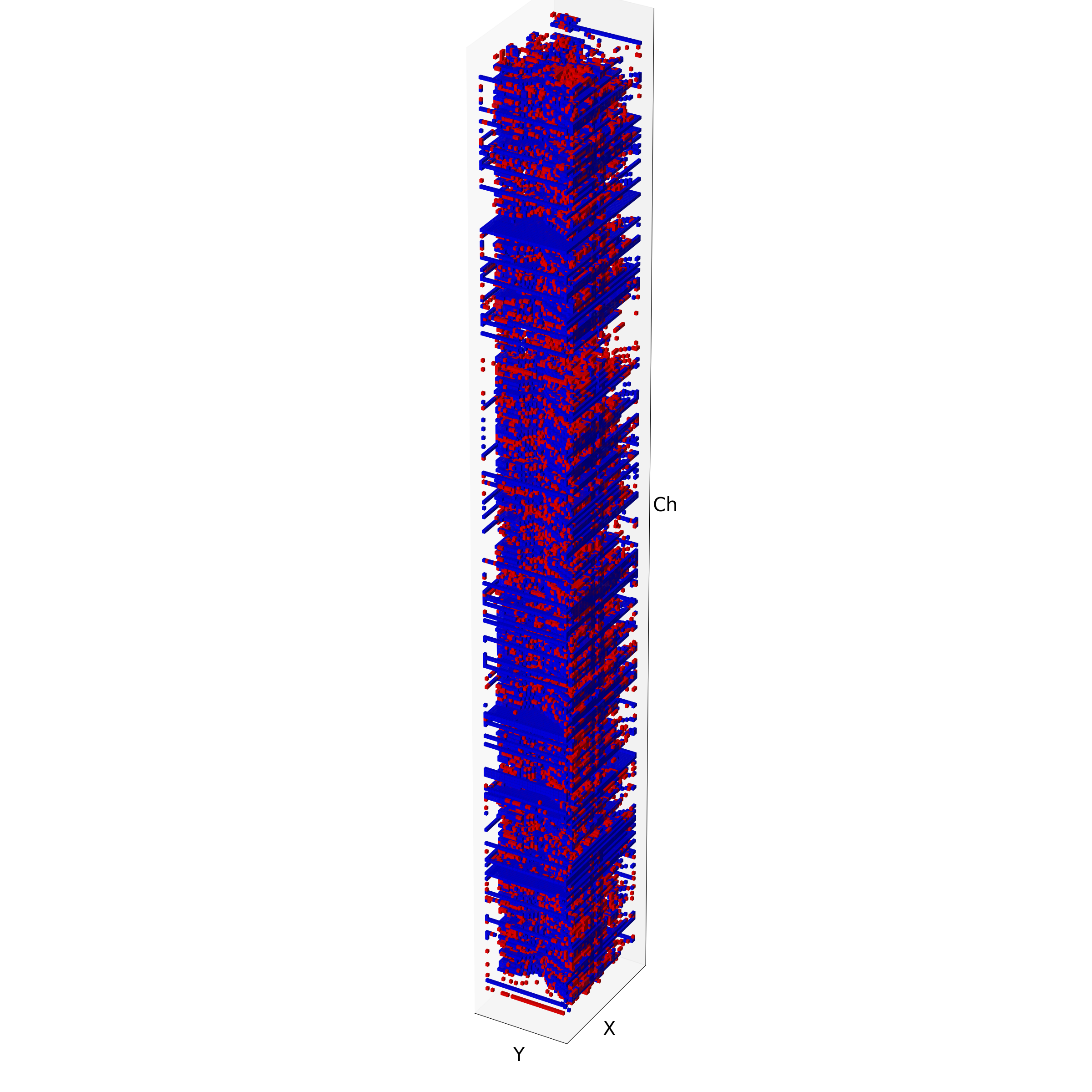}
        \vspace*{-0.7cm}
        \caption{\textit{layer 5\\21.43\%}}
    \end{subfigure}
    \begin{subfigure}{0.120\textwidth}
        \centering
        \includegraphics[width=\linewidth, clip, trim={23cm 0cm 23cm 0cm}]{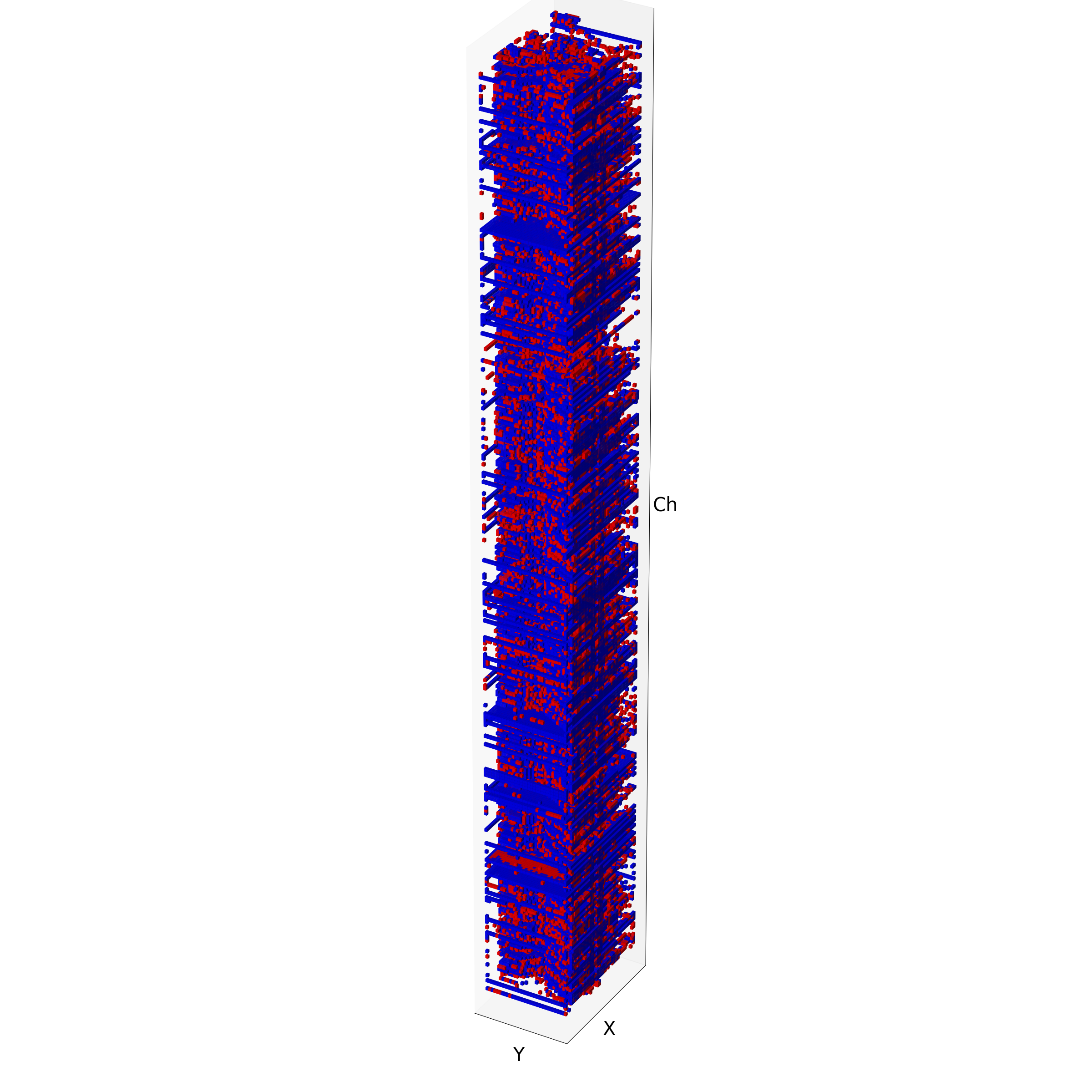}
        \vspace*{-0.7cm}
        \caption{\textit{layer 6\\29.69\%}}
    \end{subfigure}
    \begin{subfigure}{0.120\textwidth}
        \centering
        \includegraphics[width=\linewidth, clip, trim={23cm 0cm 23cm 0cm}]{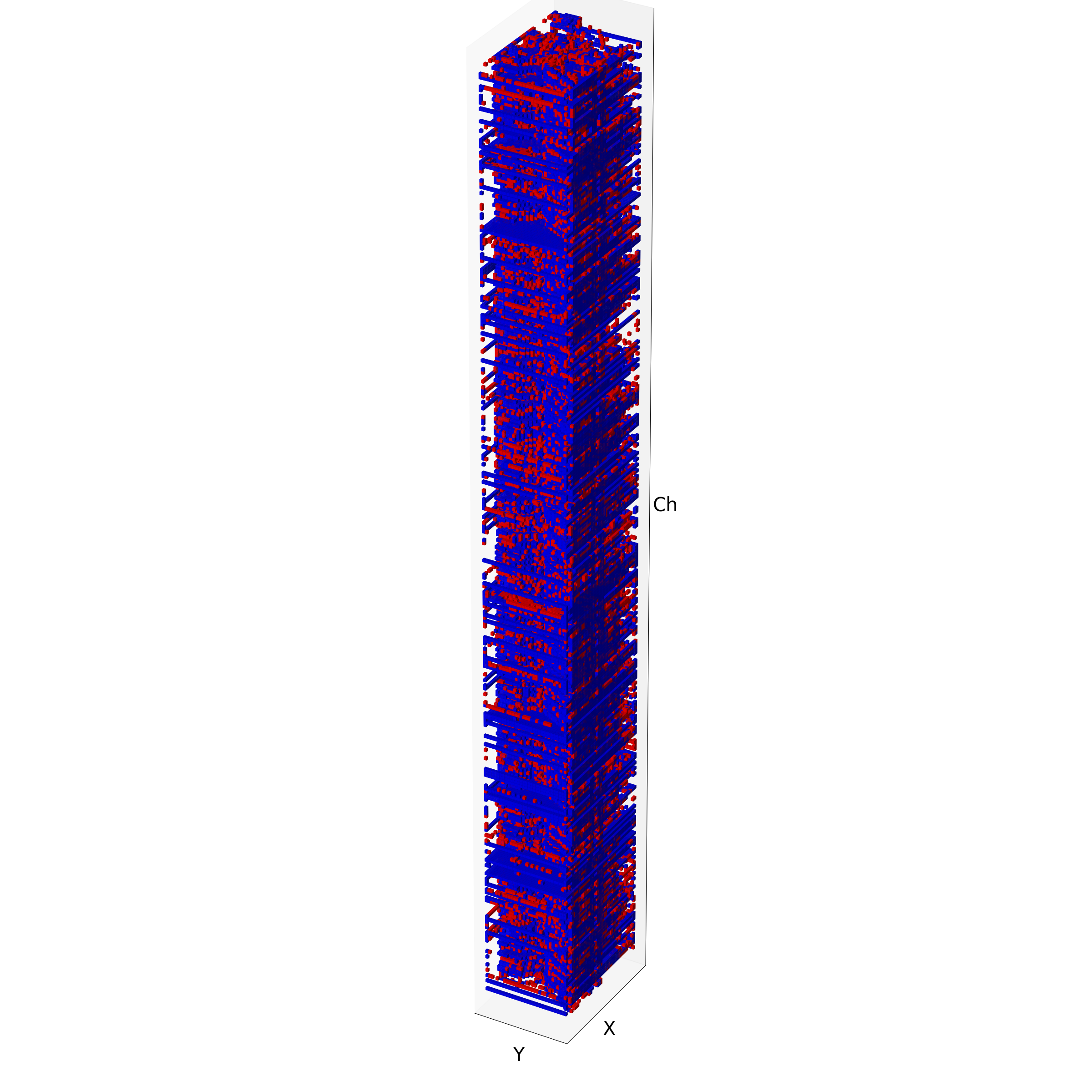}
        \vspace*{-0.7cm}
        \caption{\textit{layer 7\\39.35\%}}
    \end{subfigure}
    \begin{subfigure}{0.120\textwidth}
        \centering
        \includegraphics[width=\linewidth, clip, trim={23cm 0cm 23cm 0cm}]{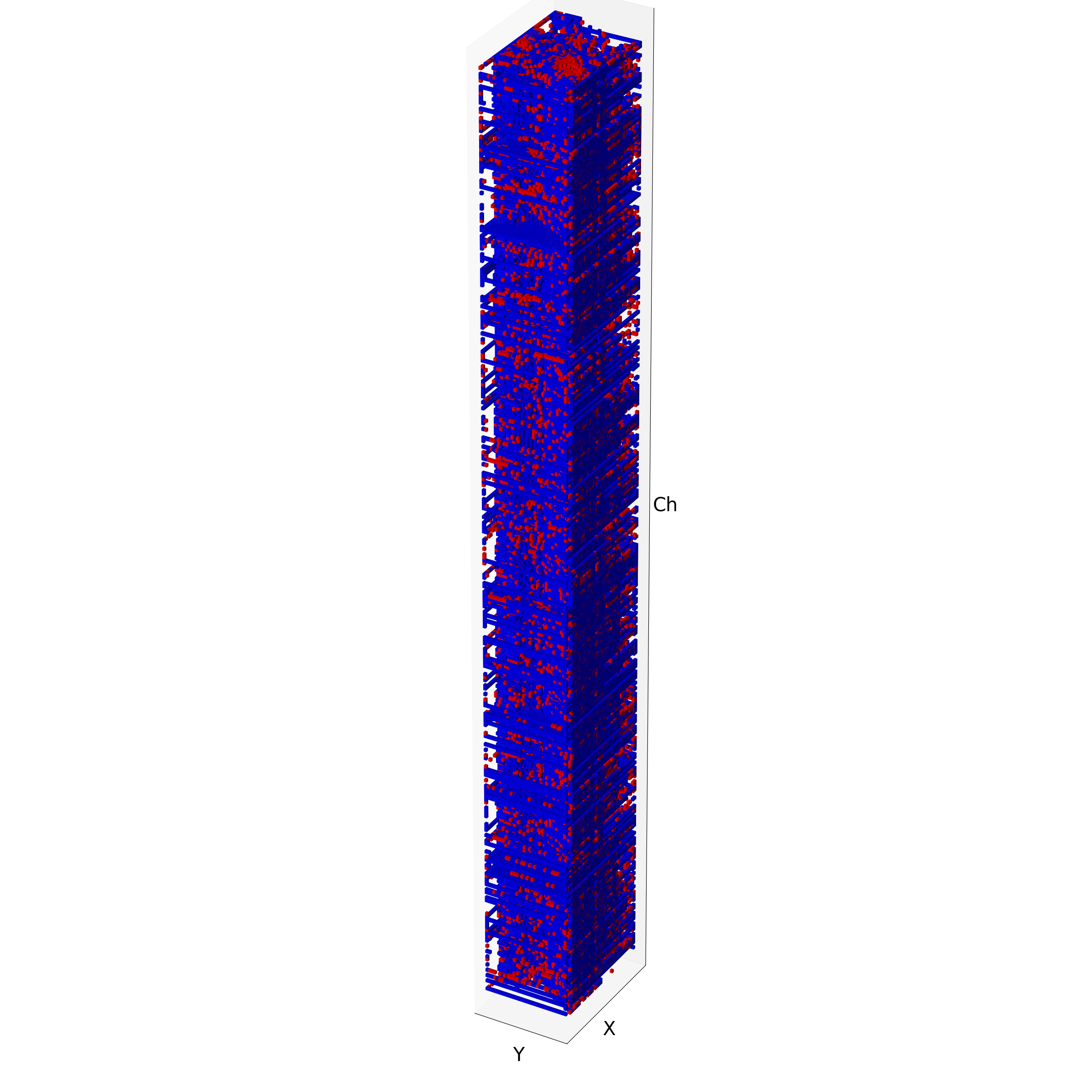}
        \vspace*{-0.7cm}
        \caption{\textit{layer 8\\47.07\%}}
    \end{subfigure}
    \vspace{-0.2cm}
    \caption{Example selection masks $m(\bm{\hat z}, l)$ for different quantization layers of our DeepHQ-MS. For each quantization layer, the newly added components ($\{1 \minus m(\bm{\hat z}, l\minus1)\} \odot m'(\bm{\hat z}, l)$ in Eq.~\ref{eq:mask_generation}) are colored with red, while the components selected for the lower quantization layer ($m(\bm{\hat z}, l\minus1)$ in Eq.~\ref{eq:mask_generation}) are colored with blue. The percentages indicate the proportion of total selected components for each quantization layer. (Top) 2-D representation (summations along the channel-axis) of $m(\bm{\hat z}, l)$.}
    \vspace{-0.5cm}
    \label{fig:mask_3d}
\end{figure*}
\section{Selective encoding of representation components}
\label{sec:selective}
Compressing all representation components regardless of a quantization layer can lead to suboptimal compression efficiency in a progressive coding model. In particular, some representation components containing fine-detail information may not necessarily be required in low-quality compression; thus, compressing these components in low-quality compression often degrades overall R-D performance. Inspired by the selective compression method~\cite{lee2022selective} in the variable-rate compression field, which has shown that component-wise selective compression according to the compression level is effective in both coding efficiency and decoding time reduction, we enable our DeepHQ to selectively compress only essential representation elements for each quantization layer as follows:
\begin{align}
\label{eq:selection_process}
\bm{\breve y}^*_l = Re(DQ(\bm{k}, &\bm{\Updelta}_l, \bm{I}_{l-1}),m(\bm{\hat z}, l)),\\
\text{with  }\bm{k} &= Q(\langle\bm{y}^*\rangle_l, \bm{\Updelta}_l, \bm{I}_{l-1}), \ \ 
\langle\bm{y}^*\rangle_l = M(\bm{y}^*, m(\bm{\hat z}, l)),\notag
\end{align}
where $m(\bm{\hat z}, l)$ is the 3-D binary mask generated from the quantized hyperprior representation $\hat{\bm z}$ for the \textit{l}-th quantization layer, indicating which representation elements of $\bm y^*$ are selected for compression, $M(\cdot)$ is the representation selection operator that extracts only the representation components indicated by $m(\bm{\hat z}, l)$, $\langle\bm{y}^*\rangle_l$ is the set of the selected components of $\bm{y}^*$ for \textit{l}-th quantization layer, and $Re(.)$ is an operator that reshapes $\langle\bm{y}^*\rangle_l$, which is in 1-D shape, back into the original 3-D shape using the same mask $m(\bm{\hat z}, l)$. 
Note that unselected components in $\bm{\breve y}^*_l$ are filled with zeros by $Re(.)$ in Eq.~\ref{eq:selection_process}.
%, but those components eventually become $\bm{\mu} / \bm{\Updelta}_l * \bm{\Updelta}^\text{inv}_l$ in the final decoder input $\bm{\breve y}^\textit{final}_l$ via the compensation process described in Sec.~\ref{sec:overall_quantization}.
The hierarchical quantization and dequantization processes $Q(\cdot)$ and $DQ(\cdot)$ are basically the same as those described in Sec.~\ref{sec:hierarchical_quantization}, except that $LB_{l,i}{=}LB_{1,i}$ and $UB_{l,i}{=}UB_{1,i}$ are used for the representation elements which are first included at the \textit{l}-th quantization layer.
In addition, because the fully generalized mask generation of the original SCR~\cite{lee2022selective} may not be appropriate for progressive coding, we propose a new mask generation method to ensure that all the elements selected in a lower quantization layer are included in a higher quantization layer, as follows:
\begin{equation}
\label{eq:mask_generation}
m(\bm{\hat z}, l) = m(\bm{\hat z}, l\minus1) + \{1 \minus m(\bm{\hat z}, l\minus1)\} \odot m'(\bm{\hat z}, l),
\end{equation}
where $m'(\bm{\hat z}, l)$ is the 3-D mask generated from $\hat{\bm z}$ without considering the hierarchical quantization according to the original SCR~\cite{lee2022selective} model (See below in this section for a detailed description of the $m'(\bm{\hat z}, l)$ generation), and $\odot$ represents the element-wise multiplication. To determine the 3-D binary mask $m(\bm{\hat z}, l)$ for our DeepHQ, we inclusively add the newly selected components $(1 \minus m(\bm{\hat z}, l\minus1)) \odot m'(\bm{\hat z}, l)$ as an update term to the mask $m(\bm{\hat z}, l\minus1)$ from the lower quantization layer $l\minus1$. By doing so, the DeepHQ can maintain the progressive (or inclusive) relation from a lower quantization layer to a higher one with further selected elements as the quantization layer gets higher. Note that for $l=1$, we use the mask $m(\bm{\hat z}, 1) = m'(\bm{\hat z}, 1)$. The representation selection process of our DeepHQ is optimized from an R-D perspective, as in SCR~\cite{lee2022selective}.
By incorporating the extended selective compression, we achieve higher coding efficiency, particularly in the low bit-rate range, by preventing the inclusion of unimportant representation elements, as will be shown in Sec.~\ref{sec:quantitative_results}. 
Fig.~\ref{fig:mask_3d} shows examples of the selected representation components for each quantization layer. For the first quantization layer ($l$=1), only 4.97$\%$ of the representation components are selected, while the selection ratio increases up to 47.07$\%$ as the final quantization layer ($l$=8). The average selection ratios for the Kodak~\cite{KODAK} dataset range from 8.04$\%$ ($l$=1) to 49.78$\%$ ($l$=8).

In addition, for better understanding, we briefly introduce the generation process of the original mask $m'(\bm{\hat z}, l)$ in the SCR~\cite{lee2022selective} model as follows: i) A single 1x1 convolution layer is applied to the output of the penultimate convolutional layer (after the activation) in the hyper-decoder network, generating a 3-D importance map $im(\bm{\hat z})$ of the same size as the $\bm y$ representation. This 3-D importance map $im(\bm{\hat z})$ represents the canonical (representative for all the compression quality levels, independently of the compression quality levels.) importance of the representation components of $\bm y$ in a component-wise manner with the values between 0 and 1. ii) To derive the target quality dedicated importance map, the canonical 3-D importance map $im(\bm{\hat z})$ is then amplified or attenuated channel-wise using a learned importance adjustment curve vector $\bm{\gamma}_l$ via $im(\bm{\hat z})^{\bm{\gamma}_l}$. The importance adjustment curve vector $\bm{\gamma}_l$ is dedicated to its corresponding $l$-th quantization layer. For example, in the case of the low-level quantization layer (low-quality compression), the 3-D importance map is attenuated in an average sense, while it is amplified for high-level quantization layers (high-quality compression). iii) The adjusted 3-D importance map $im(\bm{\hat z})^{\bm{\gamma}_l}$ is then binarized via rounding-off, resulting in the output 3-D binary mask $m'(\bm{\hat z}, l)$. 

\section{Component-wise progressive coding}
\label{sec:component-wise}
To support further fine-grained progressive coding, we enable component-wise progressive coding, as adopted in a few existing methods~\cite{Lu_progressive_2021,Lee_2022_CVPR}.
To support the component-wise progressive coding, we sort representation components according to predefined criteria and perform sequential entropy coding. The compression order is determined based on the estimated $\bm \sigma$ values where the components of higher $\bm \sigma$ values are coded earlier. It should be noted that the differences in coding efficiency among different sorting criteria are very marginal in our work
% as shown in Fig.~\ref{fig:comparision_sort_criteria} 
because the proposed DeepHQ model supports a larger number of quantization layers than the existing methods~\cite{Lee_2022_CVPR,2023_CVPR_jeon}. Therefore, we utilize the $\bm \sigma$ that does not require additional calculation.
We represent the partial compression status between two discrete quantization layers with a continuous value $l$, where the decimal part of $l$ indicates the portion of the latent representation components already encoded into the bitstream among the whole latent representation components to be hierarchically quantized and entropy-coded into the next quantization layer. For example, $l{=}3.3$ represents the status of the compressed bitstream including $30\%$ of the total selected representation components for $l{=4}$. In this case, $\bm{\breve y}^*_{3.3}$ in Eq.~\ref{eq:y_breve_final} contains representation components reconstructed (entropy-decoded and dequantized) from both quantization layers $l{=}3$ and $l{=}4$.

\begin{figure}[!t]
\begin{center}
\includegraphics[width=0.5\linewidth,clip, trim={0cm 0cm 0cm 0.26cm}]{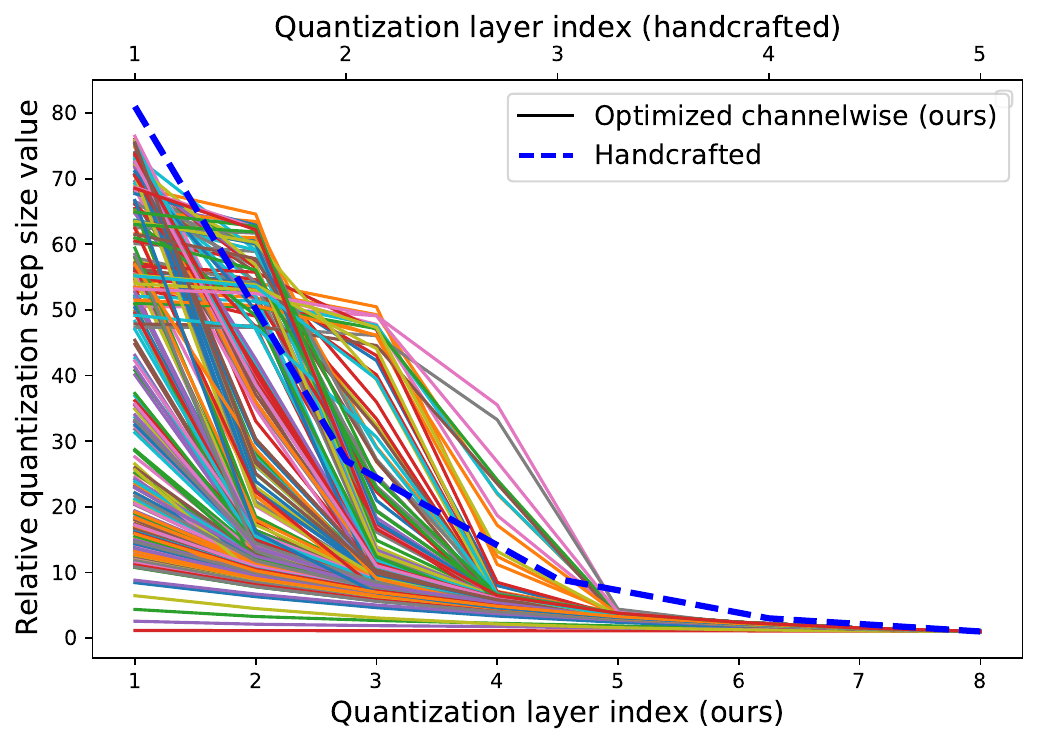}
\end{center}
\vspace{-0.3cm}\caption{Variations of learned quantization step sizes $\bm{\Updelta}_l$ of our model over 8 quantization layers. Each line corresponds to the latent representation of a channel. Normalization was performed on all quantization step sizes to represent them as ratios to the case of the highest quantization layer. The dotted bold line indicates an example of handcrafted quantization step sizes for five quantization layers (see the top $x$-axis) where the number of quantization layer \textit{l} is limited up to 5 due to the three sub-partitions of every layer when going to its next higher layer, resulting in the step size value of 81 ($=3^{5-l}$) in layer $l{=}1$, for the handcrafted methods~\cite{Lu_progressive_2021,Lee_2022_CVPR,Li2023deadzone,2023_CVPR_jeon}.
}
\label{fig:intervals}
\vspace{-0.2cm}
\end{figure}

\section{Experiments}
\label{sec:experiments}
\subsection{Training}
\label{sec:training}
The nested scheme of our DeepHQ makes dependency chains across multiple quantization layers, making their parallel training impossible. Furthermore, it entails the discontinuities in gradient calculations owing to (i) the conditional clipping in Eq.~\ref{eq:boundary_calculation}, (ii) the adaptive boundary adjustment in Eq.~\ref{eq:inteval_expansion} and (iii) the inherent characteristics of quantization, making it difficult to use gradient descent based training directly. Thus, we adopt a workaround training scheme that has recently been applied in variable rate image coding~\cite{Cui2021, lee2022selective}, focusing on the optimal quantization step sizes for various compression levels of the quantization layers from the R-D perspective.

In our training scheme, we jointly train the whole modules of DeepHQ, including the single base compression model (composed of the encoder, decoder, hyper-encoder, and hyper-decoder sub-networks), the set of whole quantization step sizes $\bm{\Updelta}=\{\bm{\Updelta}_l\ |\ l \in \{1, 2, ..., L\}\}$, and the mask generation module for selective compression, in an end-to-end manner. To this end, we employ joint optimization of rate and distortion, which is typically utilized in the learned image compression field. However, to optimize the dedicated parameters for each quantization layer ($\bm{\Updelta}$ and the set of importance adjustment curves $\bm{\gamma} = \{\bm{\gamma}_l\ |\ l \in \{1, 2, ..., L\}\}$ within the selective compression module), we construct the total loss term $\mathcal{L} = \textstyle\sum_{l}{R_l + \lambda_l \cdot D_l}$
where $R_l$, $D_l$, and $\lambda_l$ denote the rate term, the distortion term, and the balancing parameter between rate and distortion determined by $\lambda_l = 0.2 \times 2^{l-8}$, respectively, for quantization layer $l$. We determine the rate term for each quantization layer based on the entropy model as follows:
\begin{align}
\label{eq:rate_loss}
R_l = H_l&(\bm{\tilde y}^\Updelta_l \mid \bm{\tilde z}) + H(\bm{\tilde z}),\\
\text{with  } &H_l(\bm{\tilde y}^\Updelta_l \mid \bm{\tilde z}) = \frac{1}{N^{\bm x}} \sum_{i=1}^{N^l}{-\log_2{P_l({\tilde y^\Updelta_{l,i_c}}|\bm{\hat z})}}, \notag\\
&\bm{\tilde y}^\Updelta_l = \bm{y}^\Updelta_l + U(-0.5,0.5), 
\text{  }\bm{y}^\Updelta_l=\langle\bm{y} / \bm{\Updelta}_l\rangle_l, \notag
\end{align}
where $H_l(\bm{\tilde y}^\Updelta_l{\mid}\bm{\tilde z})$ and $H(\bm{\tilde z})$ represent the entropy value (estimated bits-per-pixel) for $\bm{y}^\Updelta_l$ and $\bm{z}$, respectively, $\bm{y}^\Updelta_l$ is the selected component subset of $\Updelta_l$-normalized $\bm{y}$, and $\langle\cdot\rangle_l$ denotes the component selection process in Eq~\ref{eq:selection_process}.
In calculation of $H_l(\bm{\tilde y}^\Updelta_l{\mid}\bm{\tilde z})$, we use the entropy model $P_l(\cdot)$ for each quantization layer, in which $\langle\bm \mu / \bm{\Updelta}_l \rangle_l$ and $\langle\bm \sigma / \bm{\Updelta}_l \rangle_l$ of Gaussian models are used to approximate the distribution of $\bm{y}^\Updelta_l$. Note that $\bm \mu$ and $\bm \sigma$ are canonical parameters estimated (reconstructed) from the hyper-decoder of the base compression model as shown in Figs.~\ref{fig:overall_architecture}. As in the training of most LIC methods~\cite{Balle17,Theis17,Minnen2018,Lee2019,cheng2020image,Minnen2020}, which approximate the quantization model by sampling a noisy representation and using the PDF convolved with $U(-0.5,0.5)$ as a PMF approximation, we use the noisy representation $\bm{\tilde y}^\Updelta_l$, obtained by adding the noise $U(-0.5,0.5)$, rather than directly using $\bm{y}^\Updelta_l$. Correspondingly, $P_l(\cdot)$ is given by the convolution of a Gaussian distribution and a uniform distribution: $N(\langle\bm \mu / \bm{\Updelta}_l \rangle_l, \langle\bm \sigma / \bm{\Updelta}_l \rangle_l^2) * U(-0.5,0.5)$.
In this learning scheme, which approximates rounding-based quantization, the $\Updelta_l$-based normalization in Eq.~\ref{eq:rate_loss} controls the granularity of $\bm y$ quantization, such that the quantization becomes coarser as $\Updelta_l$ increases. For $H(\bm{\tilde z})$, we use the Factorizedprior entropy model~\cite{Balle17} as in the Hyperprior model~\cite{Balle18}. For the distortion term $D$, we use the typical mean-squared-error (MSE) between the original input $\bm x$ and the reconstruction  $\bm {x'}_l$ as follows:
\begin{align}
\label{eq:distortion_loss}
D_l = \text{MSE}(\bm x, \bm{x'}_l), \text{\ \ \ \ \ with  } \bm{x'}_l = De(Re(\bm{\tilde y}^\Updelta_l,m(\bm{\hat z}, l)) \cdot \bm{\Updelta}_l),
\end{align}
where $De(\cdot)$ represent the decoding transform in Eq.~\ref{eq:y_breve_final} and $Re(\cdot)$ is the reshaping function in Eq.~\ref{eq:selection_process}. Note that since $\bm{\tilde y}^\Updelta_l$ is $\bm{\Updelta}_l$-normalized representation, it's rescaled by $\bm{\Updelta}_l$ correspondingly before being fed into $De(\cdot)$.

As a result of applying the proposed training method, Fig.~\ref{fig:intervals} shows the variations in the learned quantization step sizes $\bm{\Updelta}_l$, across different quantization layers for various target qualities. As shown, our learned quantization hierarchy (DeepHQ)  shows significantly diverse variations in channel-wise quantization step sizes for the representation $\bm y$ of 320 channels in 8 different quantization layers, compared to the handcrafted quantization hierarchy with a layer-wise fixed size and number of quantization intervals. These results demonstrate that our DeepHQ effectively leverages various learned channel-wise and quantization-layer-wise step sizes, improving overall coding efficiency. 

\subsection{Implementation}
\label{sec:implementation}
We implement two different versions of DeepHQ: i) One is DeepHQ-MS based on the Mean-scale~\cite{Minnen2018} network architecture, to compare it with the competing models on the same architecture; ii) The other is the DeepHQ-TCM model, based on TCM~\cite{liu2023tcm}, one of the most recent LIC models that demonstrates state-of-the-art coding efficiency in the field. We adapt the TCM model to use a single slice of channels for compatibility with the proposed DeepHQ, which naturally makes the model non-autoregressive. Our models are implemented using \textit{CompressAI}~\cite{begaint2020compressai} and \textit{PyTorch}~\cite{paszke2019pytorch}. For entropy encoding and decoding of representations, we used \textit{torchac}~\cite{TORCH_AC,mentzer2019practical}.

To reduce the total training time, we adopt a step-wise training as follows: (i) In the first step, the highest-quality ($\lambda$=0.2) non-progressive compression model is trained; (ii) In the second step, we learn the optimal quantization step sizes for the $L$=8 quantization layers via training with the encoder, decoder, hyper-encoder, and hyper-decoder networks in an end-to-end manner using the non-progressive model pre-trained in the first step; (iii) In the third step, we train the full models (DeepHQ-MS and DeepHQ-TCM) including our selective compression using the pre-trained models (DeepHQ without selective compression) in the second step. For the second and third steps, we use the total loss $\mathcal{L} = \sum_{l}{R_l + \lambda_l*D_l}$ as described in Sec.~\ref{sec:training} and 8 different $\lambda_l$ values are set to $0.2\cdot2^{l-8}$ for $l$=1 to 8. Note that the range of these $\lambda_l$ values is commonly used in the LIC field to span from training low-quality (< 30dB in Peak Signal-to-Noise Ratio (PSNR)) compression models to high-quality (> 40dB in PSNR) compression models. In the third training step, the selective compression processes in Eqs.~\ref{eq:selection_process} and \ref{eq:mask_generation} are further applied. Our models are trained using the ADAM ~\cite{ADAM} optimizer. We set the number of epochs for the three training steps to 50, 20, and 20, respectively, and use $65,257$ non-overlapping patches of size $256\times256$ cropped from the full CLIC~\cite{CLIC} training dataset. For each version of our model, the corresponding base model (MS~\cite{Minnen2018} or TCM~\cite{liu2023tcm}) was used as the pretrained model in the first stage. We set the batch sizes to 8, 2, and 2, respectively, for three training steps. For the first and second training steps, we initially set the learning rate to $1\times10^{-4}$, and during the last 4 epochs, the first half utilizes a learning rate of $2\times10^{-5}$, while the second half uses a learning rate of $4\times10^{-6}$. For the third training step, we set the varying learning rates to $1\times10^{-5}$, $2\times10^{-6}$, and $4\times10^{-7}$, respectively.

As an additional technical detail, in practice, we adopt an asymmetric inverse scaling scheme~\cite{Cui2021}, which enables the asymmetric inverse scaling of representations by additionally employing different sets of step sizes, $\bm{\Updelta}^\text{inv}$. We found that this asymmetric inverse scaling scheme provides a slight additional performance improvement. Note that this asymmetric scheme is independent of the hierarchical quantization/dequantization mechanism proposed in this paper. It contributes only partially to the fine-tuning of the final inputs to the decoder network; thus, our two key elements, $Q(\cdot)$ and $DQ(\cdot)$, still operate solely based on $\bm{\Updelta}$ and do not incorporate $\bm{\Updelta}^\text{inv}$ at all. Further details on the specific implementation of the asymmetric inverse scaling scheme in our DeepHQ are provided in Appendix~\ref{sec:asymmetric}.

\subsection{Experimental setup}
\label{sec:experimental_setup}
\begin{figure*}[!t]
\captionsetup[subfigure]{labelformat=empty}
\captionsetup[subfigure]{font=scriptsize,labelfont=scriptsize, aboveskip=2pt, belowskip=3pt}
\setlength\columnsep{0pt}
\captionsetup{belowskip=3pt}
\centering
% ---- table ----
\begin{minipage}{\textwidth}
\centering
\small
\captionof{table}{Model sizes, average rate savings against the BPG~\cite{BPG} codec, and average decoding times of various models.}
\label{tab:experimental_results}
\begin{tabular}{@{}l r rr rr rr@{}}
    \toprule
    & & \multicolumn{2}{c}{Kodak~\cite{KODAK}} & \multicolumn{2}{c}{CLIC~\cite{CLIC}} & \multicolumn{2}{c}{Tecnick~\cite{tecnick}} \\
    \cmidrule(lr){3-4} \cmidrule(lr){5-6} \cmidrule(lr){7-8}
    Method & Param. (M) & r-saving$\uparrow$ & d-time (s)$\downarrow$ & r-saving$\uparrow$ & d-time (s)$\downarrow$ & r-saving$\uparrow$ & d-time (s)$\downarrow$ \\
    \midrule
Mean-scale~\cite{Minnen2018} (Non-PIC) & 17.559 & 2.83\% & 0.097 & 2.37\% & 0.359 & 2.80\% & 0.237\\
    DPICT\_MS & 19.849 & -14.44\% & 0.072 & -20.04\% & 0.351 & -20.50\% & 0.228\\
    DPICT~\cite{Lee_2022_CVPR} & 29.058 & -8.83\% & 0.066 & -15.53\% & 0.281 & -9.51\% & 0.191\\
    DPICT~\cite{Lee_2022_CVPR} (w/ post) & 75.416 & -4.16\% & 0.141 & -10.54\% & 0.587 & -5.32\% & 0.391\\
    CTC~\cite{2023_CVPR_jeon} & 399.003 & 5.71\% & 2.869 & 0.27\% & 5.570 & 4.95\% & 4.131\\
    DeepHQ-MS (ours) & \bf{17.720}  & -1.32\% &\bf{0.018} & -2.40\% & \bf{0.100} & -2.60\% & \bf{0.072}\\
    DeepHQ-TCM (ours) & 56.631  & \bf{14.33\%} & 0.107 & \bf{17.33}\% & 0.641 & \bf{18.50\%} & 0.451\\
    \midrule
    DeepHQ-MS (w/o SC, ablation) & 17.564  & -5.50\% &0.043 & -6.94\% & 0.236 & -6.94\% & 0.169\\
    \bottomrule
\end{tabular}
\end{minipage}

\vspace{0.2cm}

% ---- figure ----
\begin{minipage}{\textwidth}
\centering
\begin{subfigure}{0.32\textwidth}
    \includegraphics[width=\linewidth, clip, trim={1.7cm 5.8cm 1.7cm 6.5cm}]{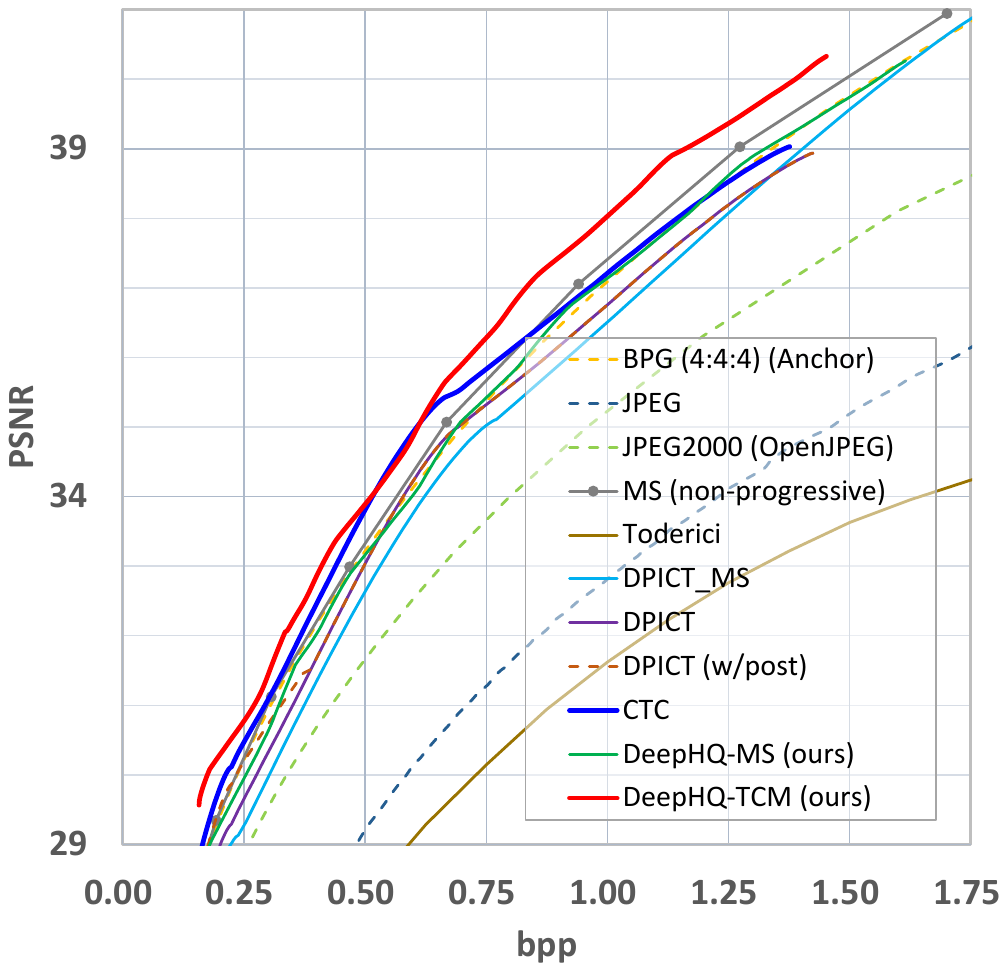}
    \caption{(a) \textit{Kodak}}
\end{subfigure}
\begin{subfigure}{0.33\textwidth}
    \includegraphics[width=\linewidth, clip, trim={1.7cm 6.0cm 1.7cm 6cm}]{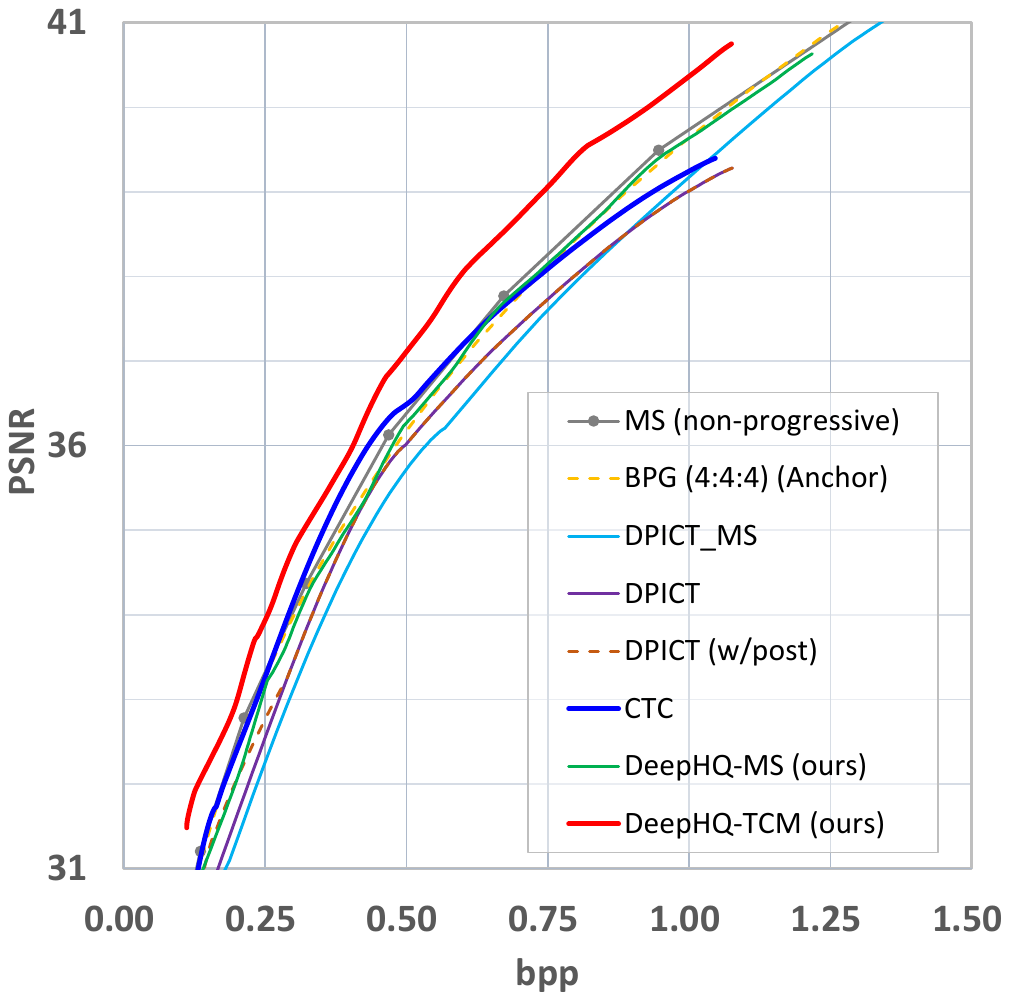}
    \caption{(b) \textit{CLIC}}
\end{subfigure}
\begin{subfigure}{0.32\textwidth}
    \includegraphics[width=\linewidth, clip, trim={1.7cm 5.6cm 1.7cm 6.6cm}]{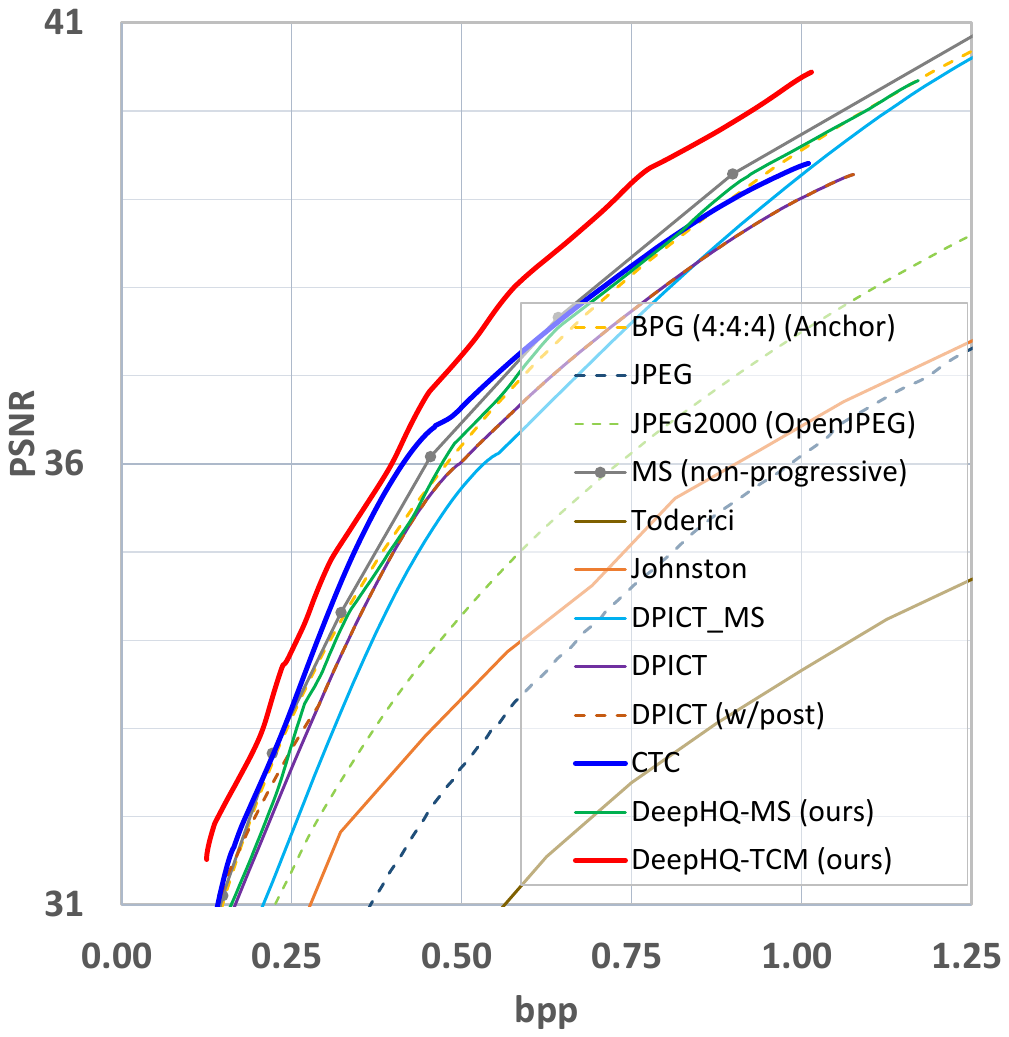}
    \caption{(c) \textit{Tecnick}}
\end{subfigure}
\vspace{-0.3cm}\caption{R-D curves of our DeepHQ models and other progressive compression models, showing bits-per-pixel (bpp) versus Peak Signal-to-Noise Ratio (PSNR). Two competing models (our DeepHQ-TCM and CTC~\cite{2023_CVPR_jeon}) are highlighted with bold lines.}
\label{fig:curves}
\end{minipage}
\vspace{-0.3cm}
\end{figure*}

We compare the proposed DeepHQ with the existing progressive neural image codecs\footnote{The recent approaches to learned progressive coding,  Lu~\etal\cite{Lu_progressive_2021} and Liu~\etal\cite{Li2023deadzone}, could not be compared with our DeepHQ because their source codes are not publicly available. Nevertheless, in our visual comparison on the R-D curves, we observed that our DeepHQ has shown much superior results in coding efficiency.}~\cite{Toderici17,Johnston2018,Lee_2022_CVPR,2023_CVPR_jeon} as well as the traditional codecs, JPEG~\cite{Penn92} and JPEG2000~\cite{Taubman2001}, with optional progressive coding modes. Because our DeepHQ achieves significantly superior coding efficiency compared to the conventional codecs~\cite{Penn92,Taubman2001} (See Fig.~\ref{fig:curves}), we particularly focus on comparison with the DPICT~\cite{Lee_2022_CVPR} and the CTC~\cite{2023_CVPR_jeon}, which are state-of-the-art learned PIC methods. It should be noted that the "DPICT" indicates the version of DPICT~\cite{Lee_2022_CVPR} without using the post-processing networks, while we refer to another version of it with the multiple post-processing networks as "DPICT (w/ post)".
For the DPICT~\cite{Lee_2022_CVPR} and CTC~\cite{2023_CVPR_jeon} models, we utilized the publicly available official source codes and their pre-trained models shared by the authors, but we excluded the padded areas in calculation of rate and distortion for fair and more precise comparison\footnote{The open source code~\cite{DPICT_github} of DPICT~\cite{Lee_2022_CVPR} and that~\cite{CTC_github} of CTC~\cite{2023_CVPR_jeon} includes the padded areas in calculation of rate and distortion, which can lead to boosting the PSNR values and reducing the bpp (bits-per-pixel) values.}.
%This is very important because the padded areas affect both the bpp and PSNR results. 
Also for comparison, a variant of DPICT~\cite{Lee_2022_CVPR} is implemented based on the Mean-scale~\cite{Minnen2018} architecture, which is denoted as "DPICT\_MS", where all the default settings are kept, but the same CLIC~\cite{CLIC} training dataset was used as ours. For the evaluation of our DeepHQ, we used a total of $162$ progressive compression points for each input image to closely match the $164$ points in the DPICT~\cite{Lee_2022_CVPR} models and $160$ points in the CTC~\cite{2023_CVPR_jeon} model.

\begin{figure*}[!htb]
    \captionsetup[subfigure]{labelformat=empty}
    \setlength\columnsep{0pt}
    \captionsetup{aboveskip=0pt}
    \captionsetup{belowskip=5pt}
    \centering

    \begin{subfigure}{0.02\textwidth}
        \centering
        \caption{}
        \vspace{-0.7cm}
    \end{subfigure}
    \begin{subfigure}{0.158\textwidth}
        \centering
        \caption{DPICT\_MS}
        \vspace{-0.7cm}
    \end{subfigure}
    \begin{subfigure}{0.158\textwidth}
        \centering
        \caption{DPICT~\cite{Lee_2022_CVPR} (w/post)}
        \vspace{-0.7cm}
    \end{subfigure}
    \begin{subfigure}{0.158\textwidth}
        \centering
        \caption{CTC~\cite{2023_CVPR_jeon}}
        \vspace{-0.7cm}
    \end{subfigure}
    \begin{subfigure}{0.158\textwidth}
        \centering
        \caption{DeepHQ-MS (ours)}
        \vspace{-0.7cm}
    \end{subfigure}
    \begin{subfigure}{0.158\textwidth}
        \centering
        \caption{DeepHQ-TCM (ours)}
        \vspace{-0.7cm}
    \end{subfigure}
    \begin{subfigure}{0.158\textwidth}
        \centering
        \caption{GT}
        \vspace{-0.7cm}
    \end{subfigure}

    \begin{subfigure}{0.02\textwidth}
        \rotatebox{90}{\makebox[2.6cm]{\hspace{-0.5cm}Rate point 1}}
    \end{subfigure}
    \begin{subfigure}{0.158\textwidth}
        \centering
        \includegraphics[scale=0.22]{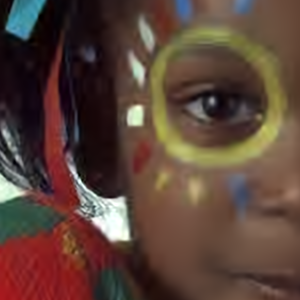}
        \vspace{-0.6cm}
        \caption{0.1017 / 28.610dB}
        \vspace{-0.7cm}
    \end{subfigure}
    \begin{subfigure}{0.158\textwidth}
        \centering
        \includegraphics[scale=0.22]{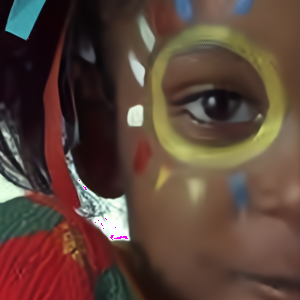}
        \vspace{-0.6cm}
        \caption{0.0959 / 29.439dB}
        \vspace{-0.7cm}
    \end{subfigure}
    \begin{subfigure}{0.158\textwidth}
        \centering
        \begin{tikzpicture}
          \node[anchor=south west, inner sep=0] (img) at (0,0)
            {\includegraphics[scale=0.22]{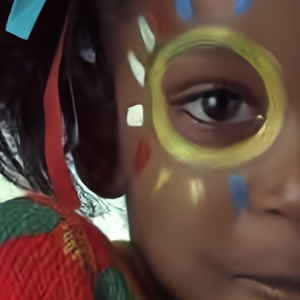}};
          \begin{scope}[x={(img.south east)}, y={(img.north west)}]
            \draw[red, line width=0.5pt] (0.58,0.63) rectangle (0.88,0.78);
            \draw[red, line width=0.5pt] (0.05,0.05) rectangle (0.29,0.38);
          \end{scope}
        \end{tikzpicture}
        \vspace{-0.6cm}
        \caption{0.0969 / 29.895dB}
        \vspace{-0.7cm}
    \end{subfigure}
    \begin{subfigure}{0.158\textwidth}
        \centering
        \includegraphics[scale=0.22]{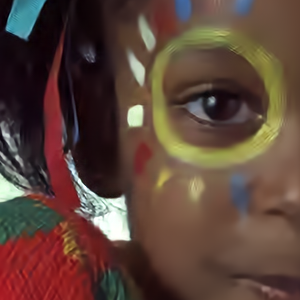}
        \vspace{-0.6cm}
        \caption{0.1020 / 29.937dB}
        \vspace{-0.7cm}
    \end{subfigure}
    \begin{subfigure}{0.158\textwidth}
        \centering
        \begin{tikzpicture}
          \node[anchor=south west, inner sep=0] (img) at (0,0)
            {\includegraphics[scale=0.22]{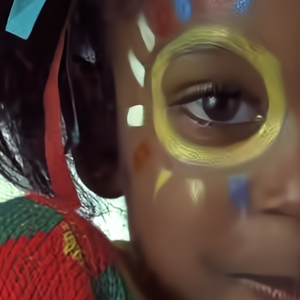}};
          \begin{scope}[x={(img.south east)}, y={(img.north west)}]
            \draw[red, line width=0.5pt] (0.58,0.63) rectangle (0.88,0.78);
            \draw[red, line width=0.5pt] (0.05,0.05) rectangle (0.29,0.38);
          \end{scope}
        \end{tikzpicture}
        \vspace{-0.6cm}
        \caption{0.0965 / \bf{31.127}dB}
        \vspace{-0.7cm}
    \end{subfigure}
    \begin{subfigure}{0.158\textwidth}
        \centering
        \includegraphics[scale=0.22]{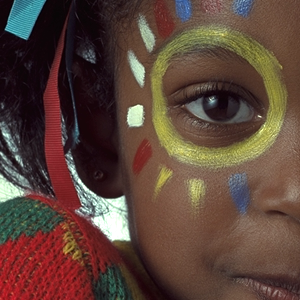}
        \vspace{-0.6cm}
        \caption{ }
        \vspace{-0.7cm}
    \end{subfigure}

    \begin{subfigure}{0.02\textwidth}
        \rotatebox{90}{\makebox[2.6cm]{\hspace{-0.5cm}Rate point 2}}
    \end{subfigure}
    \begin{subfigure}{0.158\textwidth}
        \centering
        \includegraphics[scale=0.22]{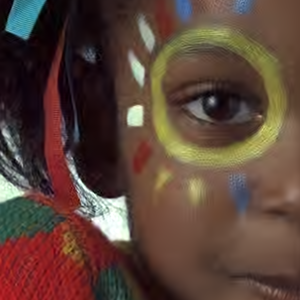}
        \vspace{-0.6cm}
        \caption{0.1781 / 30.618dB}
        \vspace{-0.7cm}
    \end{subfigure}
    \begin{subfigure}{0.158\textwidth}
        \centering
        \includegraphics[scale=0.22]{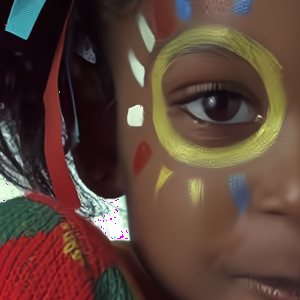}
        \vspace{-0.6cm}
        \caption{0.1903 / 31.741dB}
        \vspace{-0.7cm}
    \end{subfigure}
    \begin{subfigure}{0.158\textwidth}
        \centering
        \begin{tikzpicture}
          \node[anchor=south west, inner sep=0] (img) at (0,0)
            {\includegraphics[scale=0.22]{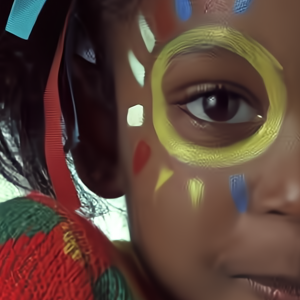}};
          \begin{scope}[x={(img.south east)}, y={(img.north west)}]
            \draw[red, line width=0.5pt] (0.53,0.53) rectangle (0.83,0.65);
            \draw[red, line width=0.5pt] (0.05,0.20) rectangle (0.29,0.38);
          \end{scope}
        \end{tikzpicture}
        \vspace{-0.6cm}
        \caption{0.1828 / 32.085dB}
        \vspace{-0.7cm}
    \end{subfigure}
    \begin{subfigure}{0.158\textwidth}
        \centering
        \includegraphics[scale=0.22]{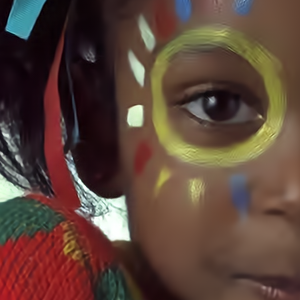}
        \vspace{-0.6cm}
        \caption{0.1871 / 31.726dB}
        \vspace{-0.7cm}
    \end{subfigure}
    \begin{subfigure}{0.158\textwidth}
        \centering
        \begin{tikzpicture}
          \node[anchor=south west, inner sep=0] (img) at (0,0)
            {\includegraphics[scale=0.22]{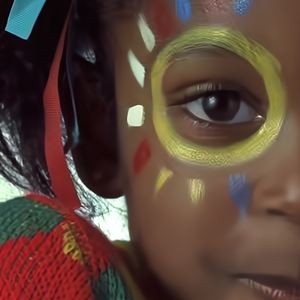}};
          \begin{scope}[x={(img.south east)}, y={(img.north west)}]
            \draw[red, line width=0.5pt] (0.53,0.53) rectangle (0.83,0.65);
            \draw[red, line width=0.5pt] (0.05,0.20) rectangle (0.29,0.38);
          \end{scope}
        \end{tikzpicture}
        \vspace{-0.6cm}
        \caption{0.1863 / \bf{32.857dB}}
        \vspace{-0.7cm}
    \end{subfigure}
    \begin{subfigure}{0.158\textwidth}
        \centering
        \vspace{-0.6cm}
        \caption{ }
        \vspace{-0.7cm}
    \end{subfigure}

    \begin{subfigure}{0.02\textwidth}
        \rotatebox{90}{\makebox[2.6cm]{\hspace{-0.5cm}Rate point 3}}
    \end{subfigure}
    \begin{subfigure}{0.158\textwidth}
        \centering
        \includegraphics[scale=0.22]{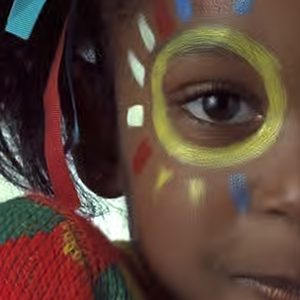}
        \vspace{-0.6cm}
        \caption{0.2644 / 32.168dB}
        \vspace{-0.7cm}
    \end{subfigure}
    \begin{subfigure}{0.158\textwidth}
        \centering
        \includegraphics[scale=0.22]{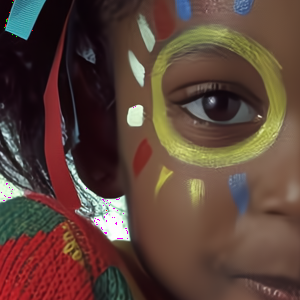}
        \vspace{-0.6cm}
        \caption{0.2732 / 32.756dB}
        \vspace{-0.7cm}
    \end{subfigure}
    \begin{subfigure}{0.158\textwidth}
        \centering
        \begin{tikzpicture}
          \node[anchor=south west, inner sep=0] (img) at (0,0)
            {\includegraphics[scale=0.22]{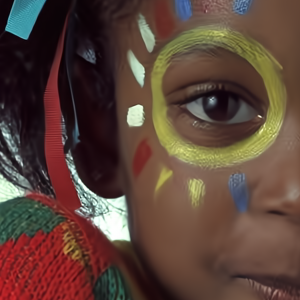}};
          \begin{scope}[x={(img.south east)}, y={(img.north west)}]
            \draw[red, line width=0.5pt] (0.48,0.3) rectangle (0.6,0.5);
            \draw[red, line width=0.5pt] (0.01,0.85) rectangle (0.2,0.99);
            \draw[red, line width=0.5pt] (0.77,0.02) rectangle (0.85,0.1);
          \end{scope}
        \end{tikzpicture}
        \vspace{-0.6cm}
        \caption{0.2707 / 33.581dB}
        \vspace{-0.7cm}
    \end{subfigure}
    \begin{subfigure}{0.158\textwidth}
        \centering
        \includegraphics[scale=0.22]{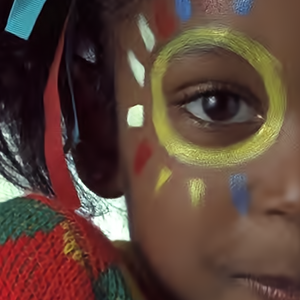}
        \vspace{-0.6cm}
        \caption{0.2732 / 33.356dB}
        \vspace{-0.7cm}
    \end{subfigure}
    \begin{subfigure}{0.158\textwidth}
        \centering
        \begin{tikzpicture}
          \node[anchor=south west, inner sep=0] (img) at (0,0)
            {\includegraphics[scale=0.22]{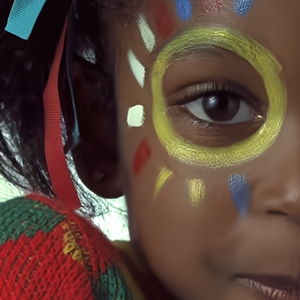}};
          \begin{scope}[x={(img.south east)}, y={(img.north west)}]
            \draw[red, line width=0.5pt] (0.48,0.3) rectangle (0.6,0.5);
            \draw[red, line width=0.5pt] (0.01,0.85) rectangle (0.2,0.99);
            \draw[red, line width=0.5pt] (0.77,0.02) rectangle (0.85,0.1);
          \end{scope}
        \end{tikzpicture}
        \vspace{-0.6cm}
        \caption{0.2699 / \bf{34.312}dB}
        \vspace{-0.7cm}
    \end{subfigure}
    \begin{subfigure}{0.158\textwidth}
        \centering
        \vspace{-0.6cm}
        \caption{ }
        \vspace{-0.7cm}
    \end{subfigure}

        \begin{subfigure}{0.02\textwidth}
        \rotatebox{90}{\makebox[2.6cm]{\hspace{-0.5cm}Rate point 1}}
    \end{subfigure}
    \begin{subfigure}{0.158\textwidth}
        \centering
        \includegraphics[scale=0.165]{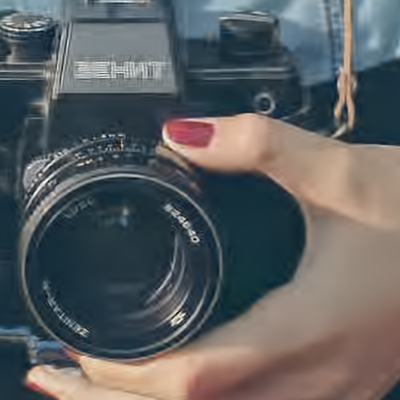}
        \vspace{-0.6cm}
        \caption{0.0864 / 29.872dB}
        \vspace{-0.7cm}
    \end{subfigure}
    \begin{subfigure}{0.158\textwidth}
        \centering
        \includegraphics[scale=0.165]{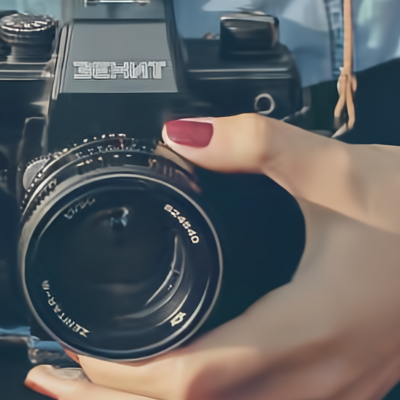}
        \vspace{-0.6cm}
        \caption{0.0896 / 30.870dB}
        \vspace{-0.7cm}
    \end{subfigure}
    \begin{subfigure}{0.158\textwidth}
        \centering
        \begin{tikzpicture}
          \node[anchor=south west, inner sep=0] (img) at (0,0)
            {\includegraphics[scale=0.165]{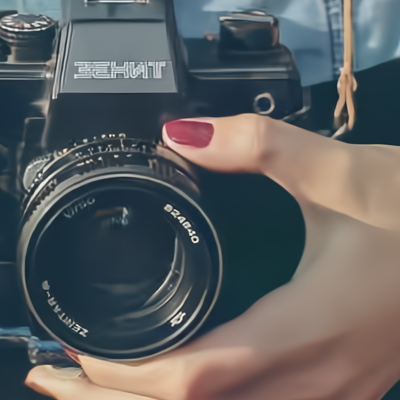}};
          \begin{scope}[x={(img.south east)}, y={(img.north west)}]
            \draw[red, line width=0.5pt] (0.26,0.60) rectangle (0.36,0.68);
            \draw[red, line width=0.5pt] (0.1,0.14) rectangle (0.23,0.31);
            \draw[red, line width=0.5pt] (0.8,0.72) rectangle (0.92,0.85);
            \draw[red, line width=0.5pt] (0.3,0.78) rectangle (0.45,0.86);
            \draw[red, line width=0.5pt] (0.05,0.51) rectangle (0.15,0.63);
            \draw[red, line width=0.5pt] (0.1,0.03) rectangle (0.23,0.11);
          \end{scope}
        \end{tikzpicture}
        \vspace{-0.6cm}
        \caption{0.0881 / 31.145dB}
        \vspace{-0.7cm}
    \end{subfigure}
    \begin{subfigure}{0.158\textwidth}
        \centering
        \includegraphics[scale=0.165]{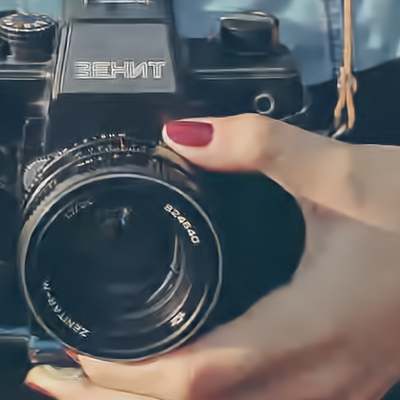}
        \vspace{-0.6cm}
        \caption{0.0899 / 31.129dB}
        \vspace{-0.7cm}
    \end{subfigure}
    \begin{subfigure}{0.158\textwidth}
        \centering
        \begin{tikzpicture}
          \node[anchor=south west, inner sep=0] (img) at (0,0)
            {\includegraphics[scale=0.165]{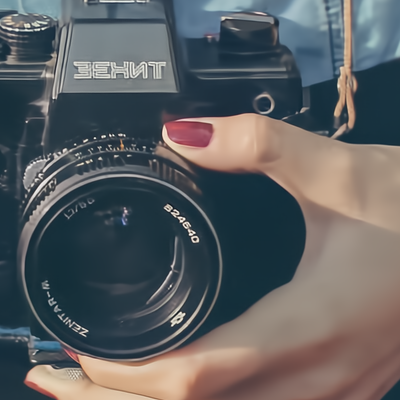}};
          \begin{scope}[x={(img.south east)}, y={(img.north west)}]
            \draw[red, line width=0.5pt] (0.26,0.60) rectangle (0.36,0.68);
            \draw[red, line width=0.5pt] (0.1,0.14) rectangle (0.23,0.31);
            \draw[red, line width=0.5pt] (0.8,0.72) rectangle (0.92,0.85);
            \draw[red, line width=0.5pt] (0.3,0.78) rectangle (0.45,0.86);
            \draw[red, line width=0.5pt] (0.05,0.51) rectangle (0.15,0.63);
            \draw[red, line width=0.5pt] (0.1,0.03) rectangle (0.23,0.11);
          \end{scope}
        \end{tikzpicture}
        \vspace{-0.6cm}
        \caption{0.0877 / \bf{32.648}dB}
        \vspace{-0.7cm}
    \end{subfigure}
    \begin{subfigure}{0.158\textwidth}
        \centering
        \includegraphics[scale=0.165]{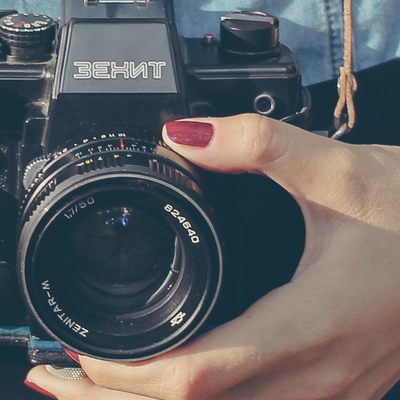}
        \vspace{-0.6cm}
        \caption{ }
        \vspace{-0.7cm}
    \end{subfigure}

    \begin{subfigure}{0.02\textwidth}
        \rotatebox{90}{\makebox[2.6cm]{\hspace{-0.5cm}Rate point 2}}
    \end{subfigure}
    \begin{subfigure}{0.158\textwidth}
        \centering
        \includegraphics[scale=0.165]{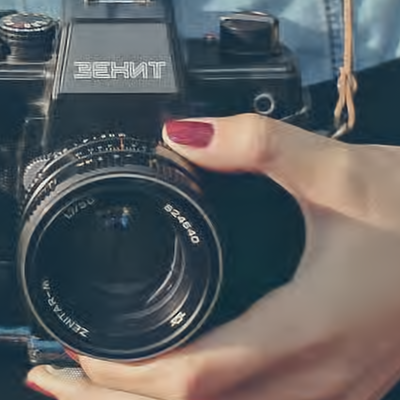}
        \vspace{-0.6cm}
        \caption{0.1697 / 32.592dB}
        \vspace{-0.7cm}
    \end{subfigure}
    \begin{subfigure}{0.158\textwidth}
        \centering
        \includegraphics[scale=0.165]{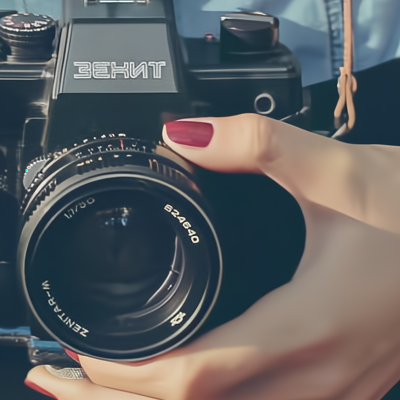}
        \vspace{-0.6cm}
        \caption{0.1701 / 33.280dB}
        \vspace{-0.7cm}
    \end{subfigure}
    \begin{subfigure}{0.158\textwidth}
        \centering
        \begin{tikzpicture}
          \node[anchor=south west, inner sep=0] (img) at (0,0)
            {\includegraphics[scale=0.165]{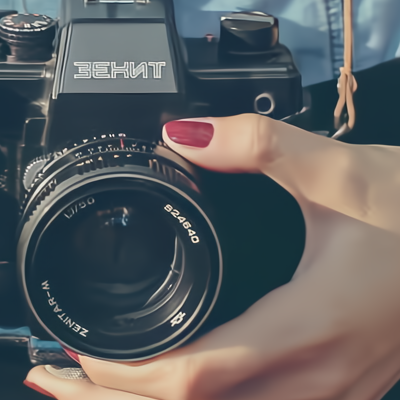}};
          \begin{scope}[x={(img.south east)}, y={(img.north west)}]
            \draw[red, line width=0.5pt] (0.1,0.03) rectangle (0.23,0.11);
            \draw[red, line width=0.5pt] (0.6,0.55) rectangle (0.73,0.73);
            \draw[red, line width=0.5pt] (0.8,0.65) rectangle (0.92,0.99);
          \end{scope}
        \end{tikzpicture}
        \vspace{-0.6cm}
        \caption{0.1650 / 33.598dB}
        \vspace{-0.7cm}
    \end{subfigure}
    \begin{subfigure}{0.158\textwidth}
        \centering
        \includegraphics[scale=0.165]{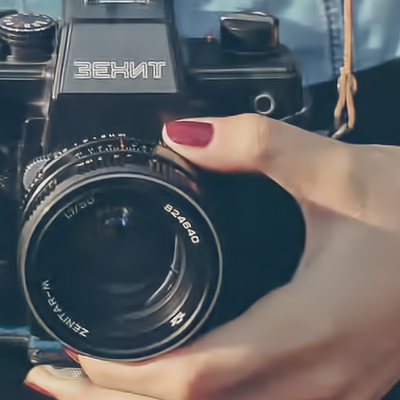}
        \vspace{-0.6cm}
        \caption{0.1638 / 33.310dB}
        \vspace{-0.7cm}
    \end{subfigure}
    \begin{subfigure}{0.158\textwidth}
        \centering
        \begin{tikzpicture}
          \node[anchor=south west, inner sep=0] (img) at (0,0)
            {\includegraphics[scale=0.165]{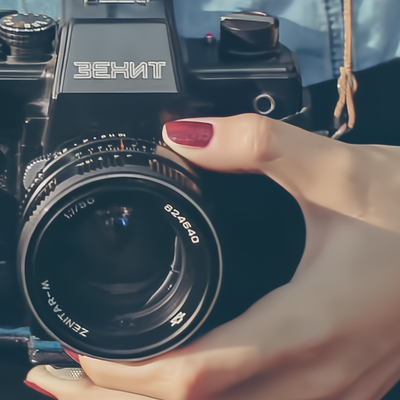}};
          \begin{scope}[x={(img.south east)}, y={(img.north west)}]
            \draw[red, line width=0.5pt] (0.1,0.03) rectangle (0.23,0.11);
            \draw[red, line width=0.5pt] (0.6,0.55) rectangle (0.73,0.73);
            \draw[red, line width=0.5pt] (0.8,0.65) rectangle (0.92,0.99);
          \end{scope}
        \end{tikzpicture}
        \vspace{-0.6cm}
        \caption{0.1642 / \bf{34.700}dB}
        \vspace{-0.7cm}
    \end{subfigure}
    \begin{subfigure}{0.158\textwidth}
        \centering
        \caption{ }
        \vspace{-0.7cm}
    \end{subfigure}

    \begin{subfigure}{0.02\textwidth}
        \rotatebox{90}{\makebox[2.6cm]{\hspace{-0.5cm}Rate point 3}}
    \end{subfigure}
    \begin{subfigure}{0.158\textwidth}
        \centering
        \includegraphics[scale=0.165]{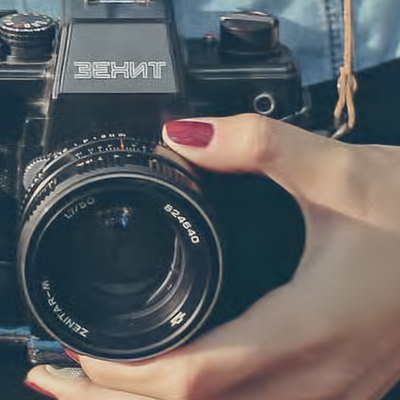}
        \vspace{-0.6cm}
        \caption{0.2265 / 34.058dB}
        \vspace{-0.7cm}
    \end{subfigure}
    \begin{subfigure}{0.158\textwidth}
        \centering
        \includegraphics[scale=0.165]{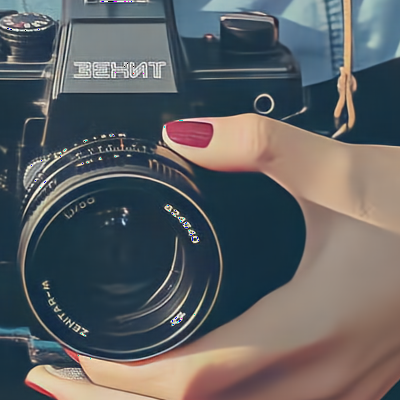}
        \vspace{-0.6cm}
        \caption{0.2214 / 34.164dB}
        \vspace{-0.7cm}
    \end{subfigure}
    \begin{subfigure}{0.158\textwidth}
        \centering
        \begin{tikzpicture}
          \node[anchor=south west, inner sep=0] (img) at (0,0)
            {\includegraphics[scale=0.165]{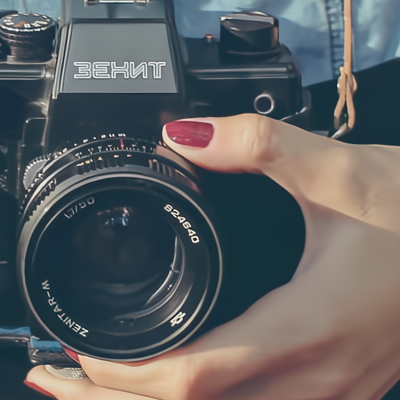}};
          \begin{scope}[x={(img.south east)}, y={(img.north west)}]
            \draw[red, line width=0.5pt] (0.1,0.03) rectangle (0.23,0.11);
            \draw[red, line width=0.5pt] (0.43,0.5) rectangle (0.5,0.6);
            \draw[red, line width=0.5pt] (0.8,0.65) rectangle (0.92,0.99);
          \end{scope}
        \end{tikzpicture}
        \vspace{-0.6cm}
        \caption{0.2191 / 35.020dB}
        \vspace{-0.7cm}
    \end{subfigure}
    \begin{subfigure}{0.158\textwidth}
        \centering
        \includegraphics[scale=0.165]{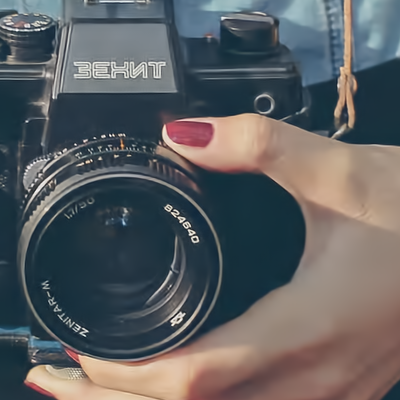}
        \vspace{-0.6cm}
        \caption{0.2204 / 34.936dB}
        \vspace{-0.7cm}
    \end{subfigure}
    \begin{subfigure}{0.158\textwidth}
        \centering
        \begin{tikzpicture}
          \node[anchor=south west, inner sep=0] (img) at (0,0)
            {\includegraphics[scale=0.165]{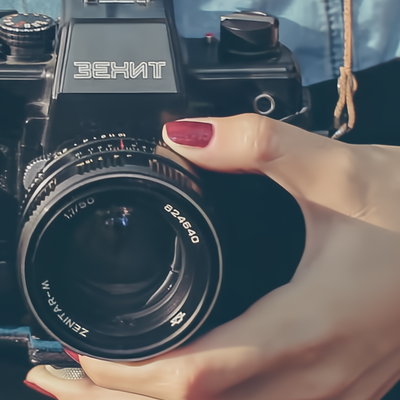}};
          \begin{scope}[x={(img.south east)}, y={(img.north west)}]
            \draw[red, line width=0.5pt] (0.1,0.03) rectangle (0.23,0.11);
            \draw[red, line width=0.5pt] (0.43,0.5) rectangle (0.5,0.6);
            \draw[red, line width=0.5pt] (0.8,0.65) rectangle (0.92,0.99);
          \end{scope}
        \end{tikzpicture}
        \vspace{-0.6cm}
        \caption{0.2186 / \bf{36.021}dB}
        \vspace{-0.7cm}
    \end{subfigure}
    \begin{subfigure}{0.158\textwidth}
        \centering
        \caption{ }
        \vspace{-0.7cm}
    \end{subfigure}
    \vspace{0.4cm}
    \caption{Cropped reconstruction samples of (top) the \textit{KODIM15} (Kodak~\cite{KODAK}) image and (bottom) the \textit{sergey-zolkin-1045} (CLIC~\cite{CLIC}) image (best viewed in zoomed-in digital format). The two numbers below each image represent bpp and PSNR, respectively. Red boxes highlight quality differences between the proposed method and the competing CTC~\cite{2023_CVPR_jeon} model.}
    \label{fig:visual_test_results}
    \vspace{-0.7cm}
\end{figure*}

\begin{figure*}[t]
    \captionsetup[subfigure]{labelformat=empty}
    \captionsetup[subfigure]{font=scriptsize,labelfont=scriptsize, aboveskip=2pt, belowskip=3pt}
    \setlength\columnsep{0pt}
    \captionsetup{belowskip=5pt}
    \setlength{\baselineskip}{0pt}
    \centering
    \begin{subfigure}{0.335\textwidth}
        \centering
        \includegraphics[width=\linewidth, clip, trim={0.0cm 0.2cm 0.0cm 0.2cm}]{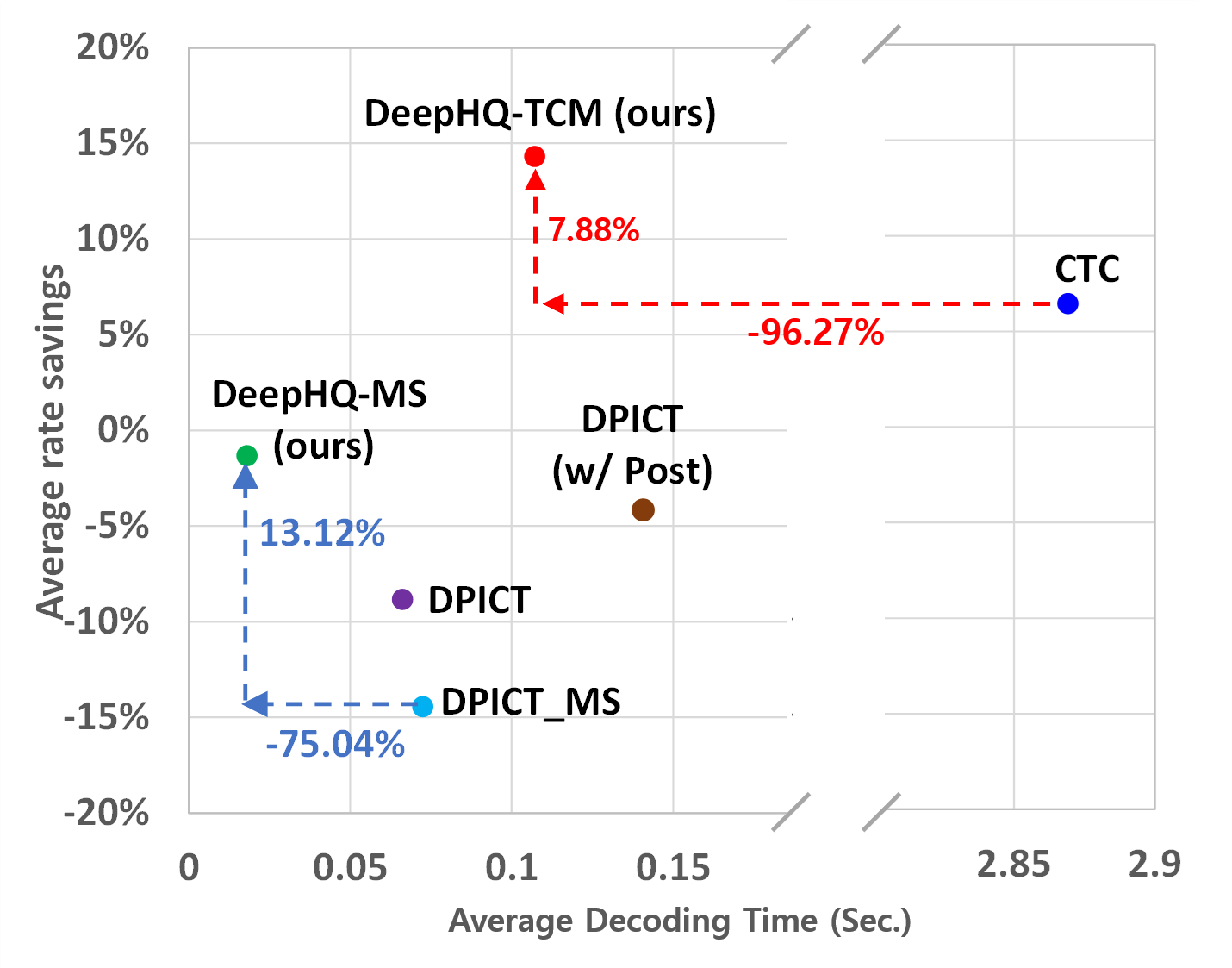}
        \caption{(a) \textit{Kodak}}
        \vspace{-0.3cm}
    \end{subfigure}
    \begin{subfigure}{0.335\textwidth}
        \centering
        \includegraphics[width=\linewidth, clip, trim={0.0cm 0.2cm 0.0cm 0.2cm}]{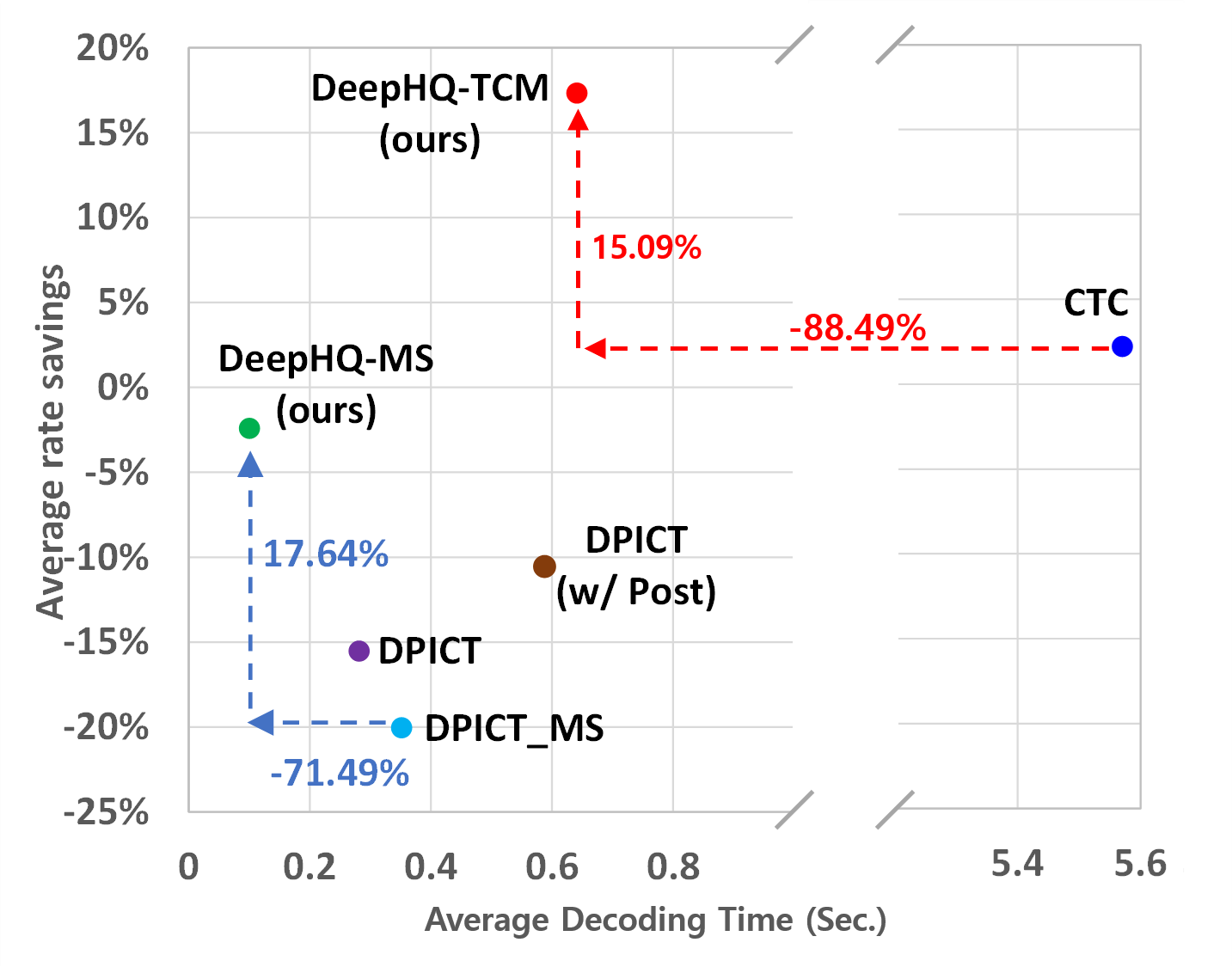}
        \caption{(b) \textit{CLIC}}
        \vspace{-0.3cm}
    \end{subfigure}
    \begin{subfigure}{0.32\textwidth}
        \centering
        \includegraphics[width=\linewidth, clip, trim={0.0cm 0.2cm 0.0cm 0.2cm}]{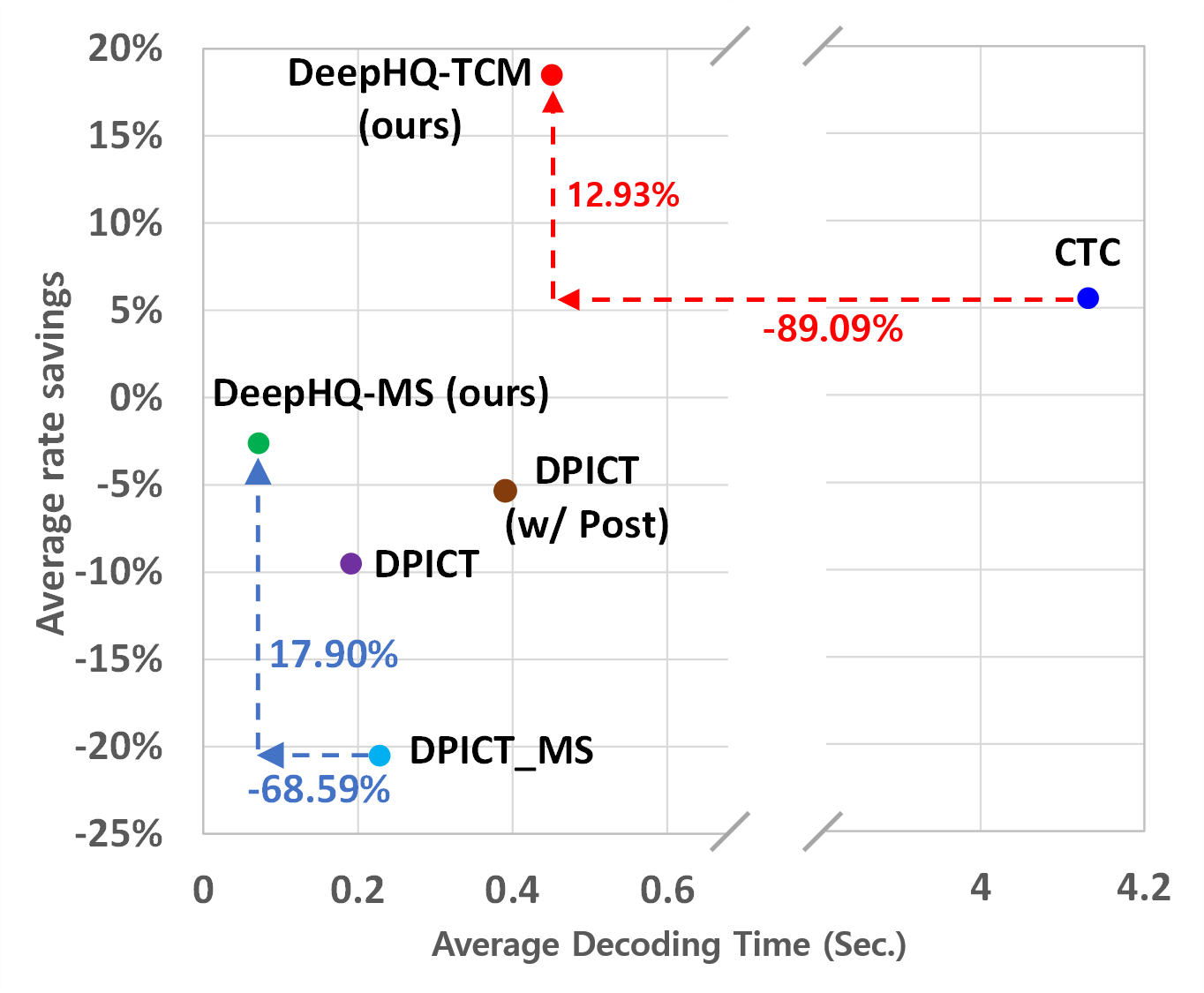}
        \caption{(c) \textit{Tecnick}}
        \vspace{-0.3cm}
    \end{subfigure}
    \caption{Average decoding time (per reconstruction) versus rate savings. 
    Decoding times are measured on RTX 8000 GPU and Xeon Gold6244 CPU @ 3.6GHz (8 Cores) x 2.
    Model loading times are excluded for all methods. "MS" denotes the Mean-scale~\cite{Minnen2018} base compression model.}
    \vspace{-0.2cm}
    \label{fig:decoding_time}
\end{figure*}

The comparison was performed on the Kodak PhotoCD (24 images)~\cite{KODAK}, CLIC \textit{professional} (41 images)~\cite{CLIC}, and Tecnick (100 images)~\cite{tecnick} datasets. To evaluate coding efficiency, we use average rate savings against BPG~\cite{BPG} rather than the BD-rate metric determined based on only four (or six) compression points. We use the BPG~\cite{BPG} as an anchor codec because it supports a wide bit-rate range enough to be compared with various other codecs. 
It should be noted that our proposed DeepHQ-TCM model supports higher-quality compression by approximately 1dB of PSNR compared to DPICT~\cite{Lee_2022_CVPR} and CTC~\cite{2023_CVPR_jeon} models. Whereas, the CTC~\cite{2023_CVPR_jeon} and DPICT~\cite{Lee_2022_CVPR} support very low-quality image compression, starting from around PSNR 20dB, but their coding efficiency in low-quality compression is significantly compromised. Therefore, including all low-quality compression cases in the experiments can potentially distort the results in favor of the proposed DeepHQ. To prevent this, we determine the PSNR range to measure the average rate savings as follows:
\begin{align}
\label{eq:PSNR_range}
\text{PSNR}^\text{high}=\text{PSNR}_\text{DeepHQ\_L}^\text{high}, \text{\ \ \ \ \ PSNR}^\text{low}=\text{PSNR}_\text{DeepHQ\_L}^\text{low} - D,
\end{align}
where $\text{PSNR}^\text{high}$ and $\text{PSNR}^\text{low}$ represent the highest and lowest PSNR values, respectively, of the PSNR comparison range, $\text{PSNR}_\text{DeepHQ\_L}^\text{high}$ $\text{PSNR}_\text{DeepHQ\_L}^\text{low}$ are the highest and lowest PSNR values, respectively, of our DeepHQ-TCM model, and $D$ is the absolute difference between the highest PSNR values of the CTC and DeepHQ\_LARGE models. Additionally, we comprehensively evaluate the superiority between methods by considering their model sizes and decoding times together. 
Since both the proposed method and competing approaches are optimized for MSE, we report PSNR as the primary performance metric. In addition, to demonstrate the superiority of the proposed method in various evaluation criteria, we also present experimental results on Fr\'echet Inception Distance (FID)~\cite{heusel2017gans} and Learned Perceptual Image Patch Similarity (LPIPS)~\cite{zhang2018unreasonable}, two metrics widely adopted in recent LIC studies that aim to improve perceptual quality. 
Since the FID metric measures the distributional discrepancy between entire datasets, we use the DIV2K~\cite{Agustsson_2017_CVPR_Workshops} high-resolution dataset containing 900 images, consisting of 800 training and 100 validation images.

\begin{figure*}
    \captionsetup[subfigure]{labelformat=empty}
    \captionsetup[subfigure]{font=scriptsize, labelfont=scriptsize, aboveskip=2pt, belowskip=3pt}
    \setlength\columnsep{0pt}
    \captionsetup{belowskip=5pt}
    \setlength{\baselineskip}{0pt}
    \centering
    \begin{subfigure}{0.33\textwidth}
        \centering
        \includegraphics[width=\linewidth, clip, trim={0.0cm 0.2cm 0.0cm 0.2cm}]{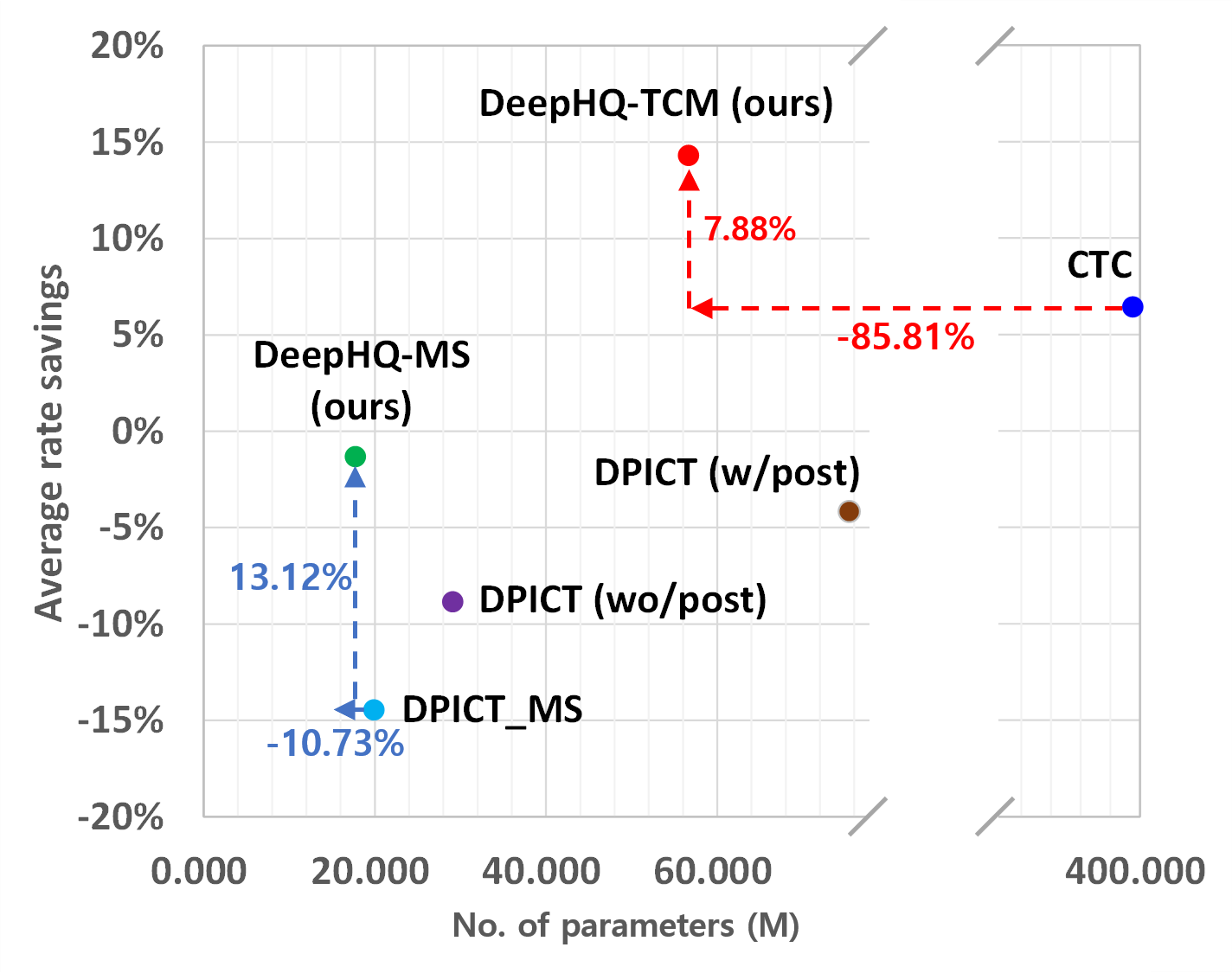}
        \caption{(a) \textit{Kodak}}
        \vspace{-0.3cm}
    \end{subfigure}
    \begin{subfigure}{0.33\textwidth}
        \centering
        \includegraphics[width=\linewidth, clip, trim={0.0cm 0.2cm 0.0cm 0.2cm}]{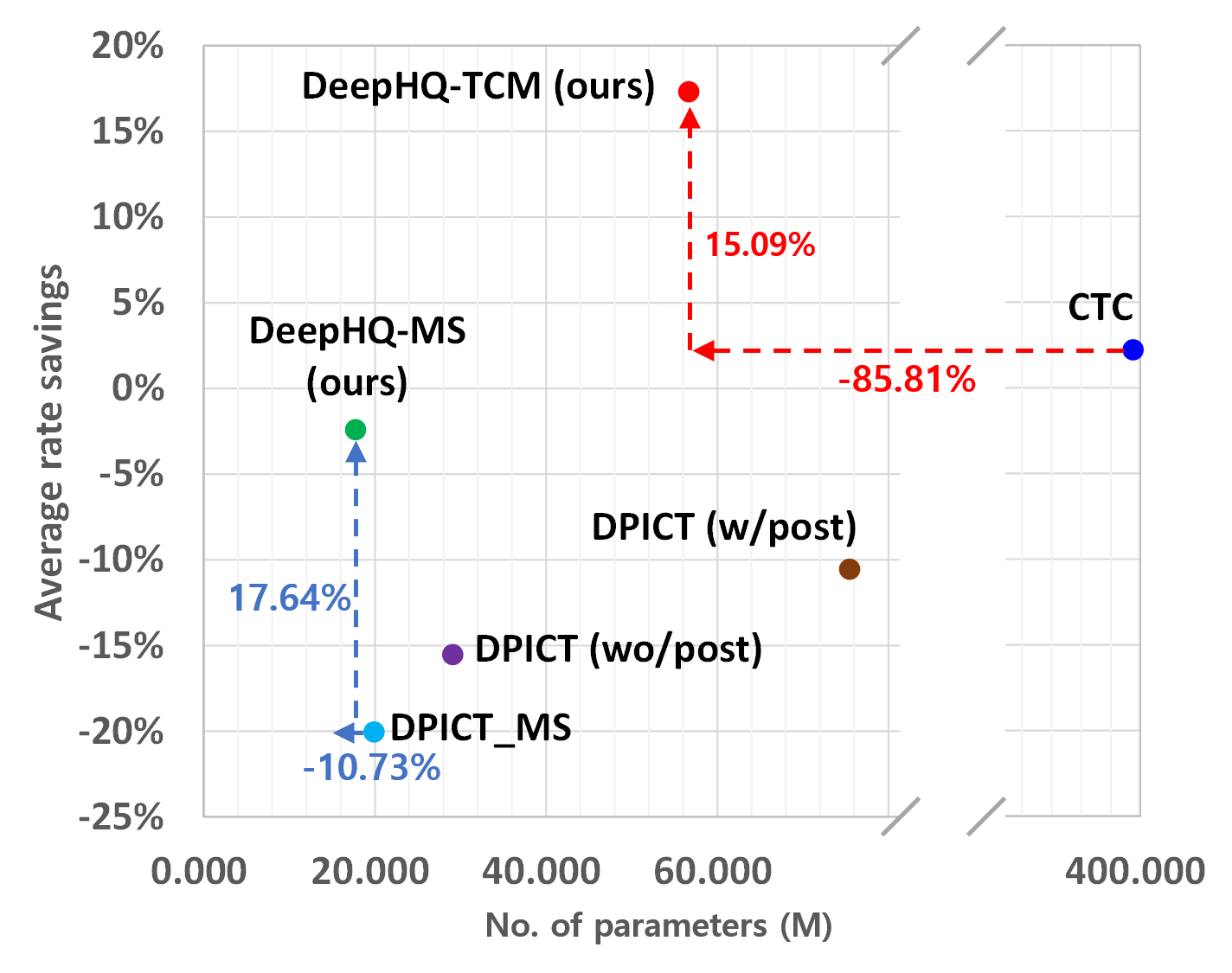}
        \caption{(b) \textit{CLIC}}
        \vspace{-0.3cm}
    \end{subfigure}
    \begin{subfigure}{0.33\textwidth}
        \centering
        \includegraphics[width=\linewidth, clip, trim={0.0cm 0.2cm 0.0cm 0.2cm}]{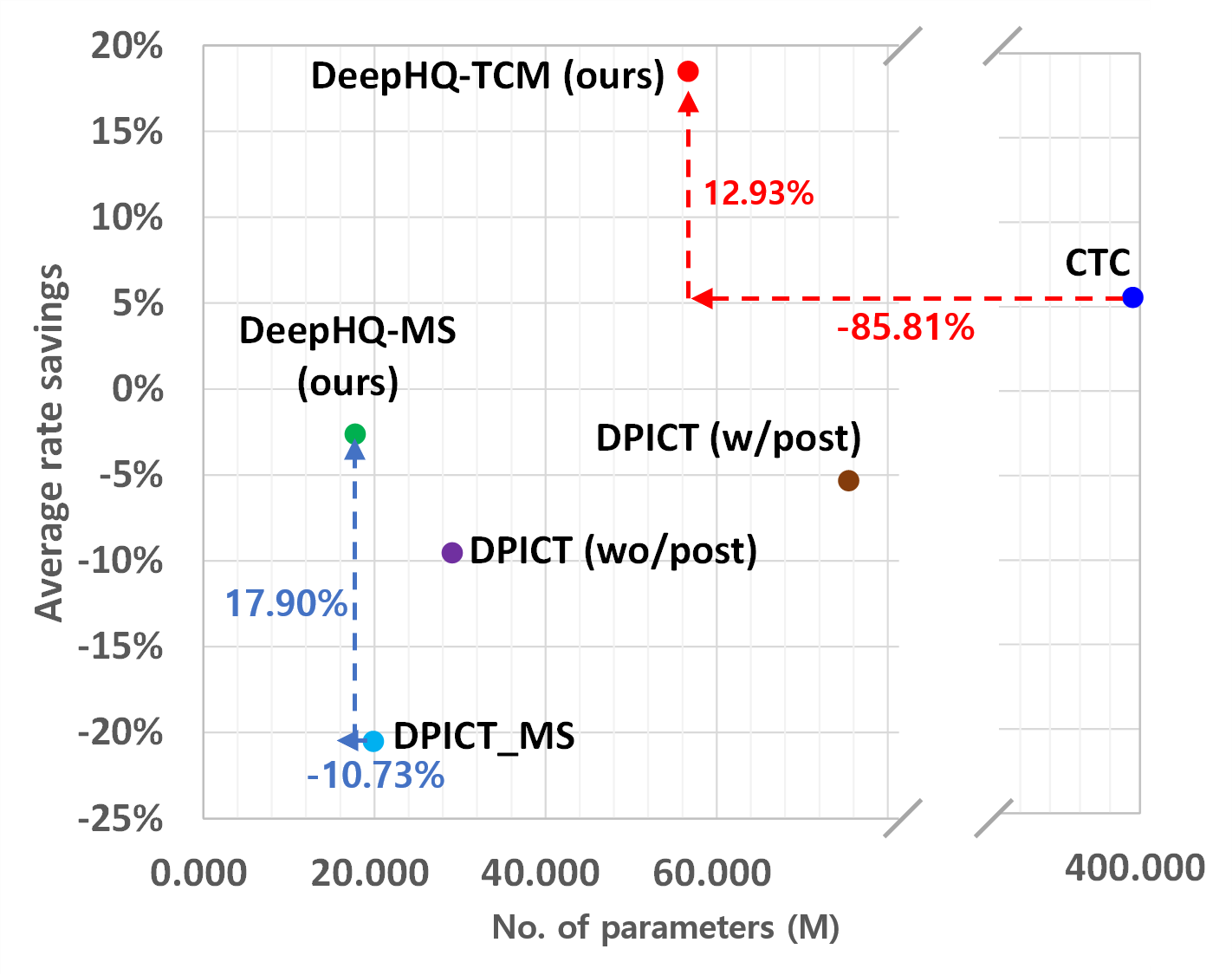}
        \caption{(c) \textit{Tecnick}}
        \vspace{-0.3cm}
    \end{subfigure}
    \caption{Model size versus rate savings. "MS" denotes the Mean-scale~\cite{Minnen2018} base compression model.}
    \vspace{-0.2cm}
    \label{fig:param_ratesavings}
\end{figure*}

\begin{figure}[!t]
\captionsetup{belowskip=5pt}
\centering
\begin{minipage}[t]{0.62\linewidth}
    \centering
    \includegraphics[width=0.7\linewidth, trim={2cm 8.8cm 2cm 10.6cm}, clip]{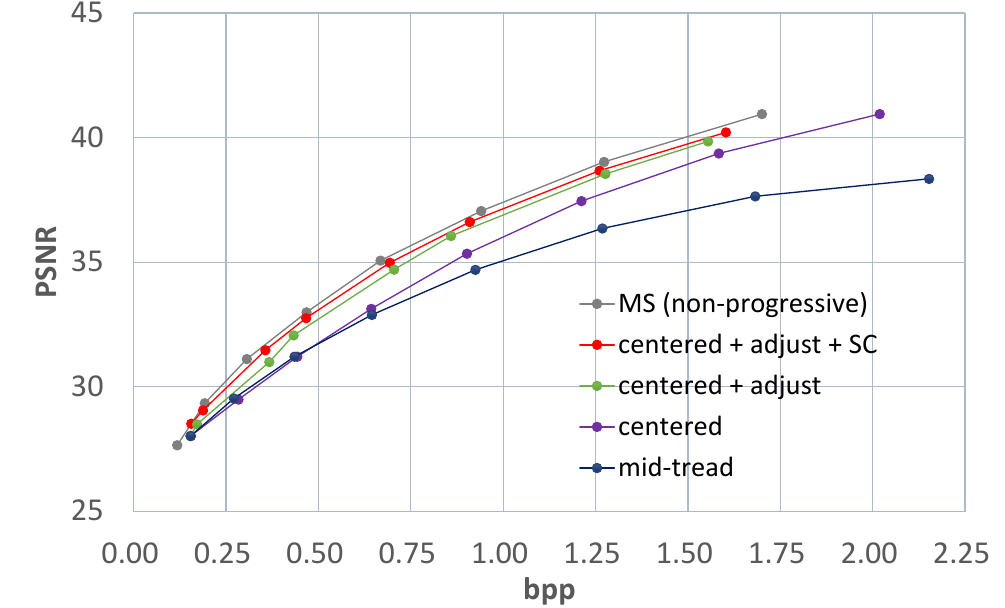}
    \vspace{-0.3cm}\caption{Changes in compression efficiency with the addition of proposed elements on the Mean-scale~\cite{Minnen2018} architecture for the Kodak~\cite{KODAK} dataset. The boundary calculation centering around the lower-layer reconstructions, boundary adjustment, and selective compression are denoted as "centered", "adjust", and "SC", respectively.}
    \label{fig:ablation}
\end{minipage}
\hfill
\begin{minipage}[t]{0.35\linewidth}
    \centering
    \includegraphics[width=\linewidth,clip, trim={1.8cm 7.8cm 1.4cm 8.2cm}]{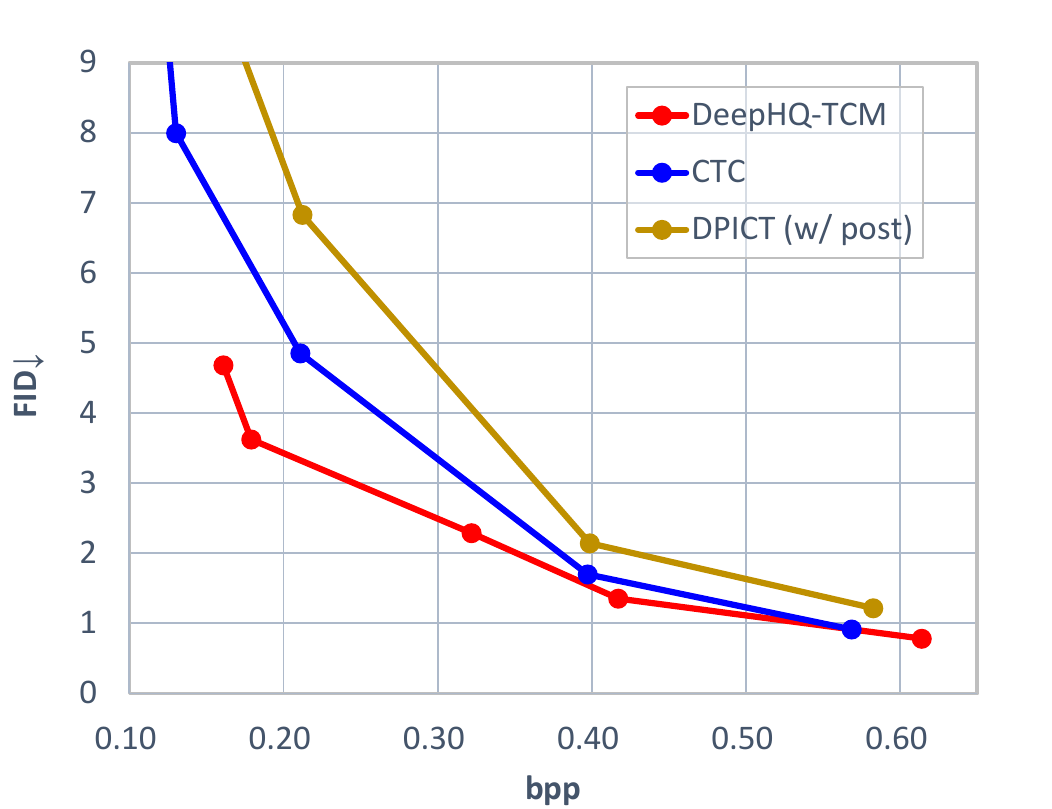}
    \vspace{-0.7cm}\caption{R-D curves of our DeepHQ-TCM, CTC~\cite{2023_CVPR_jeon}, and DPICT~\cite{Lee_2022_CVPR} on the DIV2K dataset (900 images), showing bits-per-pixel (bpp) versus FID.}
    \label{fig:fid_results}
\end{minipage}
\vspace{-0.4cm}
\end{figure}

\begin{figure}[t!]
    \captionsetup[subfigure]{labelformat=empty}
    \captionsetup[subfigure]{font=scriptsize,labelfont=scriptsize,aboveskip=2pt,belowskip=0pt}
    \setlength\columnsep{0pt}
    \captionsetup{belowskip=2pt}
    \begin{subfigure}{0.32\textwidth}
        \centering
        \includegraphics[width=\linewidth, clip, trim={1.7cm 7.5cm 1.7cm 9cm}]{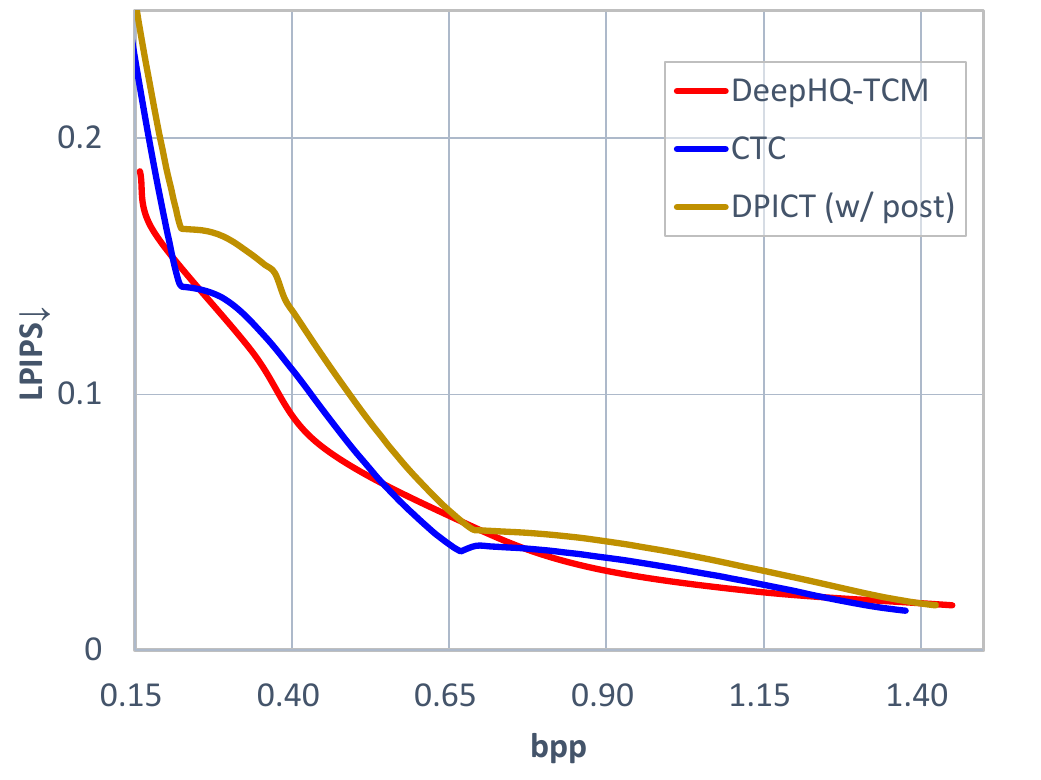}
        \caption{\hspace{0.3cm}(a) \textit{Kodak}}
    \end{subfigure}
    \begin{subfigure}{0.32\textwidth}
        \centering
        \includegraphics[width=\linewidth, clip, trim={1.7cm 7.5cm 1.7cm 9cm}]{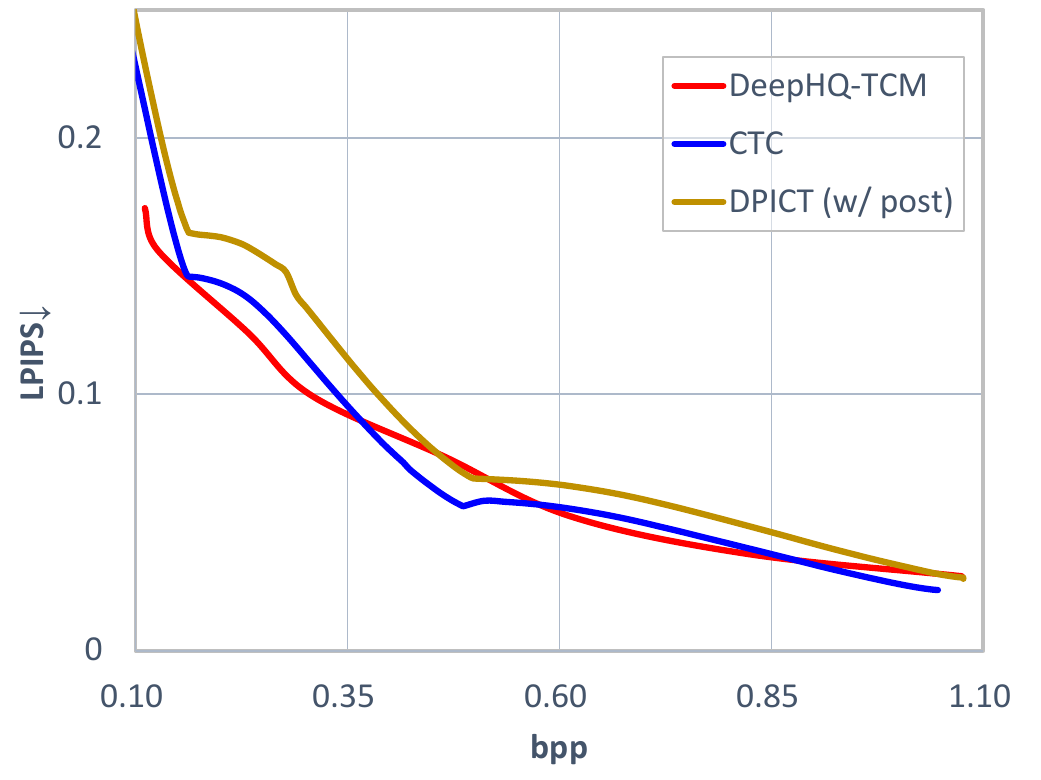}
        \caption{\hspace{0.3cm}(b) \textit{CLIC}}
    \end{subfigure}
    \begin{subfigure}{0.32\textwidth}
        \centering
        \includegraphics[width=\linewidth, clip, trim={1.7cm 7.5cm 1.7cm 9cm}]{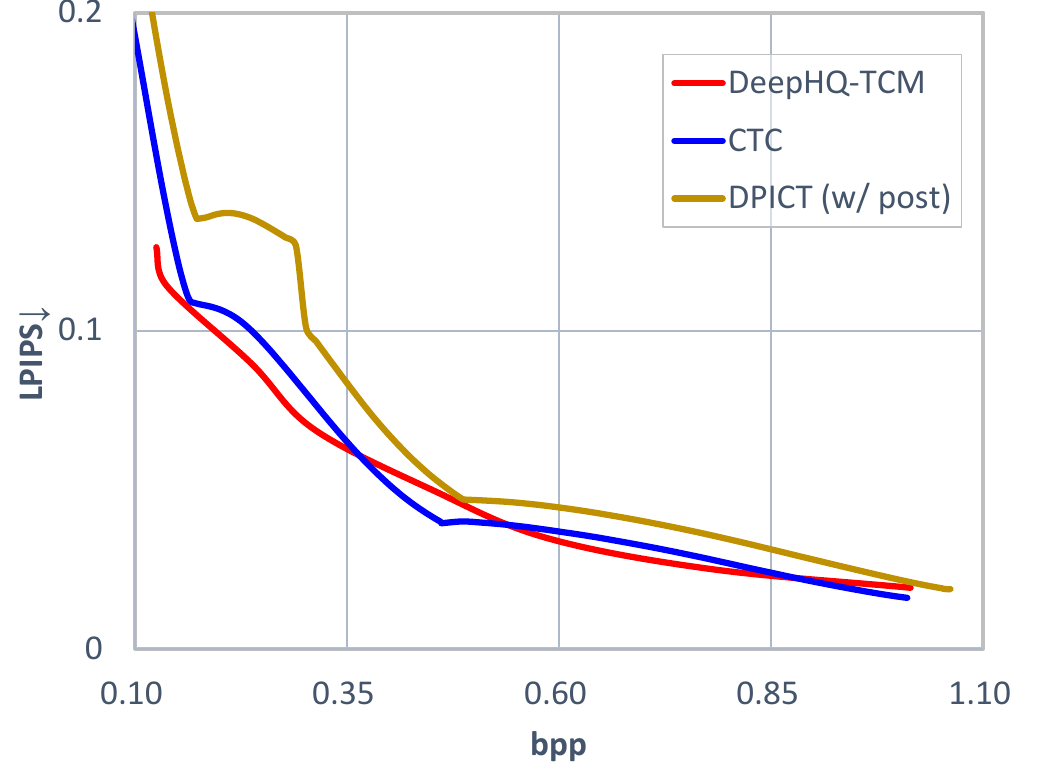}
        \caption{\hspace{0.3cm}(c) \textit{Tecnick}}
    \end{subfigure}
    \vspace{-0.3cm}\caption{R-D curves of our DeepHQ-TCM, CTC~\cite{2023_CVPR_jeon}, and DPICT~\cite{Lee_2022_CVPR}, showing bits-per-pixel (bpp) versus LPIPS.}
    \label{fig:lpips_results}
    \vspace{-0.5cm}
\end{figure}

\subsection{Experimental results}
\label{sec:quantitative_results}
Fig.~\ref{fig:curves} shows the R-D curves of the various models, and Table~\ref{tab:experimental_results} summarizes the test results in coding efficiency and complexity. Our DeepHQ-MS model provides superior coding efficiency compared to most competing models, utilizing the smallest model size and achieving the fastest decoding speed. Similarly, the DeepHQ-TCM model achieves, on average, 11.97\% higher rate savings than the current state-of-the-art (SOTA) method, CTC~\cite{2023_CVPR_jeon}, with significantly fewer (only 14.19\%) parameters and faster (avg. 11.47$\times$) decoding time. It should also be noted that the DeepHQ-MS (w/o SC) model, which only utilizes the learned quantization step sizes without selective compression, significantly outperforms DPICT\_MS implemented on the same base compression architecture.
This demonstrates that our quantization hierarchy utilizing learned step sizes offers significant advantages in terms of coding efficiency compared to the existing handcrafted quantization hierarchy. Also, when comparing our DeepHQ-MS models with and without selective compression, the selective compression significantly further improves both coding efficiency and decoding time. Fig.~\ref{fig:visual_test_results} shows a few reconstruction examples of various models.

The proposed DeepHQ-TCM model achieves larger performance gains over CTC~\cite{2023_CVPR_jeon} on the CLIC and Tecnick datasets (15.09\% and 12.93\%, respectively) compared to its gain on the Kodak dataset (7.88\%). This result may be owing to the use of Swin transformer-based attention (SWAtten) modules in TCM~\cite{liu2023tcm}, the base architecture of DeepHQ-TCM. Since the Swin Transformer~\cite{liu2021swin} architecture can capture and leverage a spatially broader context compared to CNNs, DeepHQ-TCM is likely to achieve greater compression efficiency for datasets with larger images, such as CLIC (average resolution: 1789.4$\times$1189.3) and Tecnick (1200$\times$1200), than for the Kodak dataset (768$\times$512 or 512$\times$768).

For the decoding time versus coding efficiency, our DeepHQ models outperform other methods, as shown in Fig.~\ref{fig:decoding_time}. Owing to the selective compression and minimal computational overhead, our DeepHQ models significantly lower decoding time compared to the other models of similar coding efficiency. In addition, as shown in Fig.~\ref{fig:param_ratesavings}, our DeepHQ utilizes significantly fewer model parameters than the other models while providing similar coding efficiency. In Figs.~\ref{fig:decoding_time} and \ref{fig:param_ratesavings}, we highlight the difference in performance between our DeepHQ-TCM model and the current SOTA method, CTC~\cite{2023_CVPR_jeon}, with the red dotted arrows. In addition, we also emphasize the difference in performance between our DeepHQ-MS model and the DPICT\_MS model with the blue dotted arrows because they can represent how our proposed technical components can improve the progressive coding performance on the same base compression architecture. Particularly, as mentioned in Section~\ref{sec:relatedwork}, the current SOTA method, CTC~\cite{2023_CVPR_jeon}, divides the entire bitrate range into three segments and utilizes two kinds of dedicated refinement networks (CRR and CDR) for each bitrate range, resulting in an excessively large model and significantly slower decoding times.
Similarly, the DPICT~\cite{Lee_2022_CVPR} (w/ post-processing) model, with two dedicated post-processing networks trained separately in their target bitrate ranges, significantly increases the model size. In contrast, our DeepHQ uses only a single model for the entire bitrate range, thus avoiding a model size explosion and keeping our model size similar to that of the base compression model. When using the Mean-scale~\cite{Minnen2018} model as the base compression model, our DeepHQ-MS causes only 0.59\% of the parameter overhead.

Fig.~\ref{fig:ablation} shows the changes in coding efficiency by adding the proposed elements: (i) the symmetric construction of the subinterval boundaries with the lower-layer reconstructions centered (Sec.~\ref{sec:boundary_calculation}), (ii) the boundary adjustment (Sec.~\ref{sec:boundary_calculation}), and (iii) the selective compression (Sec.~\ref{sec:selective}), which together yield significant coding gains.

\begin{figure*}[!t]
    \captionsetup[subfigure]{labelformat=empty}
    \captionsetup[subfigure]{font=scriptsize, labelfont=scriptsize}
    \setlength\columnsep{0pt}
    \captionsetup{belowskip=20pt}
    \centering

    % First row (labels)
    \begin{subfigure}[b]{0.02\textwidth}
        \caption{}
        \vspace{-0.7cm}
    \end{subfigure}
    \begin{subfigure}[b]{0.22\textwidth}
        \centering
        \caption{Reconstruction $\bm{x'}_l$}
        \vspace{-0.7cm}
    \end{subfigure}
    \begin{subfigure}[b]{0.24\textwidth}
        \centering
        \caption{MSQE of $\bm{\breve y}^\textit{final}_l$ (log scale)}
        \vspace{-0.7cm}
    \end{subfigure}
    \begin{subfigure}[b]{0.24\textwidth}
        \centering
        \caption{Selected $\bm{y}$ elements}
        \vspace{-0.7cm}
    \end{subfigure}
    \begin{subfigure}[b]{0.24\textwidth}
        \centering
        \caption{Estimated additional bit rate}
        \vspace{-0.7cm}
    \end{subfigure}
    
    % Layer 1
    \begin{subfigure}[b]{0.02\textwidth}
        \rotatebox{90}{\makebox[2.8cm]{Layer 1}}
    \end{subfigure}
    \begin{subfigure}[b]{0.22\textwidth}
        \centering
        \includegraphics[width=0.97\linewidth, trim=0 -2.2cm 0 0]{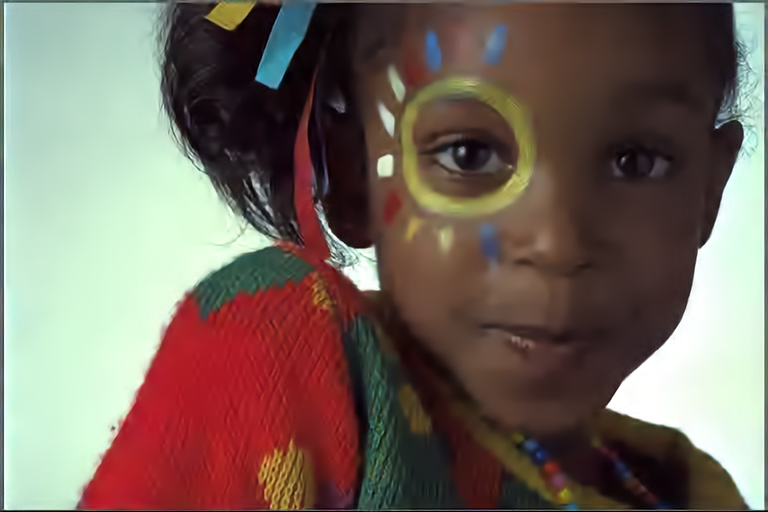}
        % \vspace{-1.4cm}
    \end{subfigure}
    \begin{subfigure}[b]{0.24\textwidth}
        \centering
        \includegraphics[width=\linewidth]{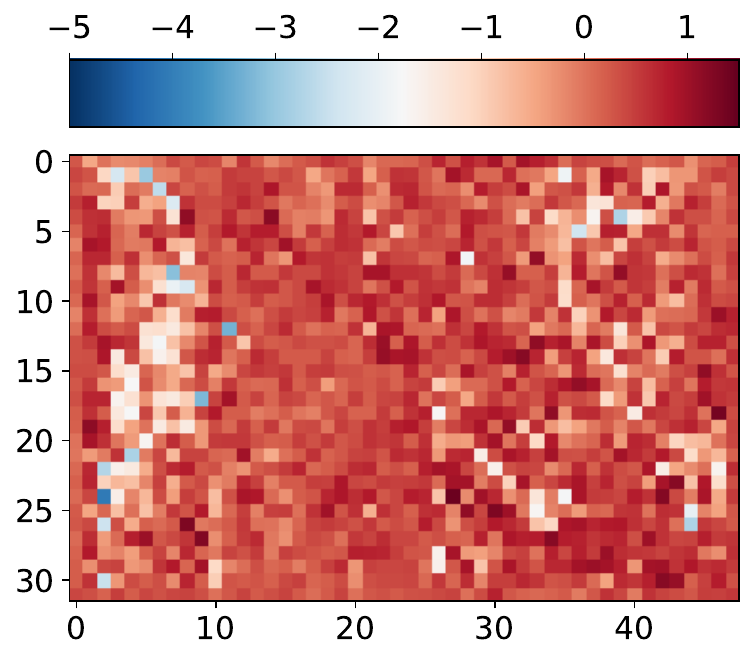}
        % \vspace{-1.8cm}
    \end{subfigure}
    \begin{subfigure}[b]{0.24\textwidth}
        \centering
        \includegraphics[width=\linewidth]{figures/overall_visual_results/Sum_selection_Q1_I0.3_mmseFalse.pdf}
        % \vspace{-1.8cm}
    \end{subfigure}
    \begin{subfigure}[b]{0.24\textwidth}
        \centering
        \includegraphics[width=\linewidth]{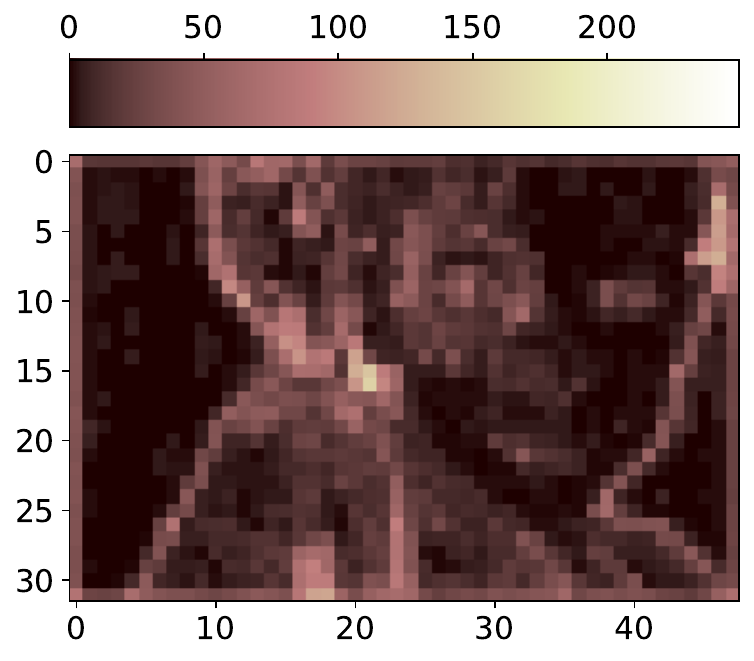}
        % \vspace{-1.8cm}
    \end{subfigure}

    \vspace{-0.4cm}
    
    % Layer 3
    \begin{subfigure}[b]{0.02\textwidth}
        \rotatebox{90}{\makebox[2.8cm]{Layer 3}}
    \end{subfigure}
    \begin{subfigure}[b]{0.22\textwidth}
        \centering
        \includegraphics[width=0.97\linewidth, trim=0 -2.2cm 0 0]{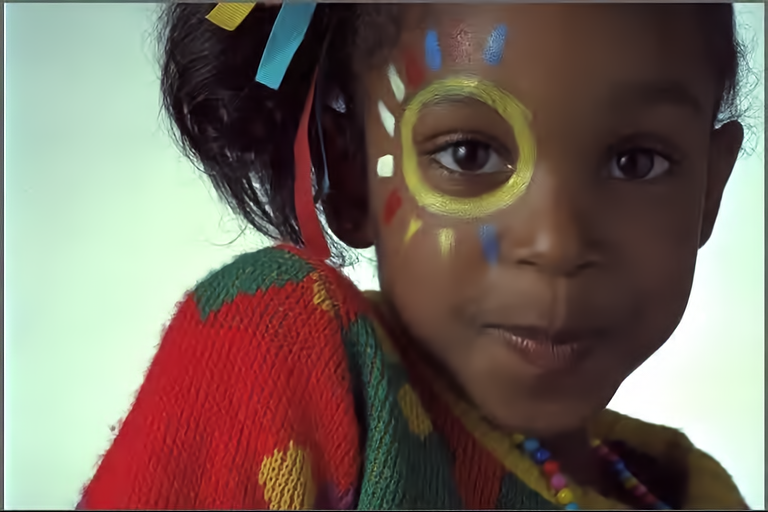}
        % \vspace{-1.4cm}
    \end{subfigure}
    \begin{subfigure}[b]{0.24\textwidth}
        \centering
        \includegraphics[width=\linewidth]{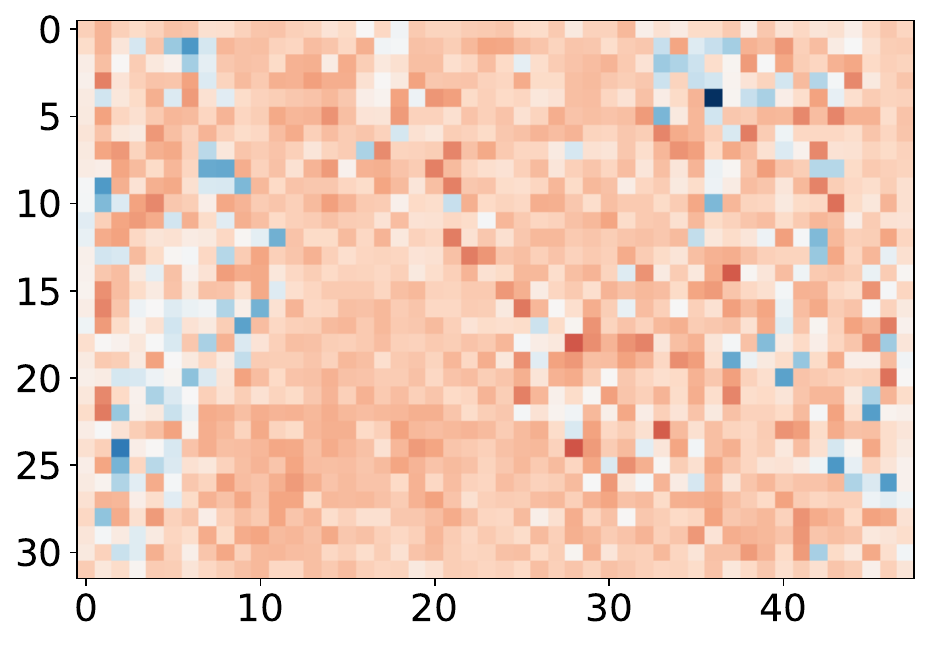}
        % \vspace{-1.8cm}
    \end{subfigure}
    \begin{subfigure}[b]{0.24\textwidth}
        \centering
        \includegraphics[width=\linewidth]{figures/overall_visual_results/Sum_selection_Q3_I0.3_mmseFalse.pdf}
        % \vspace{-1.8cm}
    \end{subfigure}
    \begin{subfigure}[b]{0.24\textwidth}
        \centering
        \includegraphics[width=\linewidth]{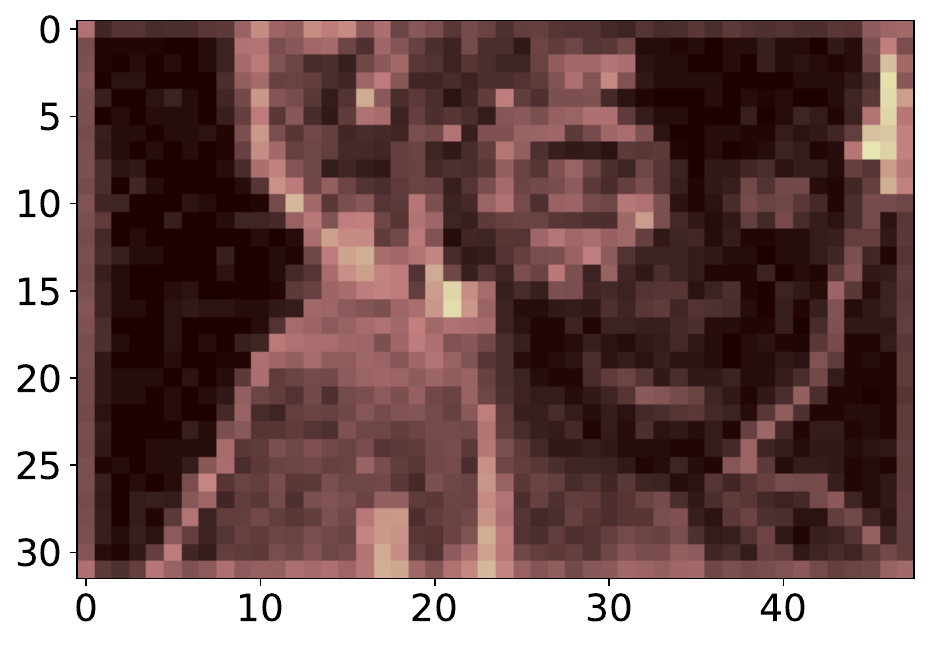}
        % \vspace{-1.8cm}
    \end{subfigure}
    \vspace{-0.4cm}
    
    % Layer 5
    \begin{subfigure}[b]{0.02\textwidth}
        \rotatebox{90}{\makebox[2.8cm]{Layer 5}}
    \end{subfigure}
    \begin{subfigure}[b]{0.22\textwidth}
        \centering
        \includegraphics[width=0.97\linewidth, trim=0 -2.2cm 0 0]{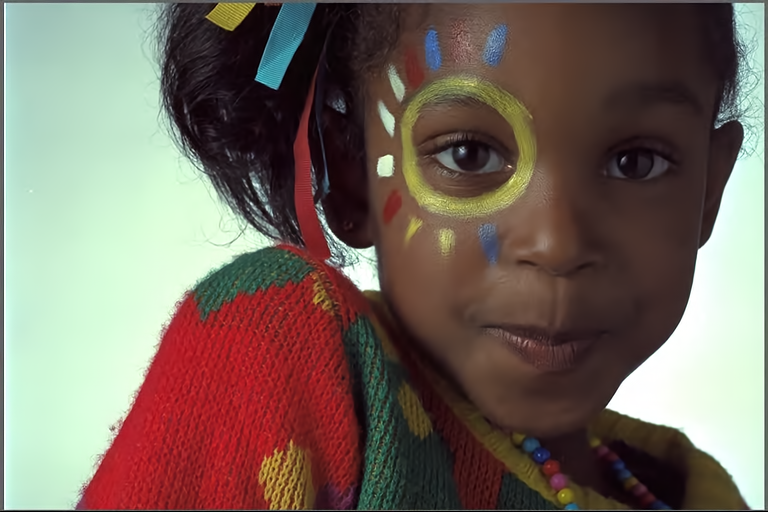}
        % \vspace{-0.7cm}
    \end{subfigure}
    \begin{subfigure}[b]{0.24\textwidth}
        \centering
        \includegraphics[width=\linewidth]{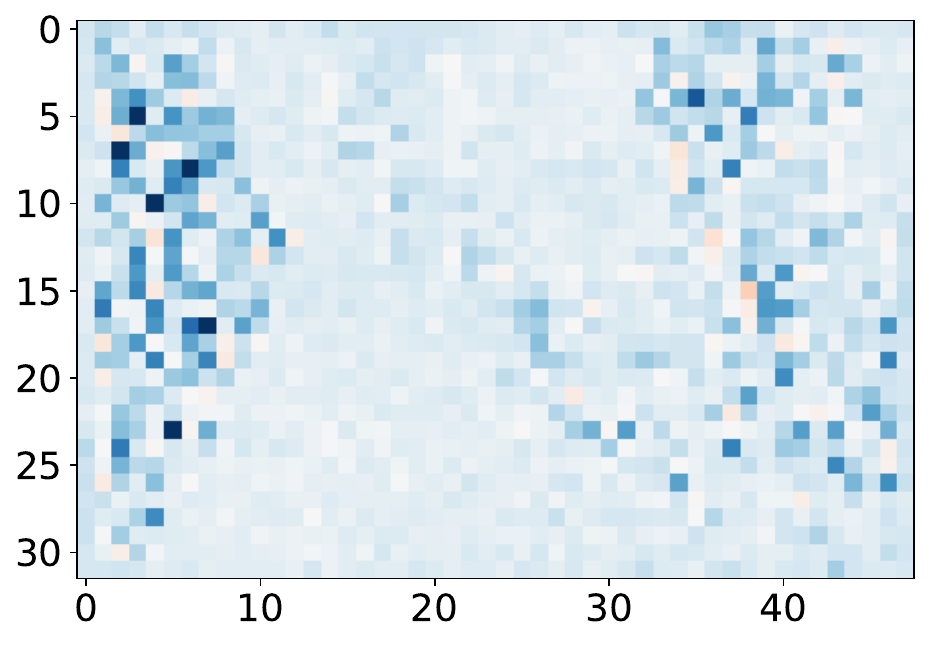}
        % \vspace{-1.1cm}
    \end{subfigure}
    \begin{subfigure}[b]{0.24\textwidth}
        \centering
        \includegraphics[width=\linewidth]{figures/overall_visual_results/Sum_selection_Q5_I0.3_mmseFalse.pdf}
        % \vspace{-1.1cm}
    \end{subfigure}
    \begin{subfigure}[b]{0.24\textwidth}
        \centering
        \includegraphics[width=\linewidth]{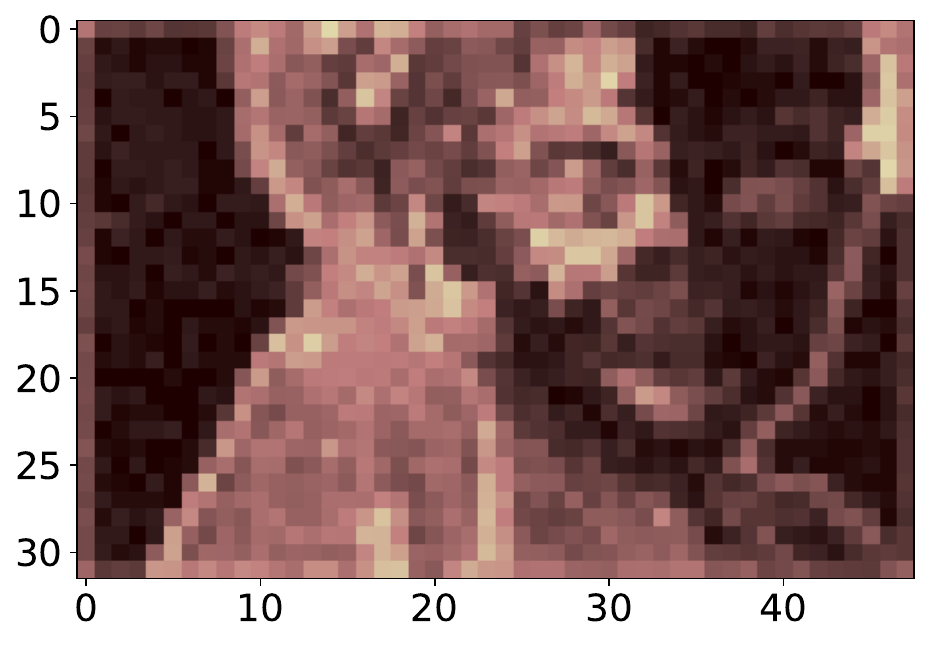}
        % \vspace{-1.1cm}
    \end{subfigure}
    \vspace{-0.4cm}
    
    % Layer 7
    \begin{subfigure}[b]{0.02\textwidth}
        \rotatebox{90}{\makebox[2.8cm]{Layer 7}}
    \end{subfigure}
    \begin{subfigure}[b]{0.22\textwidth}
        \centering
        \includegraphics[width=0.97\linewidth, trim=0 -2.2cm 0 0]{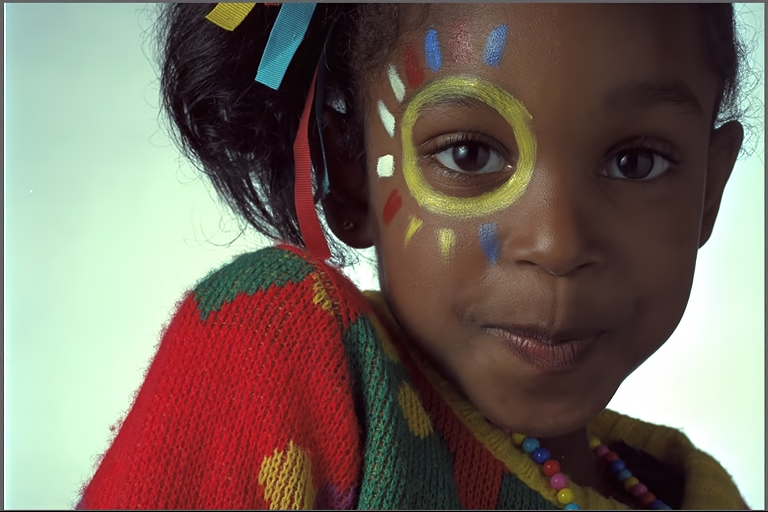}
    \end{subfigure}
    \begin{subfigure}[b]{0.24\textwidth}
        \centering
        \includegraphics[width=\linewidth]{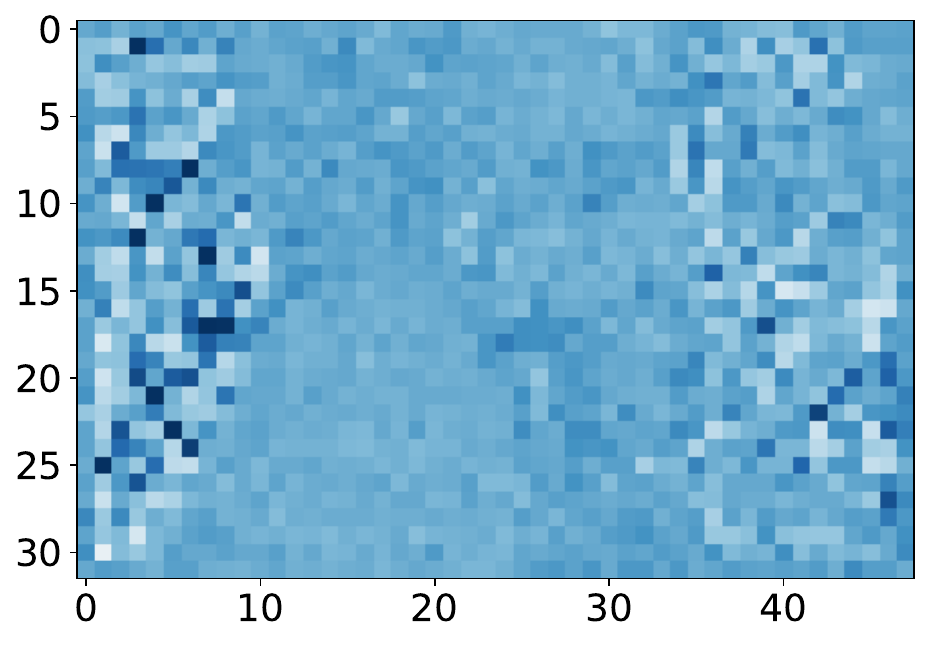}
    \end{subfigure}
    \begin{subfigure}[b]{0.24\textwidth}
        \centering
        \includegraphics[width=\linewidth]{figures/overall_visual_results/Sum_selection_Q7_I0.3_mmseFalse.pdf}
    \end{subfigure}
    \begin{subfigure}[b]{0.24\textwidth}
        \centering
        \includegraphics[width=\linewidth]{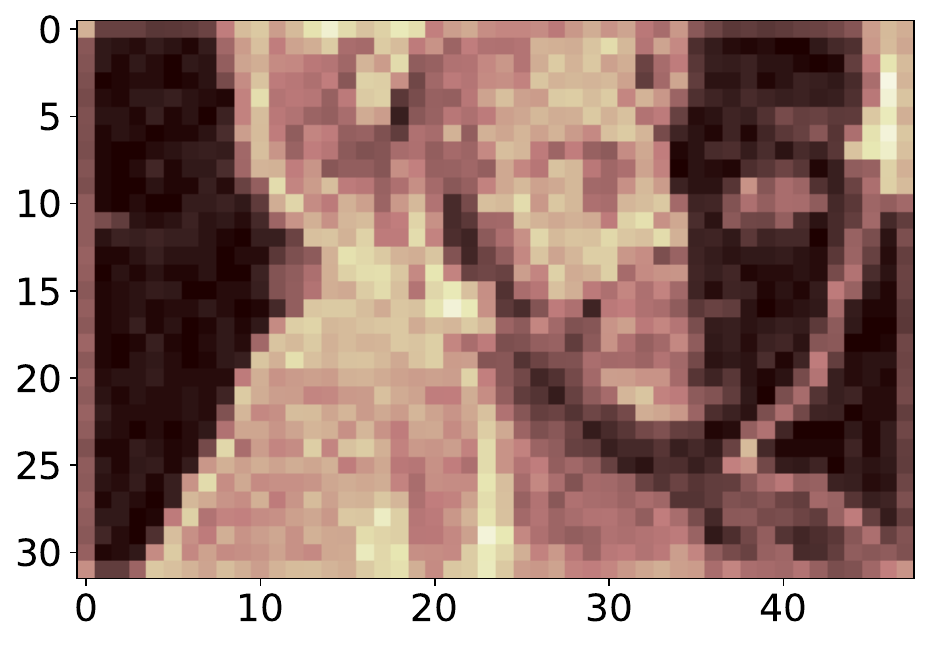}
    \end{subfigure}
    \vspace{-0.3cm}\caption{Examples of log-scale mean squared quantization errors (MSQE), selected representation elements, and estimated additional bit rates ($\minus \log_2 P$) for each quantization layer. Only 4 layers out of the total 8 layers are shown for brevity.}
    \label{fig:overall_visual_results}
    \vspace{-0.7cm}
\end{figure*}

%FID LPIPS
When evaluated using the FID metric, as shown in Fig.~\ref{fig:fid_results}, the proposed DeepHQ-TCM model achieves a 16.04\% performance gain over the CTC~\cite{2023_CVPR_jeon} model. In terms of LPIPS, as shown in Fig.~\ref{fig:lpips_results}, the proposed DeepHQ-TCM model attains rate savings of 4.33\%, 4.59\%, and 5.74\% over the CTC model on the Kodak, CLIC, and Tecnick datasets, respectively. Although the CTC model yields better results at certain rates in the LPIPS experiments, the proposed DeepHQ provides higher coding efficiency across most of the rate range.
Fig.~\ref{fig:overall_visual_results} shows an example illustrating how the quantization errors, the number of selected representation components, and the layer-wise bit rate consumption change as the number of quantization layers increases. The quantization errors are gradually reduced in response to the increase in the number of the quantization layers. The selected representation components increase with the growth of quantization layers, and the areas of greater complexity in the image tend to involve more representation components. The bit consumption for lower-quality compression is primarily allocated around contours. In contrast, as the compression quality improves, the portion of allocated bits tends to shift towards textured regions.

\section{Conclusion}
In this paper, we introduced a learned progressive image compression method, called DeepHQ, based on learned quantization step sizes for each quantization layer. Additionally, we proposed an extended selective compression method that compresses only the essential representation elements for each quantization layer, thus further improving the compression efficiency. Our DeepHQ achieves significantly higher coding efficiency than the best state-of-the-art performance in the learned progressive coding research field with significantly fewer model parameters and within a much faster decoding time. In addition, our DeepHQ can stably support the fine-grained component-wise progressive coding. In future work, we will study a method that fully and jointly trains our hierarchical quantization scheme within the entire model.

\begin{acks}
This work was supported by an internal fund/grant of Electronics and Telecommunications Research Institute (ETRI). [	
25YC1100, Development of fundamental technology for next-generation media coding and transmission standards]
\end{acks}

\bibliographystyle{ACM-Reference-Format}
\bibliography{egbib}

\appendix
\begin{figure}[!t]
\captionsetup{aboveskip=2pt}
\centering
\begin{minipage}[t]{0.45\linewidth}
    \centering
    \includegraphics[width=0.9\linewidth,clip, trim={2.1cm 8.1cm 2.1cm 9.2cm}]{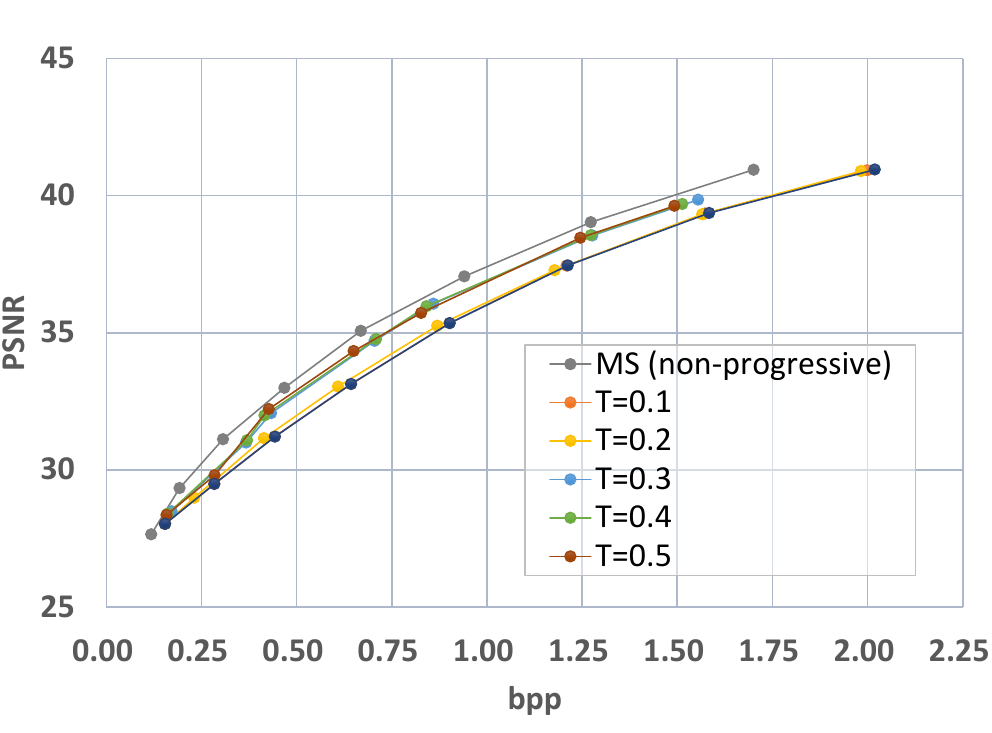}
    \caption{R-D curves with different threshold $T$ values that determine whether the boundary adjustment is performed.}
    \label{fig:T_comparison}
\end{minipage}
\hfill
\begin{minipage}[t]{0.51\linewidth}
    \centering
    \includegraphics[width=0.9\linewidth,clip, trim={0.0cm 0.1cm 0.0cm 0.2cm}]{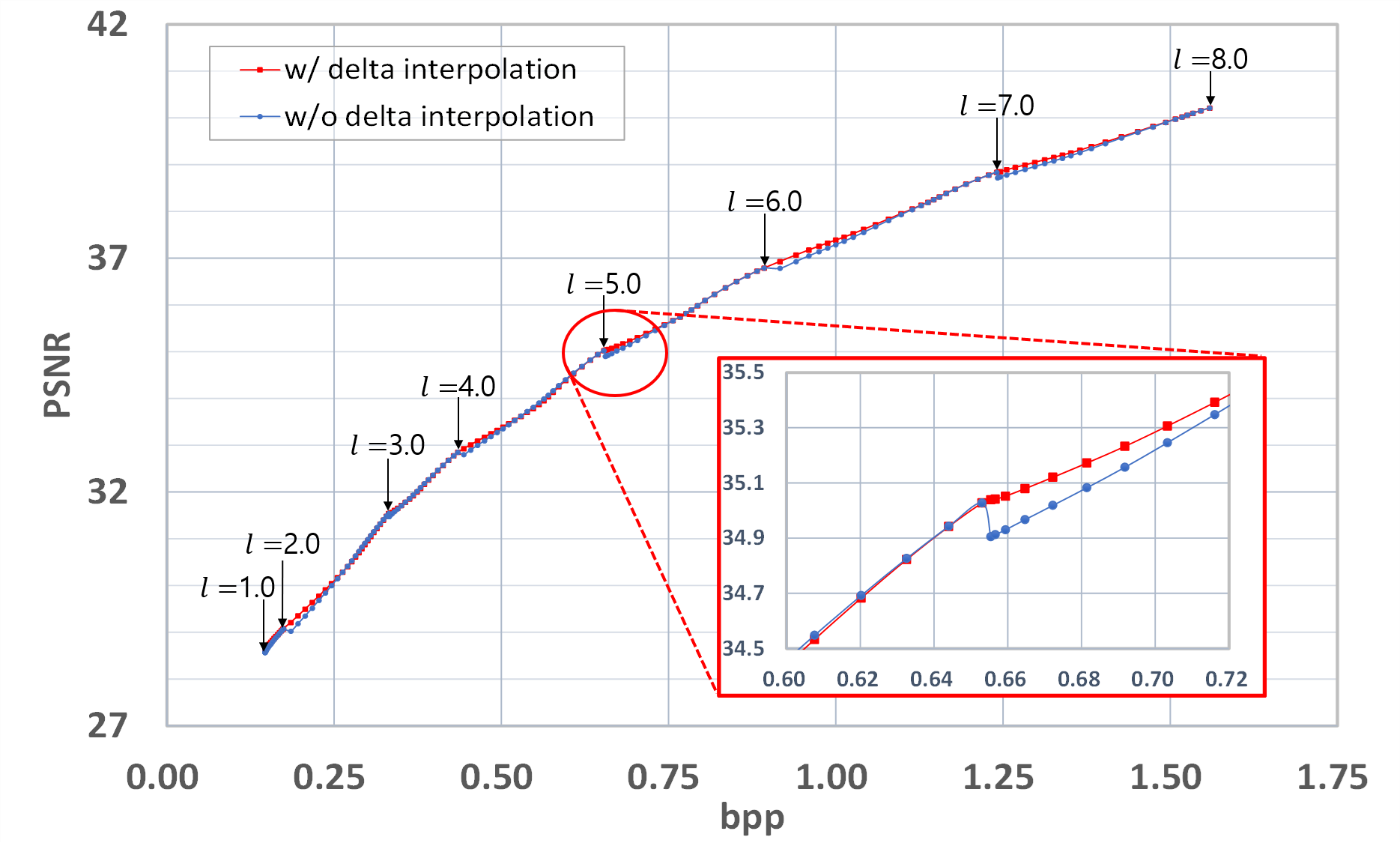}
    \caption{Progressive coding performance with and without the non-linear interpolation (Eq.~\ref{eq:delta_interpolation}) for $\bm{\Updelta}_l$ and $\bm{\Updelta}^\text{inv}_l$.}
    \label{fig:comparison_delta_interpolation}
\end{minipage}
\vspace{-0.2cm}
\end{figure}
\section{Test results on various threshold values}
\label{sec:threshold}
The proposed DeepHQ adjusts the subinterval boundaries when the ratios of the interval fragments compared to the learned quantization step $\bm{\Updelta}_l$ are below a certain threshold $T$, as we discussed in Sec.~\ref{sec:boundary_calculation}. To determine a suitable threshold $T$, we performed experiments with various values of $T$, ranging from 0.1 to 0.5. As shown in Fig~\ref{fig:T_comparison}, As shown in Fig.~\ref{fig:T_comparison}, experiments with low $T$ values exhibited poor coding efficiency, while those with high $T$ values resulted in some instability. Therefore, we selected $T=0.3$ as the optimal value.

\section{Implementation details on the asymmetric inverse scaling scheme}
\label{sec:asymmetric}
As we introduced in Sec.~\ref{sec:implementation}, we adopt an asymmetric inverse scaling scheme~\cite{Cui2021}. In this section, we provide further details on the implementation of the asymmetric scheme in our DeepHQ. As mentioned in Sec.~\ref{sec:implementation}, this asymmetric inverse scaling scheme~\cite{Cui2021} does not affect the quantization process $Q(\cdot)$, which quantizes $\bm{y}^*$ into an interval index $\bm k$, nor the dequantization process $DQ(\cdot)$, which dequantizes $\bm k$ into $\bm{\breve{y}}^*_l$. The asymmetric inverse scaling scheme~\cite{Cui2021} is only involved in the process of transforming $\bm{\breve{y}}^*_l$ into $\bm{\breve{y}}^{\textit{final}}_l$, the final input to the decoder, as follows:
\begin{align}
\label{eq:y_breve_final_appendix}
\bm{x'}_l = De(\bm{\breve y}^\textit{final}_l), \text{  with  } \bm{\breve y}^\textit{final}_l = (\bm{\breve y}^*_l + \bm \mu) / \bm{\Updelta}_l * \bm{\Updelta}^\text{inv}_l,
\end{align}
where $\bm{x'}_l$ is a reconstruction image of the \textit{l}-th quantization layer; $De(\cdot)$ is the decoding transform function (via the decoder network). Meanwhile, $\bm{\Updelta}_l$ and $\bm{\Updelta}^\text{inv}_l$ are the two asymmetric sets for the \textit{l}-th quantization layer. Note that Eq.~\ref{eq:y_breve_final_appendix} corresponds to Eq.~\ref{eq:y_breve_final} in Sec.~\ref{sec:hierarchical_quantization}. In addition, for Eq.~\ref{eq:y_breve_final_appendix}, to address the intermediate level between two discrete quantization layers in the component-wise coding introduced in Sec.~\ref{sec:component-wise}, we adopt a non-linear interpolation of the $\bm{\Updelta}_l$ ($\bm{\Updelta}^\text{inv}_l$) sets between two consecutive quantization layers, as follows:
\begin{equation}
\label{eq:delta_interpolation}
\bm{\Updelta}_l=
  {\bm{\Updelta}_{\floor l}^ {1-(l-{\floor l})}}  \cdot {\bm{\Updelta}_{\ceil l}^ {l-{\floor l}}},
\end{equation}
where $\floor \cdot$ and $\ceil \cdot$ denote the floor and ceiling operations, respectively. For instance with $l{=}3.3$, $\bm{\Updelta}_{3.3}$ is calculated by the element-wise multiplication of $\bm{\Updelta}_{3.0}^{0.7}$ and $\bm{\Updelta}_{4.0}^{0.3}$. Note that $\bm{\Updelta}^\text{inv}_l$ is determined in the same manner.
Fig.~\ref{fig:comparison_delta_interpolation} shows the comparison results between our DeepHQ models with and without the non-linear interpolation. In the ablated model, we use $\bm{\Updelta}_{\ceil l}$ and $\bm{\Updelta}^\text{inv}_{\ceil l}$.
In addition, to train the $\bm{\Updelta}^\text{inv}_l$ along with $\bm{\Updelta}_l$, we adjust the distortion loss in Eq.~\ref{eq:distortion_loss}, as follows:
\begin{align}
\label{eq:distortion_loss_appendix}
D_l = \text{MSE}(\bm x, \bm{x'}_l), \text{\ \ \ \ \ with  } \bm{x'}_l = De(Re(\bm{\tilde y}^\Updelta_l,m(\bm{\hat z}, l)) \cdot \bm{\Updelta}^\text{inv}_l).
\end{align}
In Eq.~\ref{eq:distortion_loss}, The $\bm{\Updelta}_l$ is used for rescaling, whereas, in Eq.~\ref{eq:distortion_loss_appendix}, we use the $\bm{\Updelta}^\text{inv}_l$ to enable the asymmetric inverse scaling.

\end{document}